\documentclass[preprint,12pt]{elsarticle}

\usepackage{graphics}
\usepackage{graphicx}
\usepackage{amssymb}
\usepackage{amsthm}
\usepackage{amsmath}                         %dodatkowe symbole matemat. np. kroje
\usepackage{amsbsy}
\usepackage{amsfonts}
\usepackage{placeins}
\usepackage{bbding}

\newcommand{\bdm}{\begin{displaymath}}
\newcommand{\edm}{\end{displaymath}}
\newcommand{\be}{\begin{equation}}
\newcommand{\ee}{\end{equation}}
\newcommand{\ba}{\begin{eqnarray}}
\newcommand{\ea}{\end{eqnarray}}
\newcommand{\efig}{ \end{figure}}

\journal{Physica A}

\begin{document}

\begin{frontmatter}

%% Title, authors and addresses

%% use the tnoteref command within \title for footnotes;
%% use the tnotetext command for the associated footnote;
%% use the fnref command within \author or \address for footnotes;
%% use the fntext command for the associated footnote;
%% use the corref command within \author for corresponding author footnotes;
%% use the cortext command for the associated footnote;
%% use the ead command for the email address,
%% and the form \ead[url] for the home page:
%%
%% \title{Title\tnoteref{label1}}
%% \tnotetext[label1]{}
%% \author{Name\corref{cor1}\fnref{label2}}
%% \ead{email address}
%% \ead[url]{home page}
%% \fntext[label2]{}
%% \cortext[cor1]{}
%% \address{Address\fnref{label3}}
%% \fntext[label3]{}

\title{Mesoscopic approach to minority games in herd regime \thanks{Results presented in this paper were obtained using computational grid built in the framework of the project INFO-RI-222667 {\it Enabling grids for E-science}, funded by the European Commission and Polish Ministry of Science and Higher Education in the 7th Framework Program.} }

\author[icm,ipj]{Karol Wawrzyniak\corref{cor1}} \ead{kwawrzyn@fuw.edu.pl} \ead[url]{http://agf.statsolutions.eu} \cortext[cor1]{corresponding author}
\author[ipj,icm]{Wojciech Wi\'slicki} \ead{wislicki@fuw.edu.pl}

\address[icm]{Interdisciplinary Centre for Mathematical and Computational Modelling, University of Warsaw, Pawi´nskiego 5A, PL-02-106 Warszawa}
\address[ipj]{National Centre for Nuclear Research, Ho\.za 69, PL-00-681 Warszawa, Poland}

\begin{abstract}
We study minority games in efficient regime. By incorporating the utility function and
aggregating agents with similar strategies we develop an effective mesoscale notion of state of the game.
Using this approach, the game can be represented as a Markov process with
substantially reduced number of states with explicitly computable probabilities.
For any payoff, the
finiteness of the number of states is proved. Interesting features of an extensive
random variable, called aggregated demand, viz. its strong inhomogeneity and presence of patterns
in time, can be easily interpreted. Using Markov theory and
quenched disorder approach, we can explain important macroscopic characteristics of the game:
behavior of variance per capita and predictability of the aggregated demand. We prove that in
case of linear payoff many attractors in the state space are possible.
\end{abstract}

\begin{keyword}
%% keywords here, in the form: keyword \sep keyword
Minority game, adaptive system, Markov process, mesoscopic scale
%% MSC codes here, in the form: \MSC code \sep code
%% or \MSC[2008] code \sep code (2000 is the default)
\end{keyword}

\end{frontmatter}

%%
%% Start line numbering here if you want
%%
% \linenumbers
\section{Introduction}\label{sec:intro}
Evolution of complex systems capable to adapt to varying environments by using shared memory is
often considered as one of the fundamental dynamical problems in sciences.
But large numbers of parameters {\it a priori} needed to describe them render difficult their exact analytic
treatments.
More efficient approaches are based on computational methods and direct modelling of adaptive systems
with populations of agents.
In course of development of these models it has been soon realized that even simplified approach
with no communication between individuals, where the only dependence between them
is given by common memory resource, appears to be useful and interesting.
Among variety of multi-agent models, the minority game
(MG) provides with a particularly intuitive representation of self-adaption where individuals
reason out inductively and their rationality is limited. The MG was originally designed in
Ref.~\cite{arthur94AmerEconRev84} to account for profitability of playing in opposite to the
plurality of decision makers. The model has been subsequently formalized in
Refs.~\cite{challet97PhysicaA246,challet98PhysicaA256} and became a well-established area of
the game-theoretical, dynamical and statistical research~\cite{challet05minority}.

The MG is a typical bottom-up construct and therefore usual definitions of the game first specify
rules of behaviour for individuals. Then, piecing together microscopic variables, one defines
higher-order quantities characterizing grander systems. In some cases, however, other descriptions
are also possible, e.g. functions of state like score functions can be attributed to groups of
agents without specifying agents individually~\cite{jefferies01PhysRevE65}. And again, despite an apparent simplicity of
basic rules of taking decisions by agents, adaptive abilities and phenomenology of populations
playing MGs appear to be surprisingly non-trivial~\cite{challet05minority,coolen05OxfordPress}.
As shown in
Refs.~\cite{challet97PhysicaA246,savit99PhysRevLetters82}, phenomenology of MG depends
qualitatively on game parameters.
For example, the macroscopic quantity called \emph{aggregate
attendance}, or {\emph aggregate demand}, pooling together individual choices, identifies three regimes of the MG:
the random, cooperation and herd. After the authors of
Refs.~\cite{savit99PhysRevLetters82,cara99EurPhysJB10}, the latter case is also called {\it
efficient}, because the total number of strategies is small, compared to the number of agents, and
players have access to all available information.
In addition, in this exceptional case the relatively small number of parameters enables analytical solutions.

The very first attempt of solving MG analytically was based on the method of statistical mechanics called the \emph{replica analysis}.
In order to find a more detailed analogy between statistical physics and MG, the group of Challet and
Zhang~\cite{challet00PhysRevLetters84,marsili00PhysicaA280} limits their analysis to only two
strategies per agent where the manifestation of cooperative effects is the strongest. The agents'
choice is then treated analogously to the projection of the particle's spin on a quantization axis in space.
The aggregate demand is split into two
terms: the deterministic, forced by the quenched disorder, and a stochastic one that is further
neglected. The quenched disorder term is related to systems in statistical physics when some
parameters defining system's behavior are stationary random variables, chosen when the system is created.
Even such a simplified approach led to quite accurate analytical
solutions for variance per capita as a function of the control parameter in the random regime.
Additionally, the authors of Ref.~\cite{marsili01PhysRevE64} showed that properties of the MG in
the symmetric phase depend on the initial conditions, what was confirmed numerically in
Ref.~\cite{garrahan00PhysRevE62}. If the initial conditions (i.e. the strategies) are drawn randomly, the
system exhibits the so called quenched or frozen disorder. This theory, however,
provides little knowledge on underlying dynamics of the game, i.e. on the evolution of utilities of strategies, and on existence of time
patterns.
An also it does not explain differences of macroscopic observables for different payoffs.

Another analytical approach, based on \emph{generating functional}, is offered by Heimel and Coolen
in Ref.~\cite{heimel01PhysRevE63}. This is the second most used technique, applicable to the
statistical physics and a problem of disordered systems with random interactions. This method is in
principle exact in the limit $N \rightarrow \infty$, although generally more difficult to
apply than the replica procedure. The authors redefine the game for two strategies in such a way
that instead of two independent utility values they operate only on one variable $q$ combining
these two for each agent. As a result, the generalized MG is driven by only three equations, where
the vector $\mathbf{q} = (q_1, \ldots, q_N)$ represents the state, and $N$ is the number of
players. Then, the game is described in terms of the microscopic probability densities
$Pr(\mathbf{q})$, where the discrete-time dynamics is replaced by the continuous-time one. Since
the state depends on $N$, the behavior in the limit $N \rightarrow \infty$ can be examined.
Similarly to the replica analysis, the method does not provide any insight into the game dynamics.

Concurrently, the group of Johnson introduced the so called \emph{crowd-anti- crowd theory}
offering approximate expressions for aggregate demand~\cite{johnson99PhysA269,hart01PhysicaA298}.
Agents act as a \emph{crowd} if they use the same strategy. If there is a group of agents using simultaneously
the strategy anticorrelated to the first one, they make the opposite decisions and are considered as an
\emph{anticrowd}. There exist many different pairs of crowds and anticrowds at the same time. If
sizes of crowd and anticrowd are similar, as it is the case in the cooperation regime, then the
choices of these two groups cancel mutually and the volatility is kept small.
If the crowd dominates, the majority of agents behave in the same manner and the volatility becomes large.
It has been demonstrated that, considering fluctuations of the aggregated demand, analytical results are consistent with the
numerical ones. Following the crowd-anticrowd reasoning, Jefferies \emph{et al.} in
Ref.~\cite{jefferies01PhysRevE65} cast the game into the functional map, which reproduces the game when
iterated. Such approach has a serious advantage compared to heuristically introduced rules in
Refs.~\cite{challet97PhysicaA246,moro00minorityIntroductoryGuide}, since it does not need to keep
track of the labels for individual agents. In the definition of the functional map, those agents who hold
the same combination of strategies are grouped together. In MG, individuals with the same
strategies respond in the same way to all values of the global information set
$\mu=\{0,1,\ldots,P-1\}$ ($P$ standing for the number of possible realizations of the winning decision history $\mu$), provided that the game starts with the same initial utilities for all the
strategies. The grouping is done using the $S$-dimensional tensor, where S is the number of
strategies per agent. Assuming that the Reduced Strategy Space (RSS)~\cite{challet98PhysicaA256} is
used, rows and columns of the tensor are of length $2P$ and each entry is equal to the number of
agents holding a different combination of strategies. The concept of the state that is based on (i)
utilities of pairwise different strategies and, (ii) history of past winning decisions, is
subsequently introduced. Collecting above elements, a set of time-dependent equations, which
reproduce the essential dynamics of the minority game, is written down. The authors figured out
that MG can be interpreted as a stochastically disturbed deterministic system. To simplify the
analysis, the stochastic term is skipped and attention is paid only to the deterministic part of
the game. Then, the game is called the Deterministic MG. In the first studies of dynamics it is
observed that the microscopic dynamics is affected markedly by the choice of the payoff function.
The bahavior of the game is dictated by realization of distribution of agents over strategies and
not just by specific game parameters. Hence, without knowledge about the disorder, the game cannot
be classified to as being in either the efficient or inefficient regime. In
Ref.~\cite{hart01EurPhysJB20}, the dynamical approach is extended to the analysis of stochastic
terms. The achieved analytical results provide correct explanation of variance per capita in herd
regime, provided linear payoff, but no description of dynamics or
predictabilities is given.

There are some similarities between the crowd-anticrowd theory and our \emph{mesoscopic approach}
introduced in Ref.~\cite{wawrzyniak09ACSNo6} and further developed in this article.
We incorporated the same concept of state as in Ref.~\cite{jefferies01EurPhysJB20} for the
step-like payoff. We found however that the linear payoff requires different
definition~\cite{wawrzyniak09ACSNo6}. In the mesoscopic approach we aggregated agents playing the
same strategy into {\it fractions}, and treated the fraction as one player. Such approach allowed
us to represent the game in the herd regime as a Markov process, regardless of the payoff. We found
it crucial to incorporate the stochastic transitions in the model - otherwise it is impossible to
describe analytically the real dynamics. The mesoscopic approach was developed in stages, starting
with Ref.~\cite{wawrzyniak09ACSNo6}. First, we examined the system where fractions are of equal
sizes. The following statements were proved: (i) the utility is bounded and the number of states is
finite, (ii) the transition probabilities are both stochastic and deterministic. Incorporating
these results we worked out the methodology of how to find the Markov representation of the process.
Our analyses based on dynamics of the utility were mostly limited to the step-like payoff
function and were technically hard to generalize. In addition, some important macroscopic
observables, like demand variance {\it per capita} and predictability, were not yet analyzed and
the quenched disorder was neglected. Here, we extend the method providing the consistent theory
comprising different payoffs and quenched disorder. We start in section ~\ref{sec:observables}
where macroscopic differences between games with different payoffs are presented. The theory of how to
describe the game in terms of the Markov process is provided in section~\ref{sec:mesoscopic}. In many
cases the explanation of macroscopic observables required relaxation of the assumption about
equality of fraction sizes and we proved that such relaxation affects transition probabilities. We
found it interesting that increasing the number of players does
not make alike systems with equal and unequal fractions, even if in the latter
case distributions of sizes are symmetric. Our analysis of the attractor structure of the Markov
chain explains this and other dynamical phenomena observed in the herd
regime, viz. oscillations of the aggregate attendance, its periodicity and predictability, or
its dependence on the payoff form. These results are presented in
section~\ref{sec:mesoscopic}. The numerical studies of the periodicity in time are also found
in Refs.~\cite{cara99EurPhysJB10,zheng01PhysicaA301}. More comprehensive review of the literature
is presented in Ref.~\cite{wawrzyniak11Phd}.

\section{The Formal Definition of the Minority Game}\label{sec:definition}
At each time step $t$, the $n$-th agent out of $N$ $(n=1,\ldots,N)$ takes an action
$a_{\alpha_n}(t)$ according to some strategy $\alpha_n(t)$. The action $a_{\alpha_n}(t)$ takes
either of two values: $-1$ or $+1$. An aggregated demand is defined
\begin{eqnarray}
A(t)=\sum_{n=1}^{N}a_{\alpha_n^\prime}(t), \label{eq: A}
\end{eqnarray}
where $\alpha_n^\prime$ refers to the action according to the best strategy, as defined in
eq.~(\ref{eq: argmax}) below. Such defined $A(t)$ is the difference between numbers of agents who
choose the $+1$ and $-1$ actions. Agents do not know each other's actions but $A(t)$ is known to
all agents. The minority action $a^\ast(t)$ is determined from $A(t)$
\begin{eqnarray}
a^\ast(t)=- \mbox{sgn}\, A(t). \label{eq: aMinority}
\end{eqnarray}
Each agent's memory is limited to $m$ most recent winning, i.e. minority, decisions. Each agent has
the same number $S\ge 2$ of devices, called strategies, used to predict the next minority action
$a^\ast(t+1)$. The $s$th strategy of the $n$-th agent, $\alpha_n^s$ $(s=1,\ldots,S)$, is a function
mapping the sequence $\mu$ of the last $m$ winning decisions to this agent's action
$a_{\alpha_n^s}$. Since there is $P=2^m$ possible realizations of $\mu$, there is $2^P$ possible
strategies. At the beginning of the game each agent randomly draws $S$ strategies, according to a
given distribution function $\rho(n):n\rightarrow \Delta_n$, where $\Delta_n$ is a set consisting
of $S$ strategies for the $n$-th agent.

Each strategy $\alpha_n^s$, belonging to any of sets $\Delta_n$, is given a real-valued function
$U_{\alpha_n^s}$ which quantifies the utility of the strategy: the more preferable strategy, the
higher utility it has. Strategies with higher utilities are more likely chosen by agents.

There are various choice policies. In the popular {\it greedy policy} each agent selects the
strategy of the highest utility
\begin{eqnarray}
\alpha_n^\prime(t)=\arg \max_{s:\,\alpha_n^s \in \Delta_n} U_{\alpha_n^s}(t). \label{eq: argmax}
\end{eqnarray}
If there are two or more strategies with the highest utility then one of them is chosen randomly.
The highest-utility strategy (\ref{eq: argmax}) used by the agent is called the {\it active
strategy}, in contrast to {\it passive strategies}, unused at given moment. However, at any time
all agents evaluate all their strategies, the active and passive ones. Each strategy $\alpha_n^s$
is given the {\it payoff} depending on its action $a_{\alpha_n^s}$
\begin{eqnarray}
\Psi_{\alpha_n^s}(t)=-a_{\alpha_n^s}(t)\,g[A(t)], \label{eq: R}
\end{eqnarray}
where $g$ is an odd {\it payoff function}, e.g. the steplike $g(x)= \mbox{sgn}(x)$
\cite{challet97PhysicaA246}, proportional $g(x)=x$ or scaled proportional  $g(x)=x/N$. The learning
process corresponds to updating the utility for each strategy
\begin{eqnarray}
U_{\alpha_n^s}(t+1)=U_{\alpha_n^s}(t)+\Psi_{\alpha_n^s}(t), \label{eq:U}
\end{eqnarray}
such that every agent knows how good its strategies are.

\section{Macroscopic observables}\label{sec:observables}
%\addcontentsline{toc}{chapter}{CHAPTER II \ \ Statistical
%properties of the model}
%\setcounter{equation}{0}  % reset counter
%\setcounter{chapter}{2}
%
Macroscopic variables are understood here as random variables resulting from integration of random
variables defined for individuals, over subsets of degrees of freedom of all individuals in the
system. An example of such variable is the aggregate demand $A$, defined in the previous section.
In this section we introduce and discuss two other particulary interesting macroscopic
observables, viz. variance per capita and predictability. The variance per capita reflects the
coordination between agents and is one of the most intriguing variables due to its nonmonotonic
variation as a function of the control parameter $N/P$. Generally, variance per capita remains
insensitive to the form of payoff function. In contrast, the predictabilities that detect the
existence of patterns are susceptible to the payoff. Here, we demonstrate these phenomena paying
attention mostly to the numerical results. The detailed analytical background is given later in
section \ref{sec:mesoscopic}. Finally, time dependencies of the aggregate demand and utilities are
presented, providing an insight into the origin of time patterns.

\subsubsection{Observables as functions of the control parameter} 
\begin{figure}[h]
\begin{tabular}{cc}
\includegraphics[scale=.27]{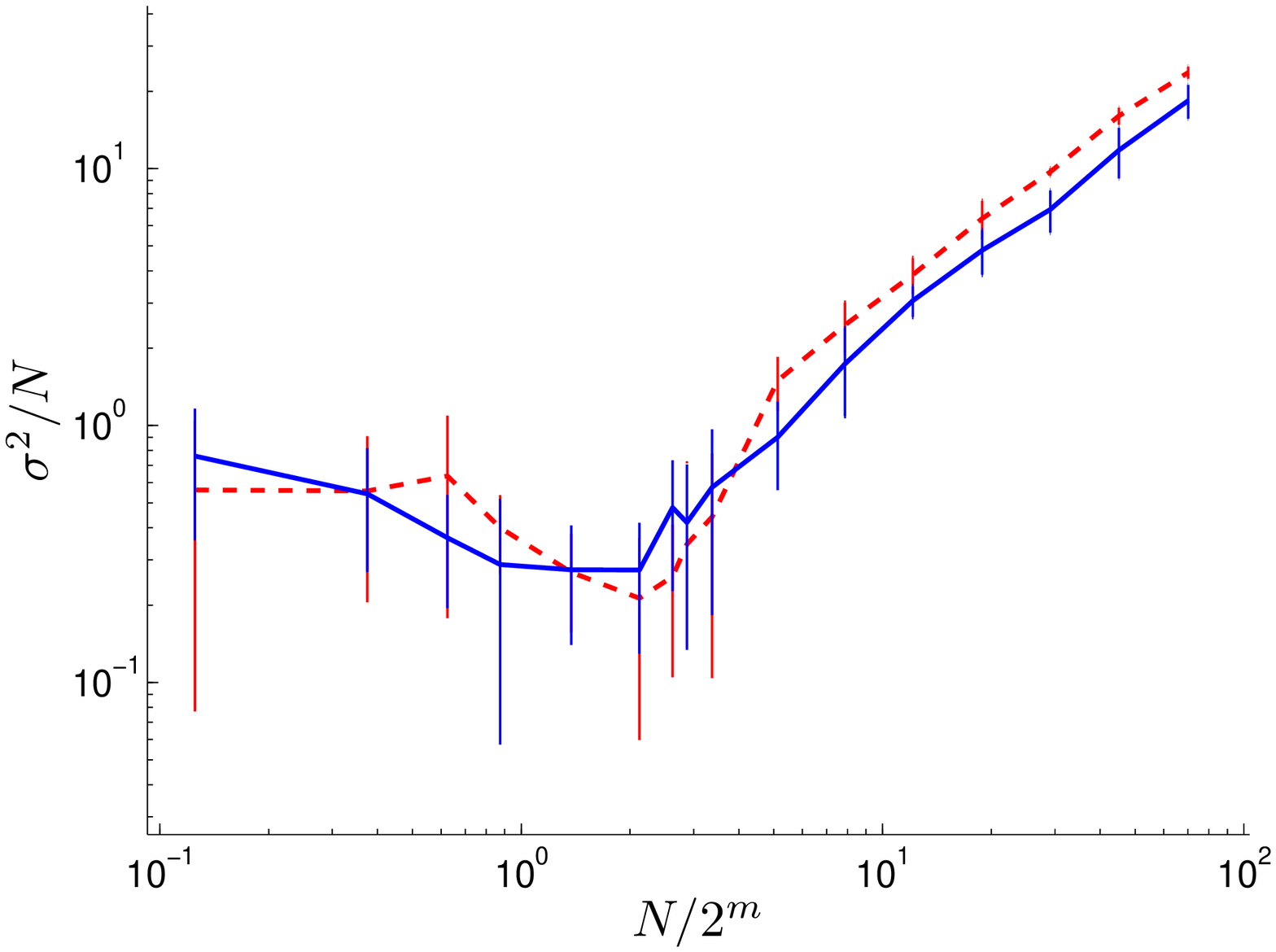} & \includegraphics[scale=.27]{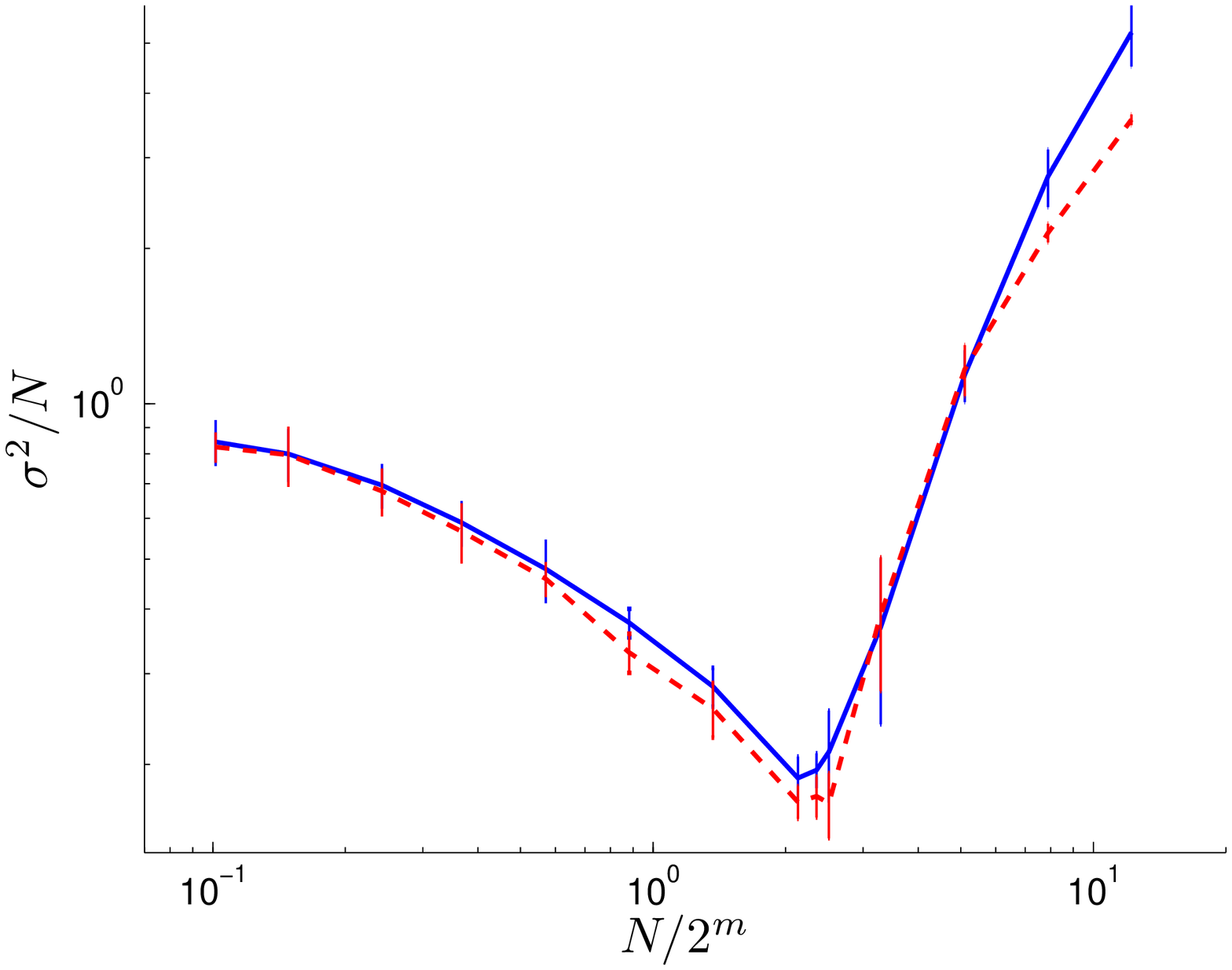}
\end{tabular}
\caption{\label{fig:1}\em Variance per capita $\sigma (A)^2/N$ as a function of $N/P$ for S = 2, m
= 3 (left) or $m=7$ (right). Two different payoff functions are used; full blue lines correspond to
$g(x)=\mbox{sgn}\,(x)$ and dashed red lines to $g(x)=x$. Each point is a mean from ten games, error
bars correspond to one standard deviation and curves are drawn to guide ones eye.}
\end{figure}

The \emph{variance per capita} for
given game is defined using sample taken in subsequent time steps during time $T$ and assuming
ergodicity of the process~\cite{challet97PhysicaA246,savit99PhysRevLetters82}:
\begin{eqnarray}
\sigma (A)^2 =\frac{1}{T}\sum_{t=0}^T A(t)^2. \label{eq:sigma}
\end{eqnarray}
The variance, considered as a function of the control parameter $N/2^m$, represents a widely
discussed result for MGs \cite{challet97PhysicaA246,savit99PhysRevLetters82}, relevant to economic
applications. For our present study it is important to note that its shape seems to be insensitive
to form of the payoff function, as it is presented for two different payoffs and $m=3$ and $m=7$ in
Fig.~\ref{fig:1}.
Similar premise for such payoff-independence is given by another macroscopic observable $H_a/N$,
called {\it predictability}, where $H_a$~\cite{challet99PhysRevE60} is defined as
\begin{eqnarray}
H_a=\frac{1}{P}\sum_{\mu=1}^P\langle a^\ast|\mu\rangle^2, \label{eq:Ha}
\end{eqnarray}
where $\langle a^\ast|\mu\rangle$ is the conditional average of $a^\ast$ given $\mu$ and the mean
is calculated over all $P$ histories.

\begin{figure}[h]
\begin{tabular}{cc}
\includegraphics[scale=.27]{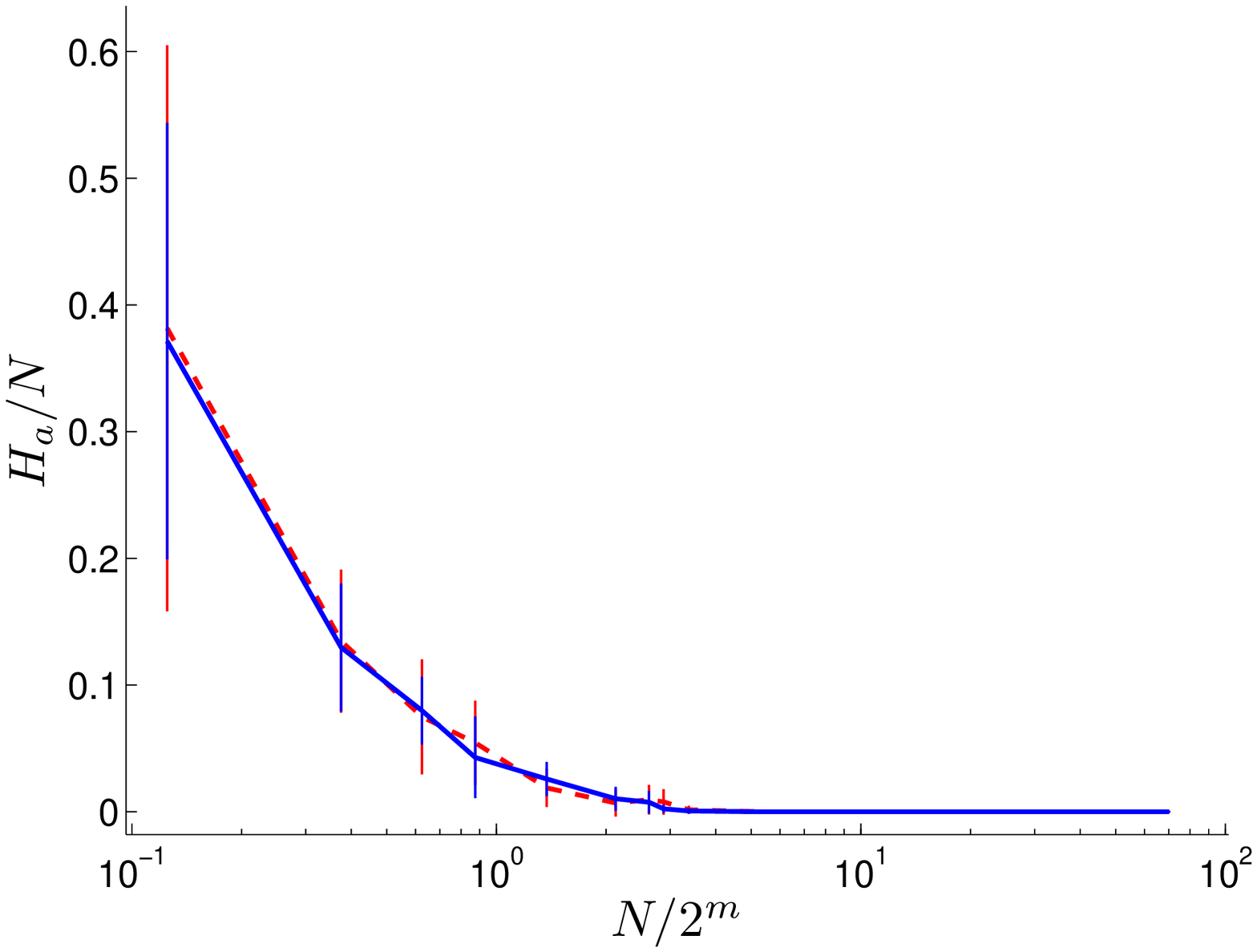} & \includegraphics[scale=.27]{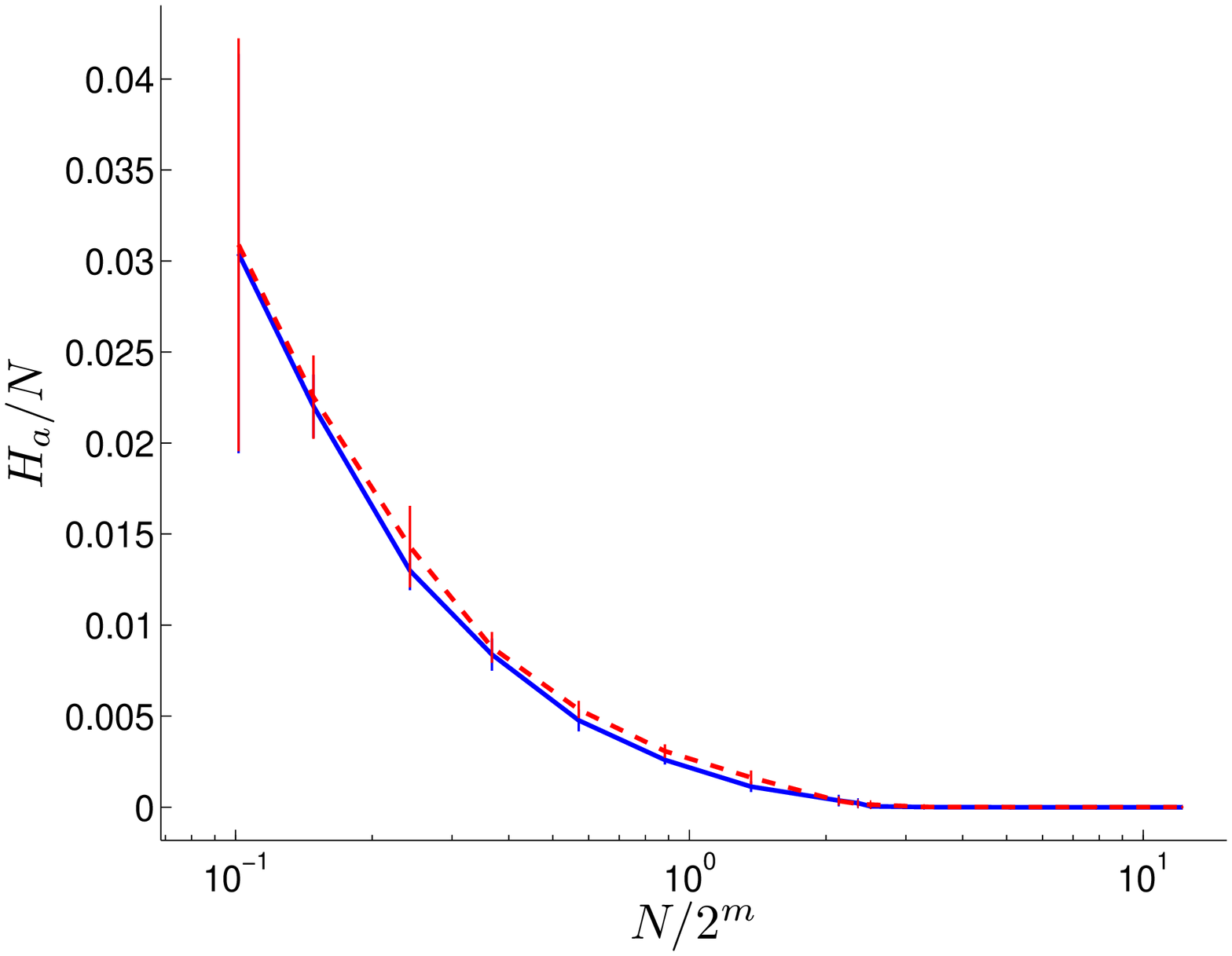}
\end{tabular}
\caption{\label{fig:sigma_m3m7}\em Predictability per capita $H_a/N$ as a function of $N/P$ for S =
2, m = 3 (left) or $m=7$ (right). Two different payoff functions are used: full blue lines
correspond to $g(x)=\mbox{sgn}\,(x)$ and dashed red lines to $g(x)=x$. Each point is a mean from ten
games, error bars correspond to one standard deviation and curves are drawn to guide ones eye.}
\end{figure}

The $H_a$ was demonstrated to be useful in detecting two interesting phases of the MG:
\begin{itemize}
\item The \emph{symmetric phase} with $H_{a} \simeq 0$, where after the particular history $\mu(t)$
both signs of $a^*(t)$ appear with the same frequency. It is often claimed in
literature~\cite{challet99PhysRevE60,moro00minorityIntroductoryGuide} that if $H_{a} = 0$ then
patterns in the time sequence do not exist. We find this condition to be the necessary but not
sufficient one to state the lack of patterns. For example, if every appearance of given $\mu$ is
followed by negative and positive minority decision alternately then $H_{a} = 0$ and the
predictable pattern exists. Indeed, such a behavior is observed for the MG in the herd regime and
for $g(x) = x$~\cite{wawrzyniak09ACSNo6}. Hence, $H_{a}$ measures disproportions in frequencies
between positive and negative minority decisions rather than detects patterns. \item The
\emph{asymmetric phase} with $H_{a} > 0$ and existing predictable patterns. In the asymmetric
phase, sign predictions significantly better than random are possible.
\end{itemize}
As presented in Figs~\ref{fig:sigma_m3m7}, plots of $H_a/N$ seem to be independent of the payoff
function, similarly to $\sigma^2/N$.

\begin{figure}[h]
\begin{tabular}{cc}
\includegraphics[scale=.27]{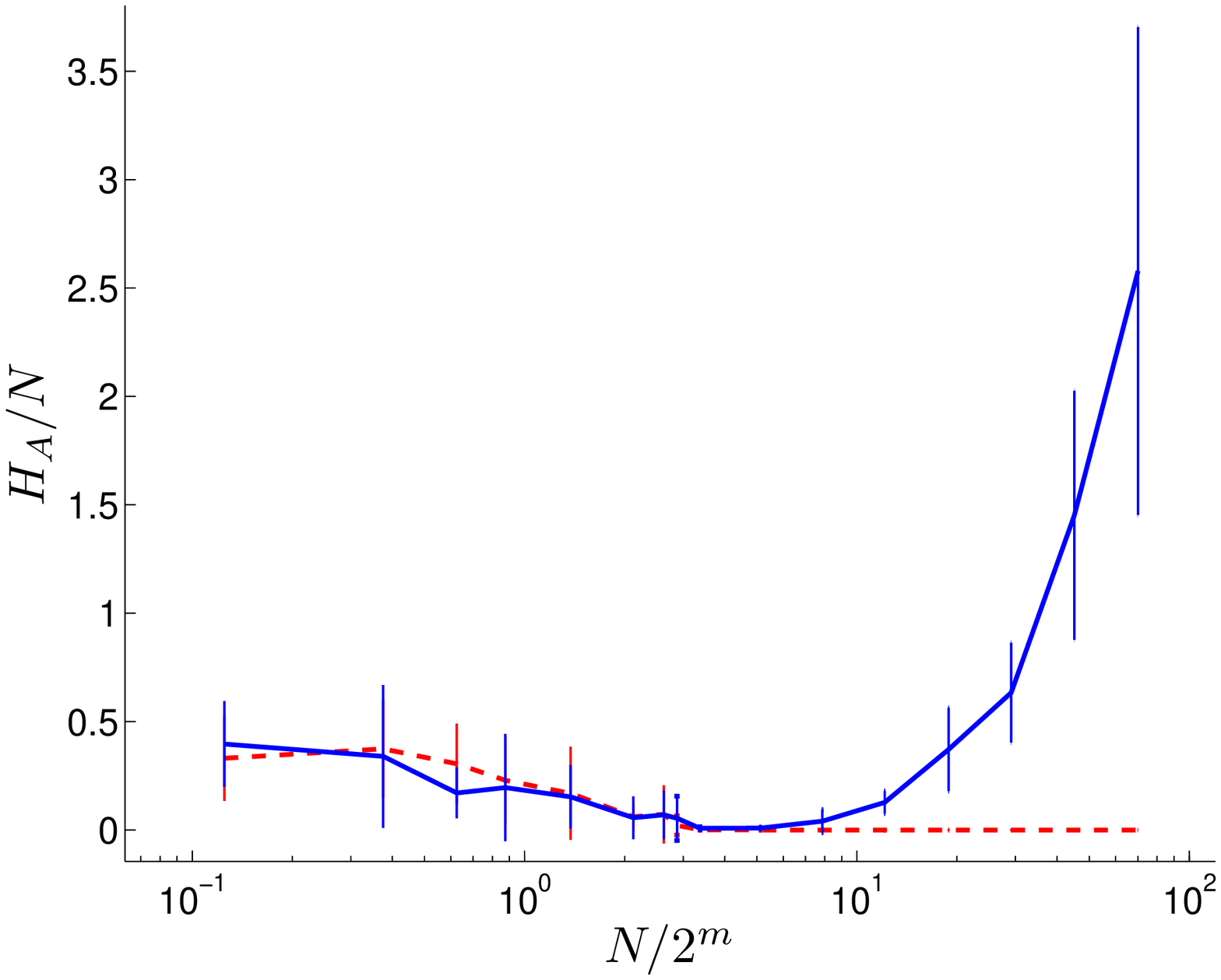} & \includegraphics[scale=.27]{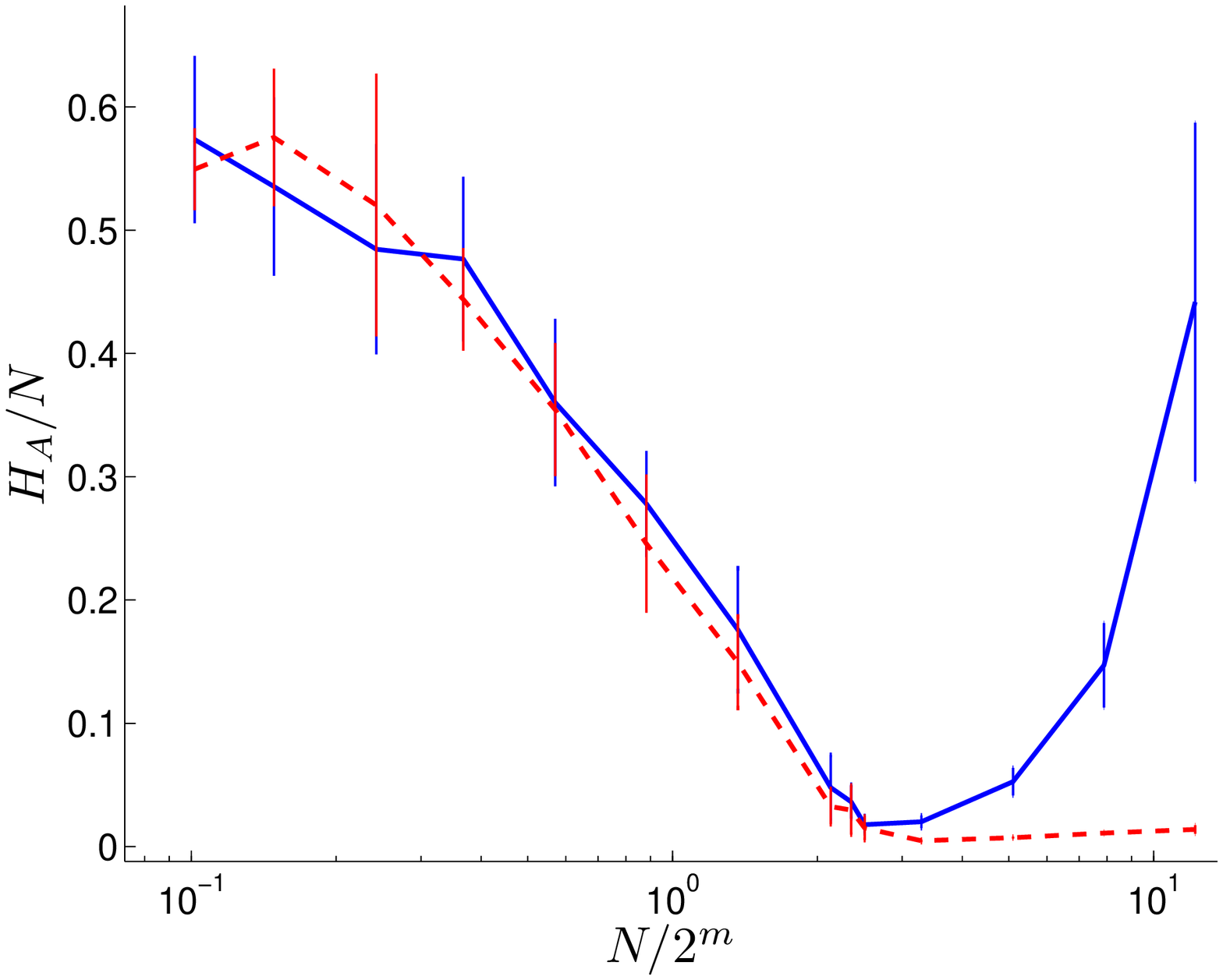}
\end{tabular}
\caption{\label{fig:Ha_m3m7}\em Demand predictability per capita $H_A/N$ as a function of $N/P$ for
S = 2, m = 3 (left) or $m=7$ (right). Two different payoff functions are used: full blue lines
correspond to $g(x)=\mbox{sgn}\,(x)$ and dashed red lines to $g(x)=x$. Each point is a mean from ten
games, error bars correspond to one standard deviation and curves are drawn to guide ones eye.}
\end{figure}

By that means it was conjectured in early literature (cf. e.g. Ref.~\cite{cavagna99PhysRevE59})
that only the payoff's evenness is relevant to the macroscopic observables. Failure of this
hypothesis is visible by analysing a modified macroscopic observable, we call {\it demand
predictability}, which may be useful for prediction of the sign of demand. This variable is defined as
\begin{eqnarray}
H_A=\frac{1}{P}\sum_{\mu=1}^P\langle A|\mu\rangle^2. \label{eq:HA}
\end{eqnarray}
Plots of $H_A/N$ (cf. Fig.~\ref{fig:Ha_m3m7}) exhibit its spectacular sensitivity to the payoff
function in the effective regime, i.e. high $N/P$, in contrast to $H_a/N$ and $\sigma^2/N$.
For further analysis we decompose the conditional expected values into components corresponding to
decisions $+1$ and $-1$:
\begin{eqnarray}
\langle a^\ast|\mu\rangle & = & \langle a_+^\ast|\mu\rangle + \langle a_-^\ast|\mu\rangle,
\label{eq:adec}
\end{eqnarray}
where, formally
\begin{eqnarray}
\langle a_{\pm}^\ast|\mu \rangle=\frac{1}{T}\sum_{t=1}^T a^\ast(t)\,\delta
\big(\mu(t),\mu \big)\,\delta \big(a^\ast(t),\pm 1 \big), \label{eq:apm}
\end{eqnarray}
$\delta(i,j)$ standing for the Kronecker symbol. Similarly,
\begin{eqnarray}
\langle A|\mu\rangle & = & \langle A_+|\mu\rangle + \langle A_-|\mu\rangle, \label{eq:Adec}
\end{eqnarray}
where
\begin{eqnarray}
\langle A_{\pm}|\mu \rangle=\frac{1}{T}\sum_{t=1}^T A(t)\,\delta \big( \mu(t),\mu
\big)\,\delta \big(\mbox{sgn} A(t),\pm 1 \big). \label{eq:Apm}
\end{eqnarray}
The case $H_a=0$ is possible if $\langle a^\ast|\mu\rangle=0$ for every $\mu$ which, as seen from
Eq.~(\ref{eq:adec}), requires $|\langle a_+^\ast|\mu\rangle|=|\langle a_-^\ast|\mu\rangle|$. This
means that the positive and negative values of $A(t)$ have to come with the same frequency.
Similarly, the case $H_A=0$ happens if $|\langle A_+|\mu\rangle|=|\langle A_-|\mu\rangle|$ for
every $\mu$, i.e. the positive and negative $A$ mutually compensate (cf. Eq.~(\ref{eq:Adec})).
Combinations like (i) $H_a=0$ and $H_A>0$, and (ii) $H_a>0$ and $H_A=0$, are also possible.
\subsection{Observables as functions of time}
In order to examine MGs in the efficient regime, we performed a series of numerical simulations
with different combinations of game parameters. We chose three representative cases:
$(m,N)=(1,401),(2,1601),(5,1601)$, all with the number of strategies per agent $S=2$. All three
games are in the efficient mode. In the first two cases the condition $NS\gg 2^P$ is fulfilled. In
the third one it is not met and consequences of this fact will become clear later in the text. In
all three experiments the full strategy space is used.

\begin{figure}[t!]
\begin{center}
\begin{tabular}{ccc}
\includegraphics[scale=.23]{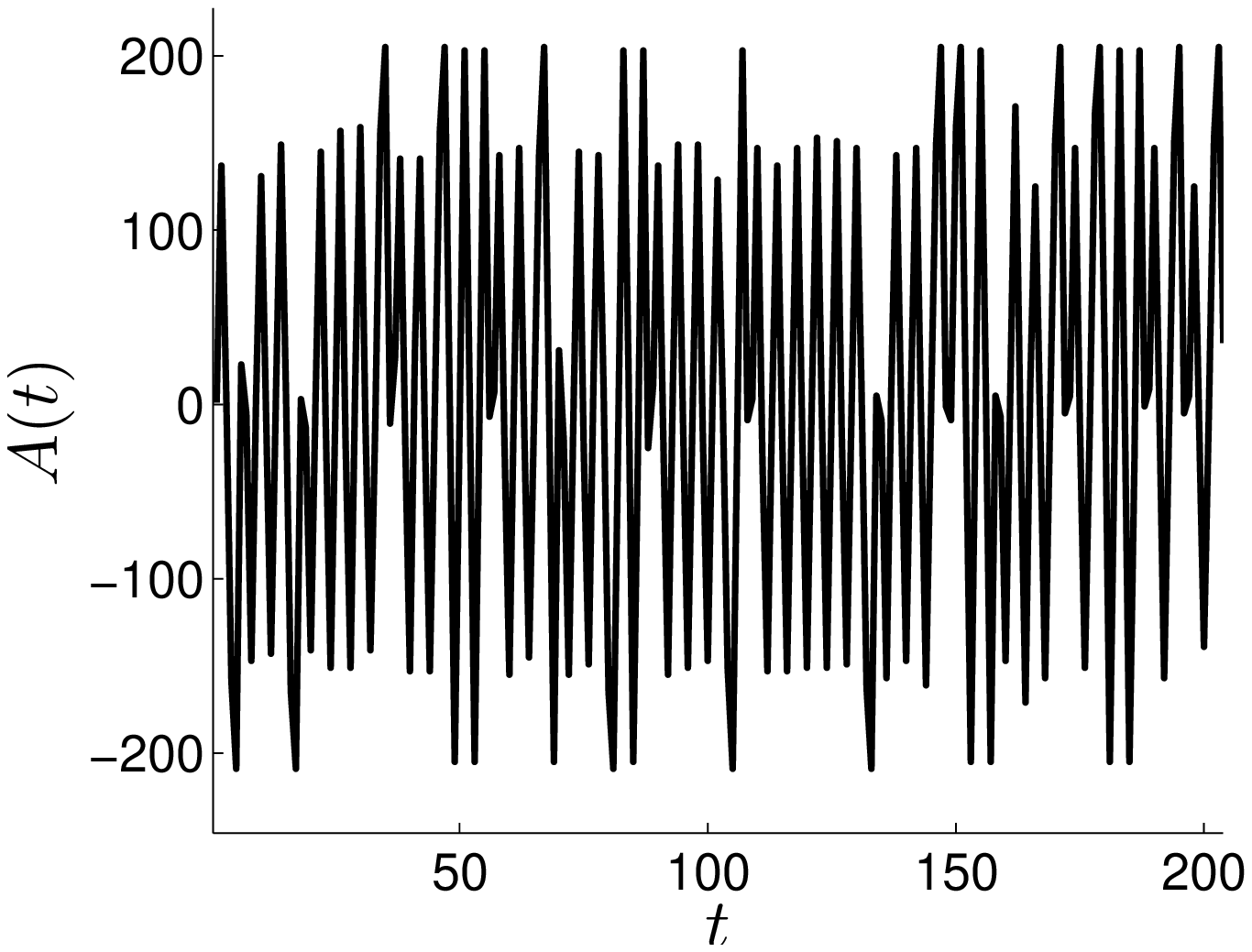} & \hspace{2mm}
\includegraphics[scale=.23]{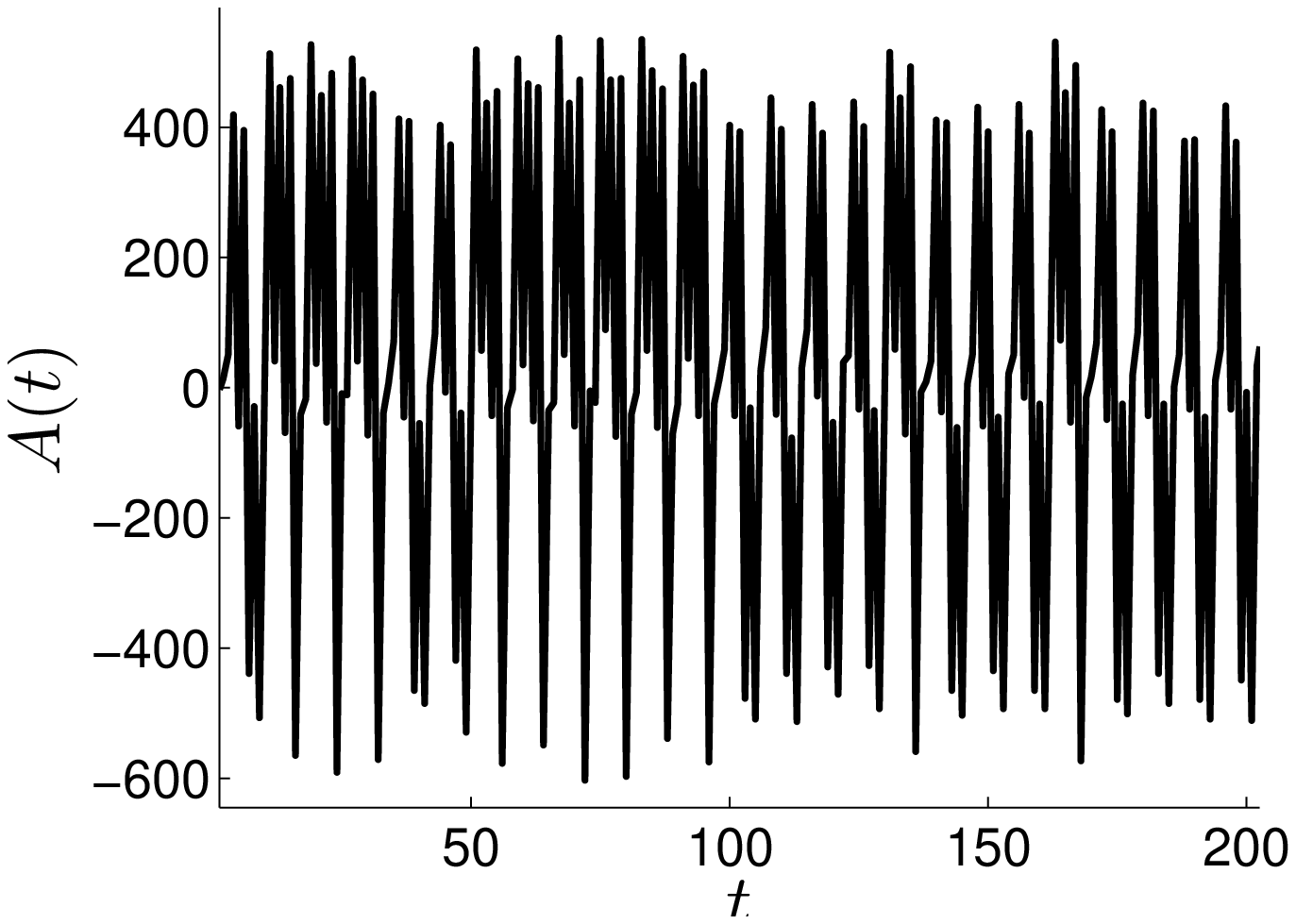} & \hspace{2mm}
\includegraphics[scale=.23]{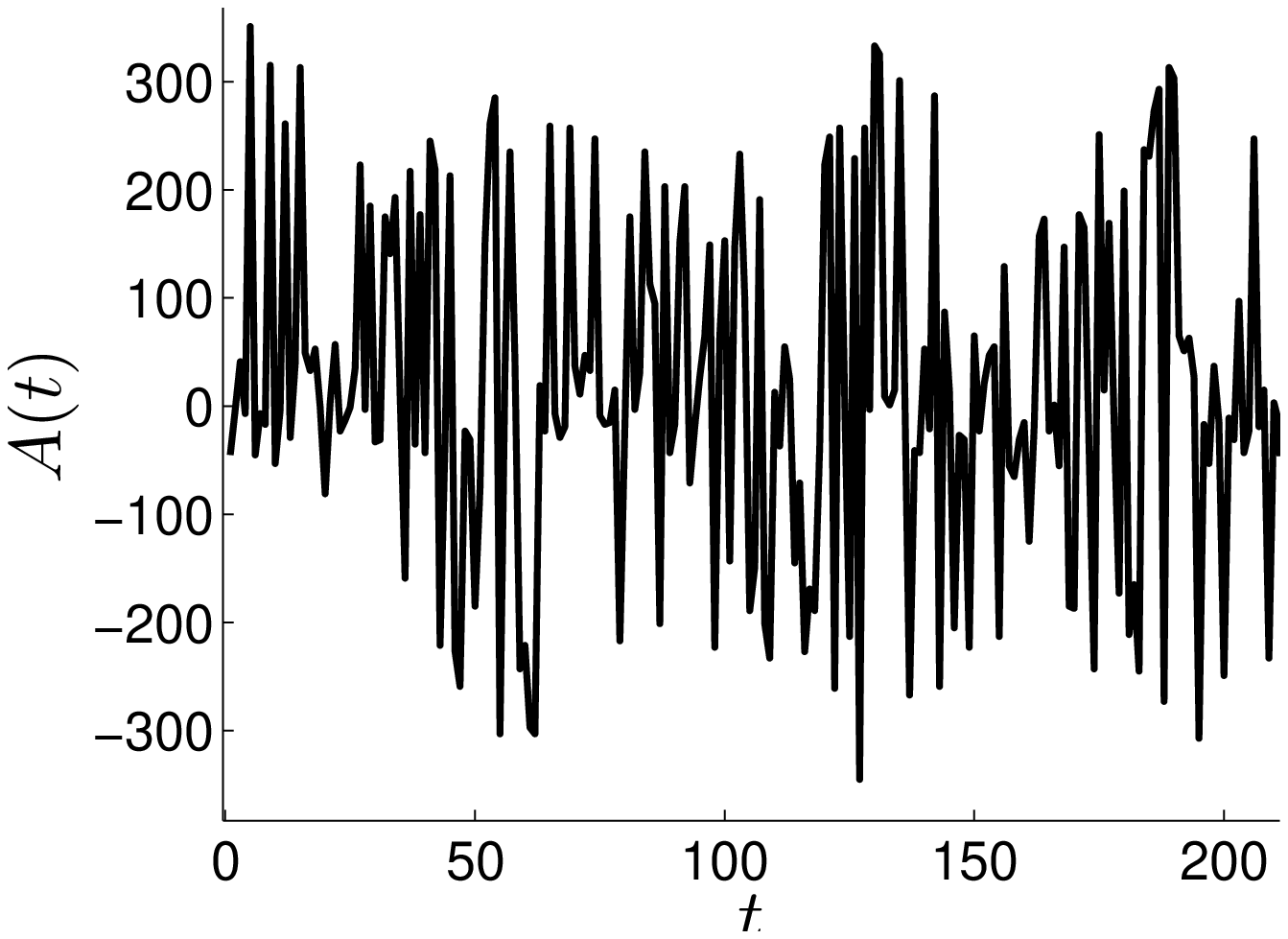}
\end{tabular}
\end{center}
\vspace*{8pt} \caption{\label{fig:A3_signx}\em Time evolution of the aggregated demand $A(t)$ for
three combinations of the population size $N$ and agent memory $m$: $N=401$, $m=1$ (left),
$N=1601$, $m=2$ (middle) and $N=1601$, $m=5$ (right). Simulations were done for $S=2$ and
$g(x)=\mbox{sgn} (x)$. Preferred values of $A$ are visible for all three games.}
%\end{figure}
%\begin{figure}[t]
\begin{center}
\begin{tabular}{ccc}
\includegraphics[scale=.23]{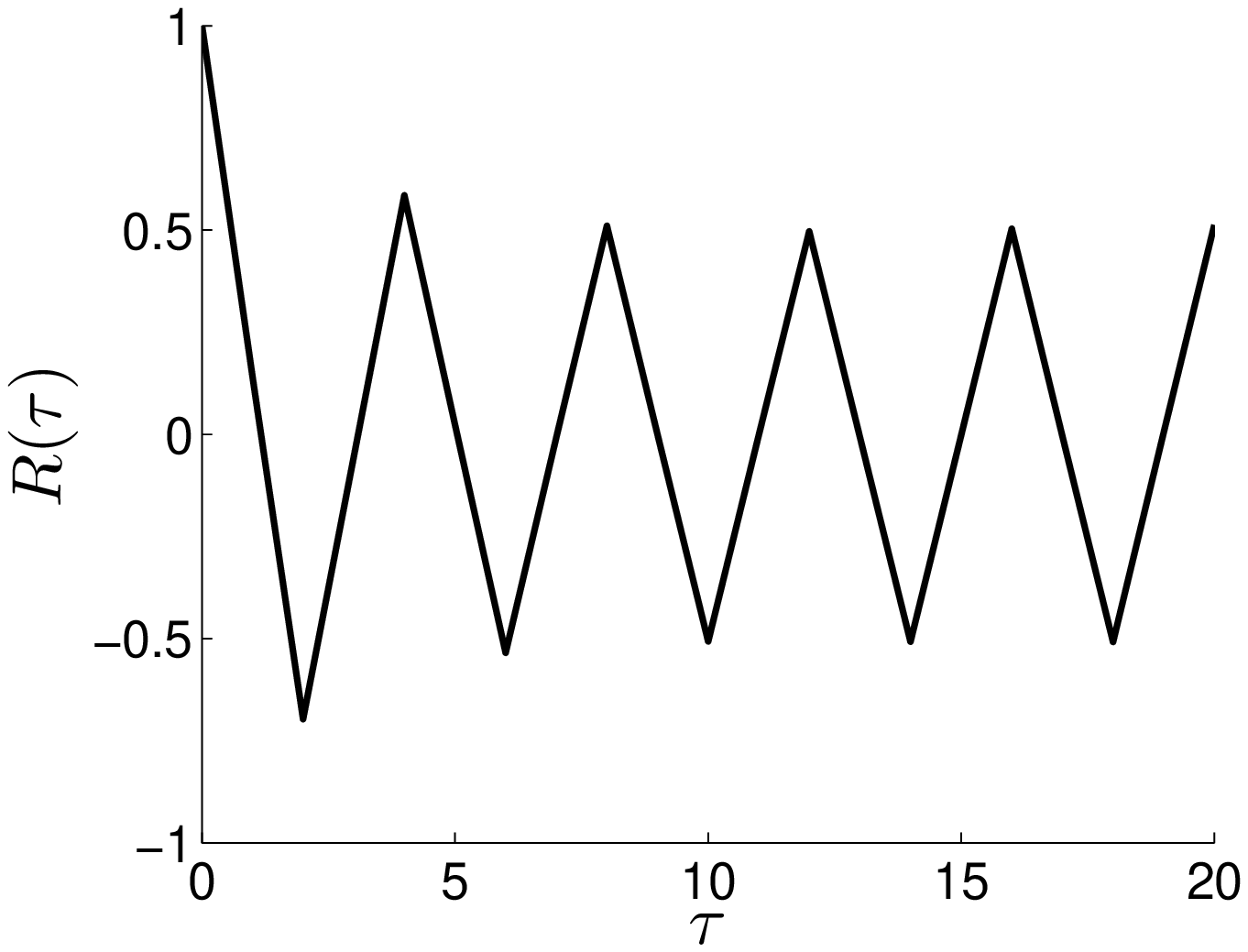} & \hspace{2mm}
\includegraphics[scale=.23]{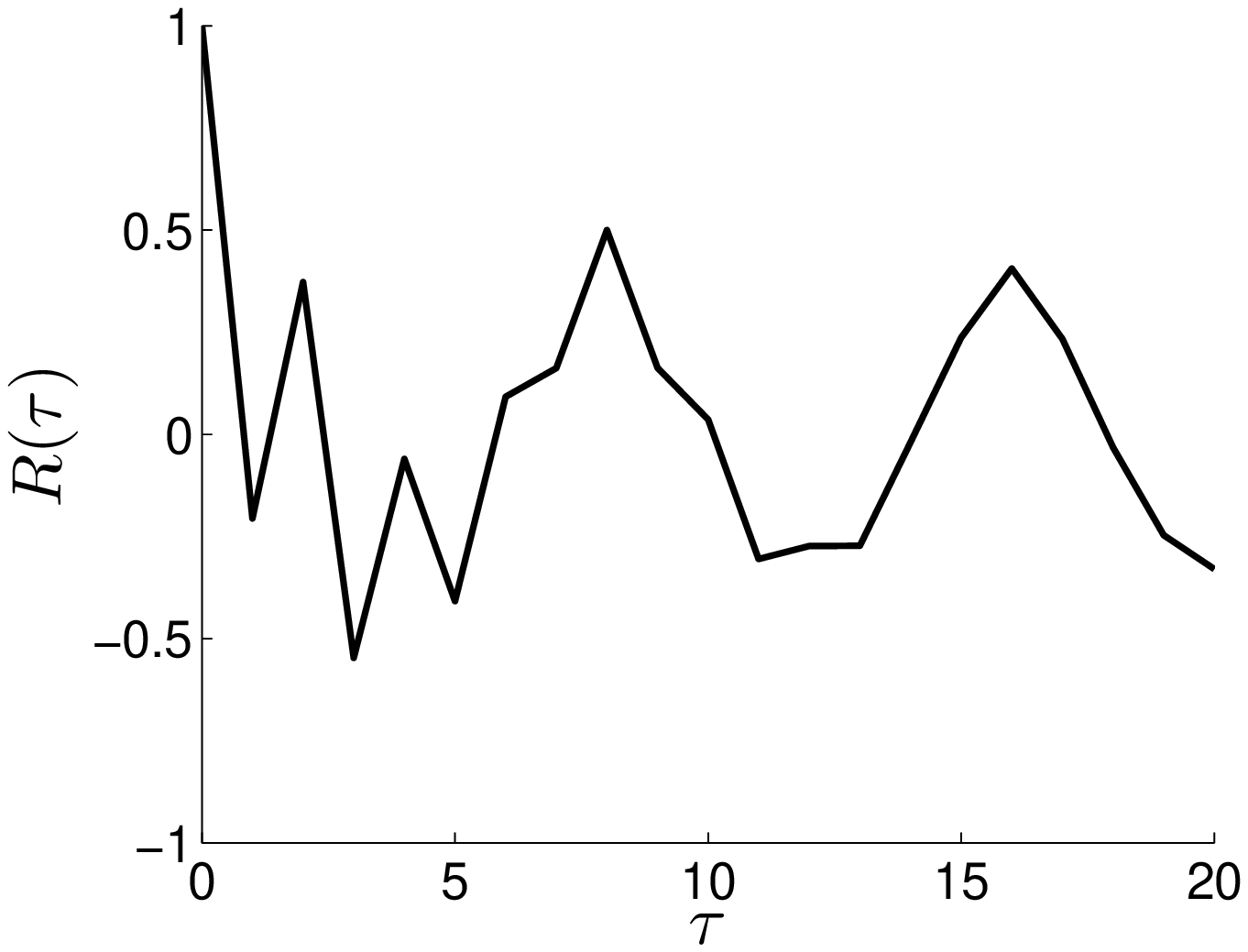} & \hspace{2mm}
\includegraphics[scale=.23]{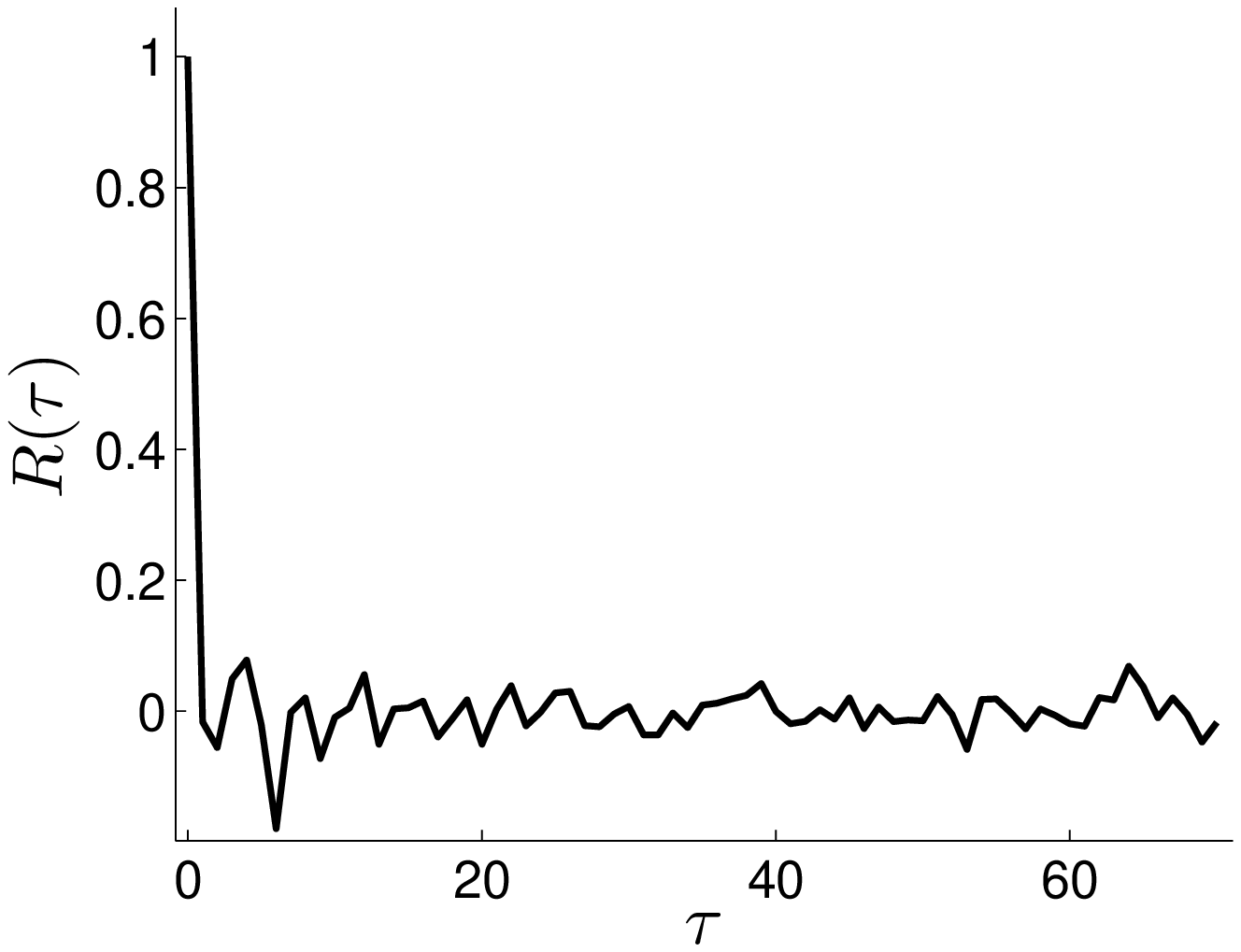}
\end{tabular}
\end{center}
\vspace*{8pt} \caption{\label{fig:R3_signx}\em Autocorrelation function $R(\tau)$ for three
combinations of the population size $N$ and agent memory $m$: $N=401$, $m=1$ (left), $N=1601$,
$m=2$ (middle) and $N=1601$, $m=5$ (right). Simulations were done for $S=2$ and $g(x)=\mbox{sgn}
(x)$. The highest values of $R$ are for $\tau = 2\cdot 2^m$, except for $\tau=0$, for all games
fulfilling the $NS\gg 2^P$ condition.}
%\end{figure}
%\begin{figure}[t]
\begin{center}
\begin{tabular}{ccc}
\includegraphics[scale=.21]{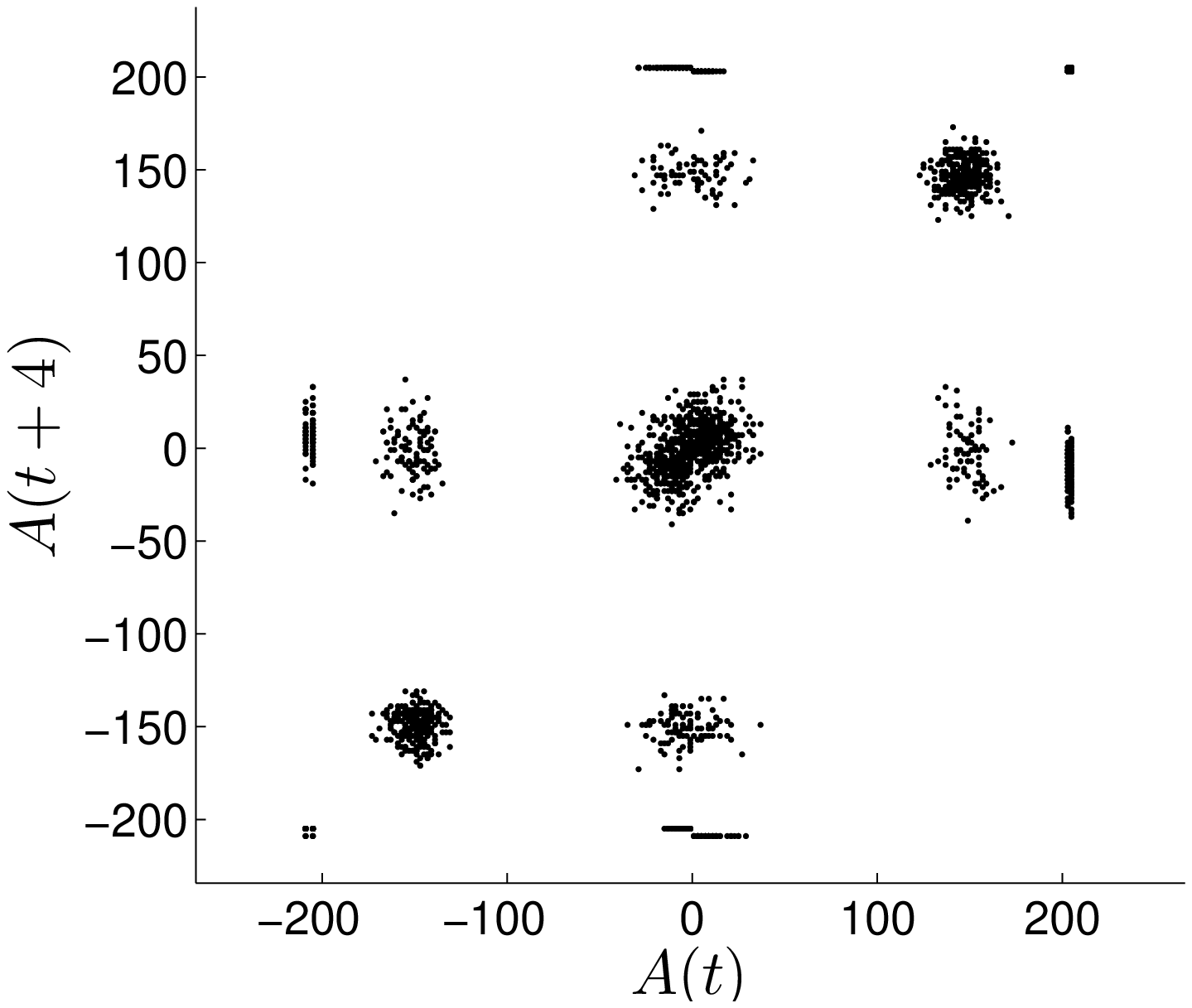} & \hspace{2mm}
\includegraphics[scale=.21]{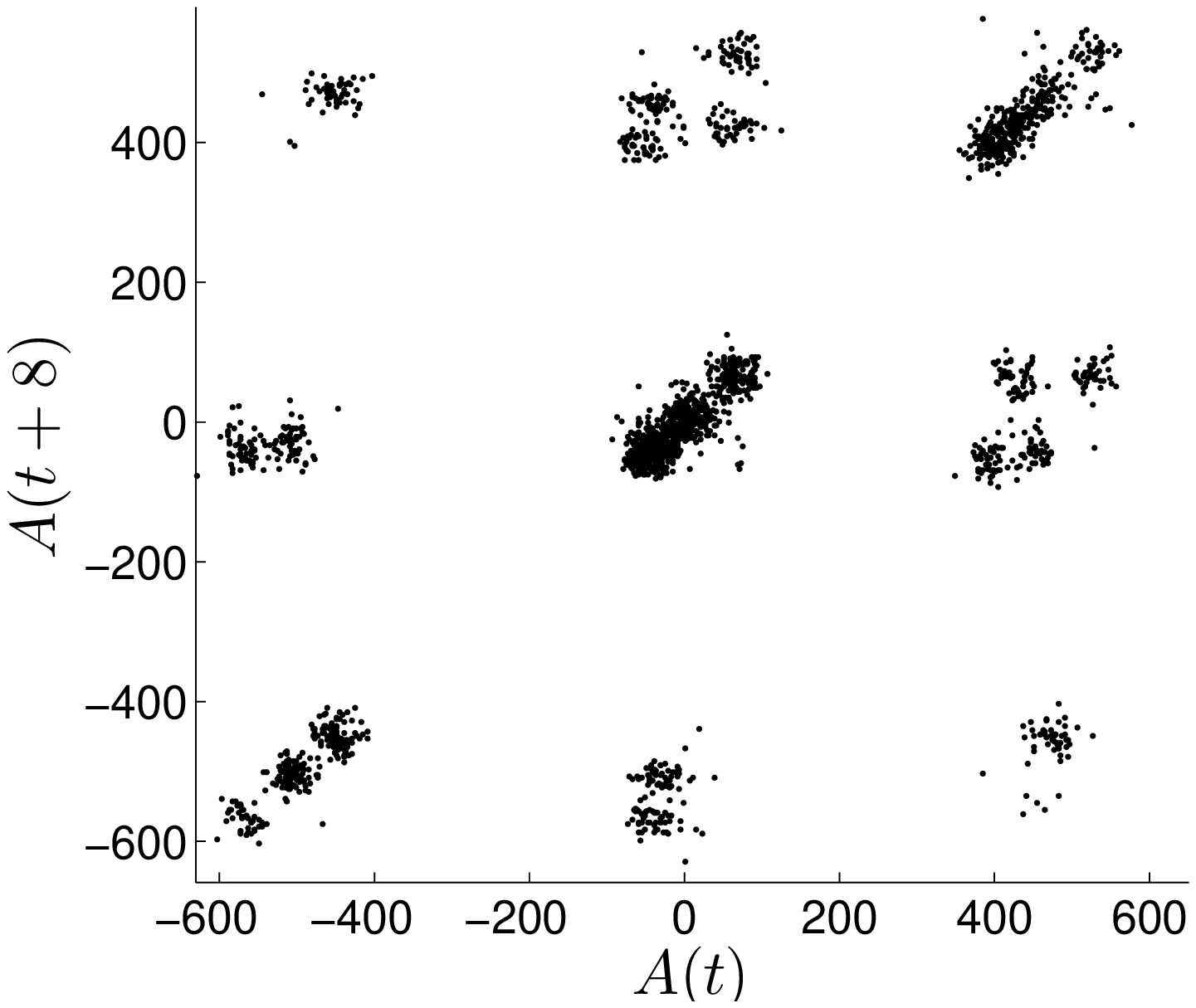} & \hspace{2mm}
\includegraphics[scale=.21]{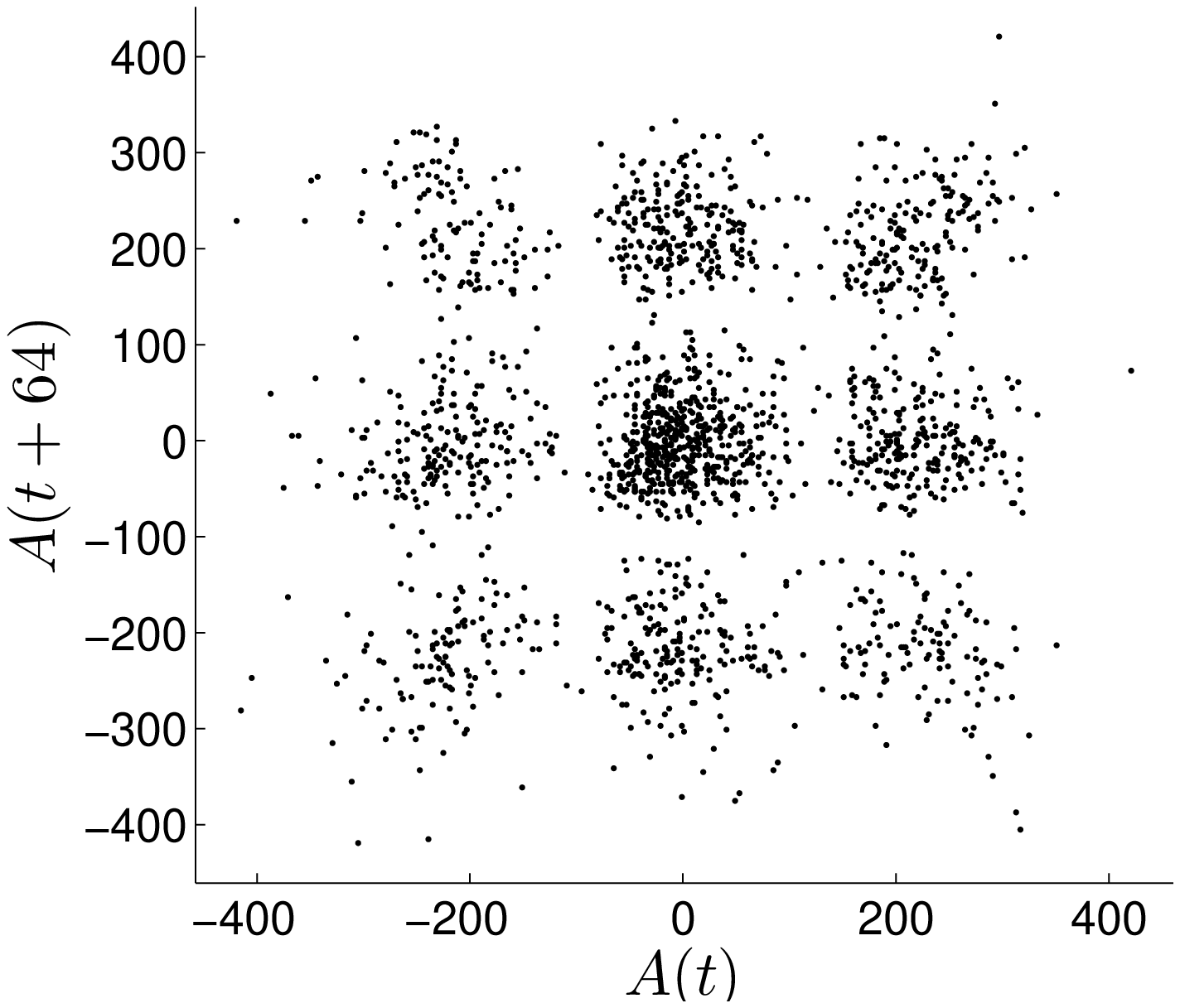}
%\hspace{-8pt} \includegraphics[scale=.2]{./fig33_a.eps} & \hspace{-8pt}
%\includegraphics[scale=.2]{./fig33_b.eps} & \hspace{-8pt} \includegraphics[scale=.3]{./fig33_c.eps}
\end{tabular}
\end{center}
\vspace*{8pt} \caption{\label{fig:AagainstA3_signx}\em Plots of the aggregated demand $A(t+2\cdot
2^m)$ vs. $A(t)$ for three combinations of the population size $N$ and agent memory $m$: $N=401$,
$m=1$ (left), $N=1601$, $m=2$ (middle) and $N=1601$, $m=5$ (right). Simulations were done for $S=2$
and $g(x)=\mbox{sgn} (x)$. Apparent preferred levels of $A(t)$ are seen as clusters of points. For
$m=1$ and $m=2$ points tend to flock around diagonals indicating positive correlation for
$\tau=2\cdot 2^m$.}
\end{figure}

\begin{figure}[t]
\begin{center}
\begin{tabular}{ccc}
\includegraphics[scale=.18]{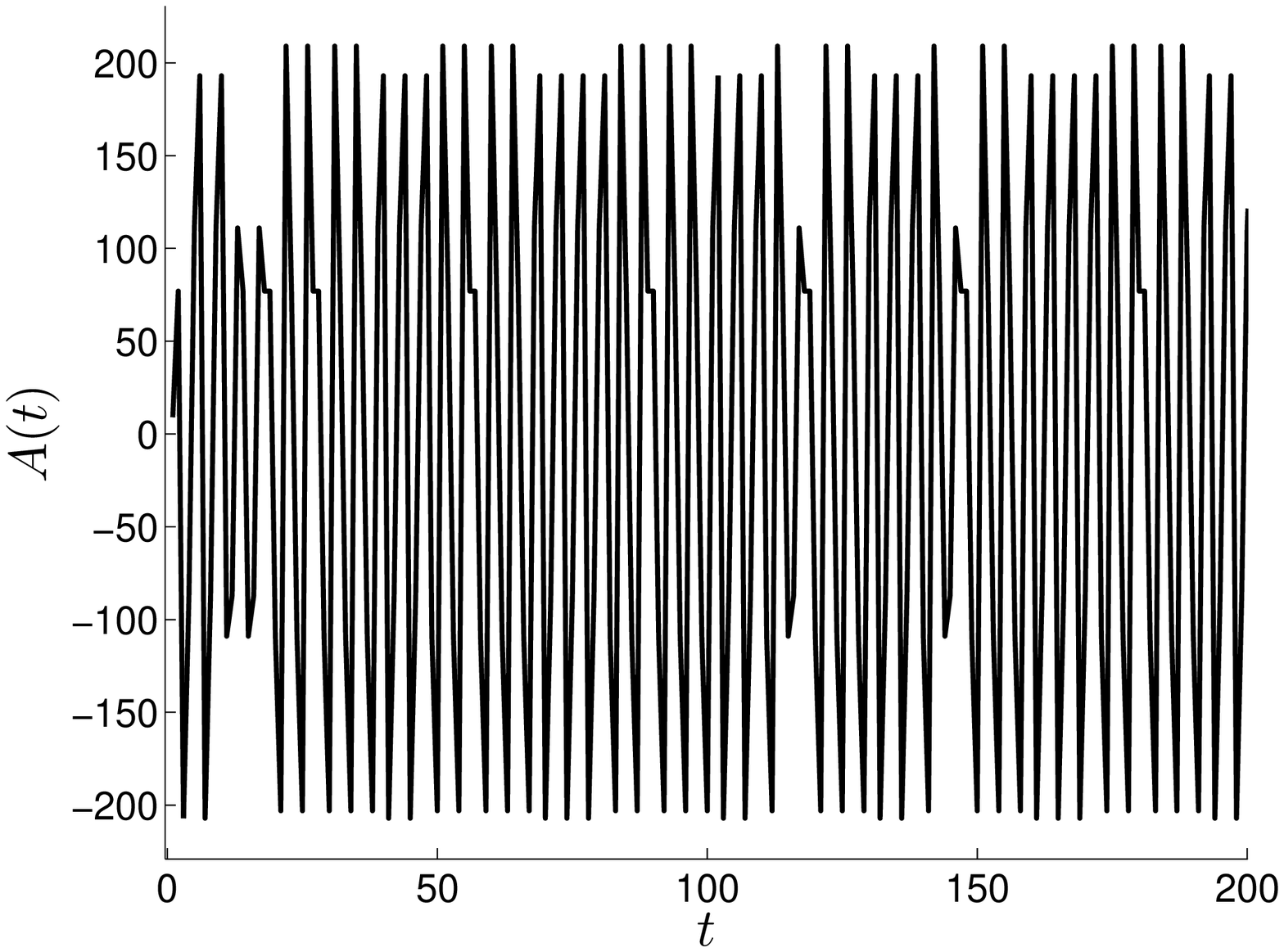} & \hspace{2mm}
\includegraphics[scale=.18]{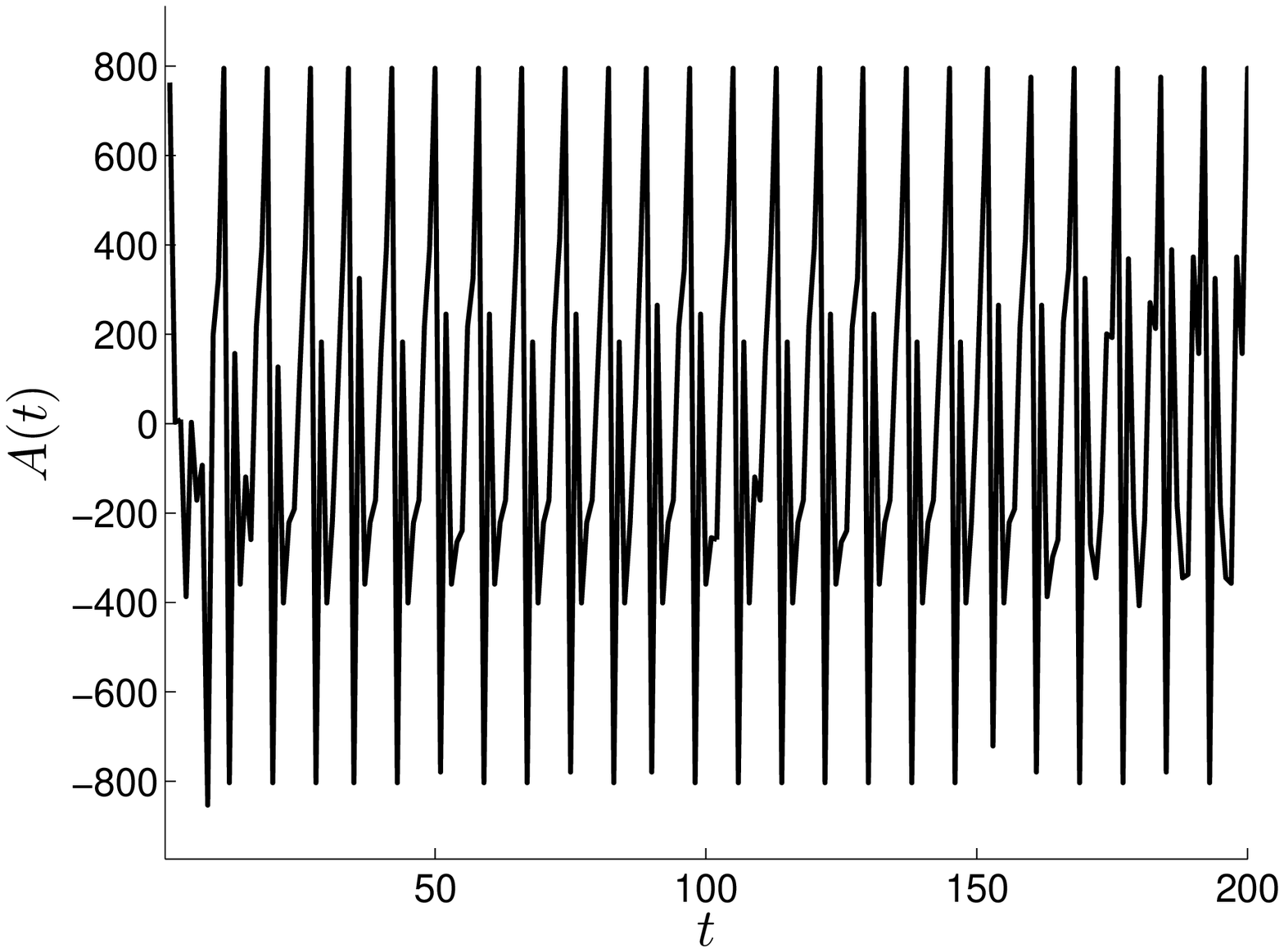} & \hspace{2mm}
\includegraphics[scale=.18]{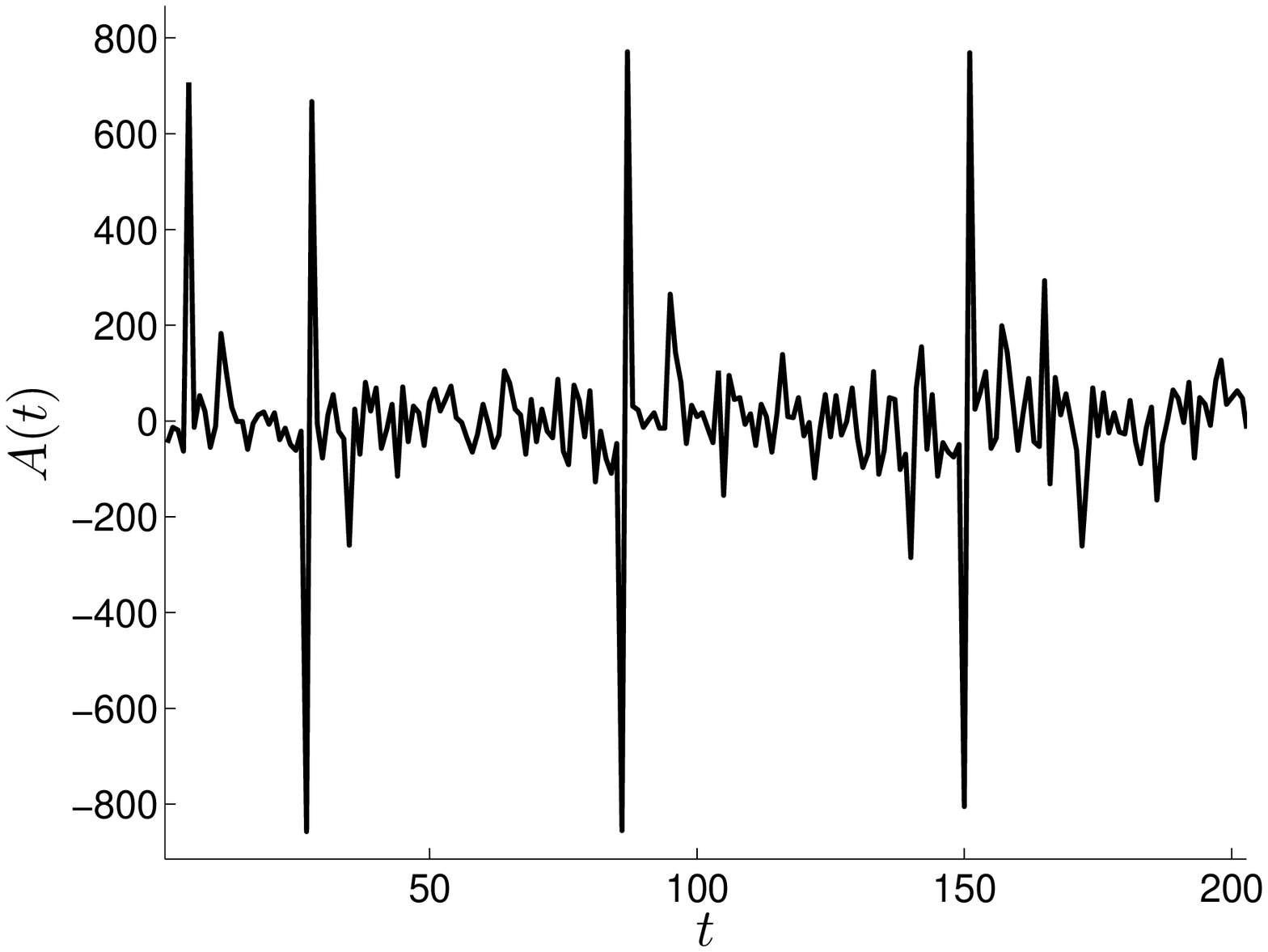}
\end{tabular}
\end{center}
\vspace*{8pt} \caption{\label{fig:A3_linear}\em Time evolution of the aggregated demand $A(t)$ for
three combinations of the population size $N$ and agent memory $m$: $N=401$, $m=1$ (left),
$N=1601$, $m=2$ (middle) and $N=1601$, $m=5$ (right). Simulations were done for $S=2$ and $g(x)=x$.
Preferred values of $A$ are visible for all three games.}
%\end{figure}
%\begin{figure}[t]
\begin{center}
\begin{tabular}{ccc}
\includegraphics[scale=.18]{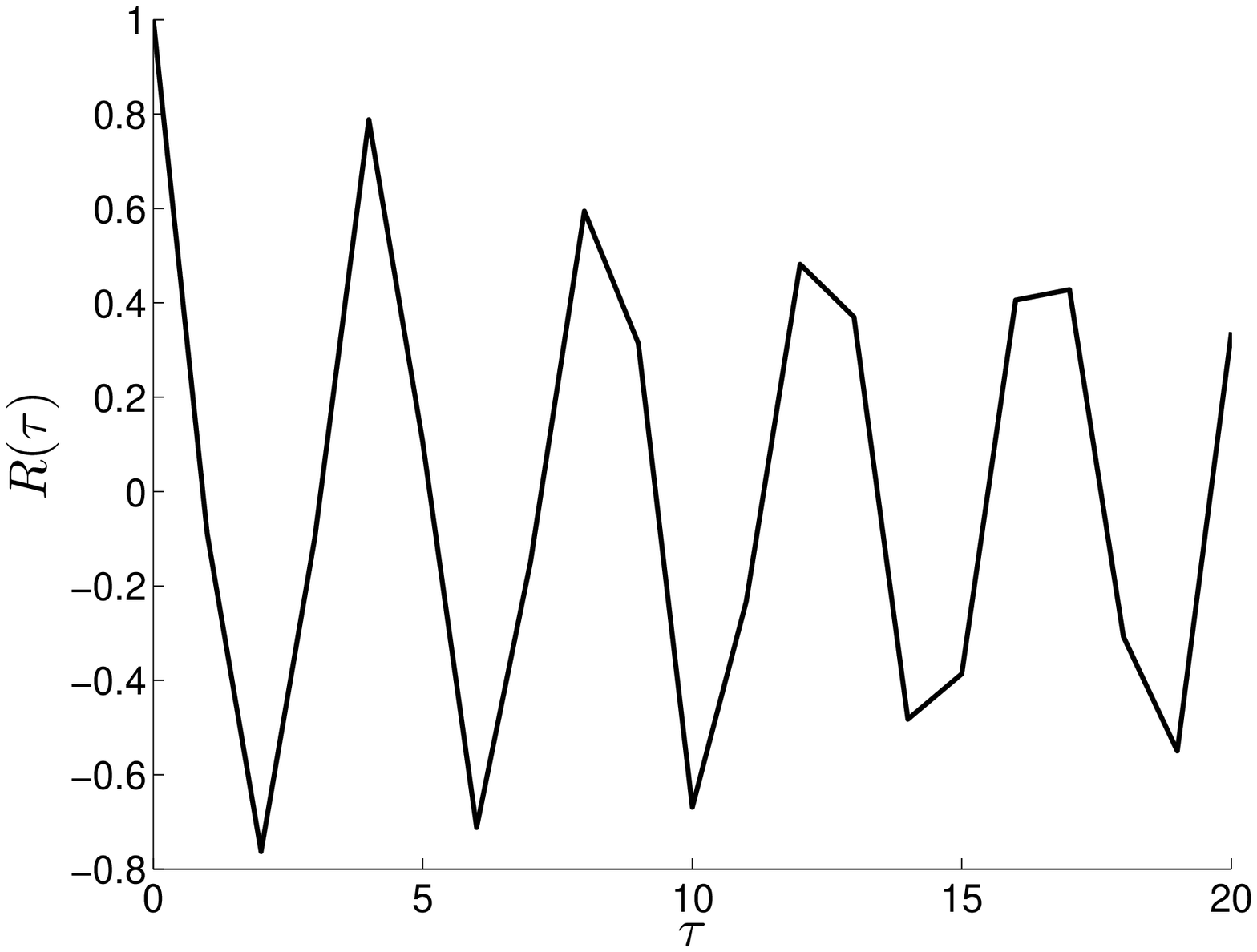} & \hspace{2mm}
\includegraphics[scale=.18]{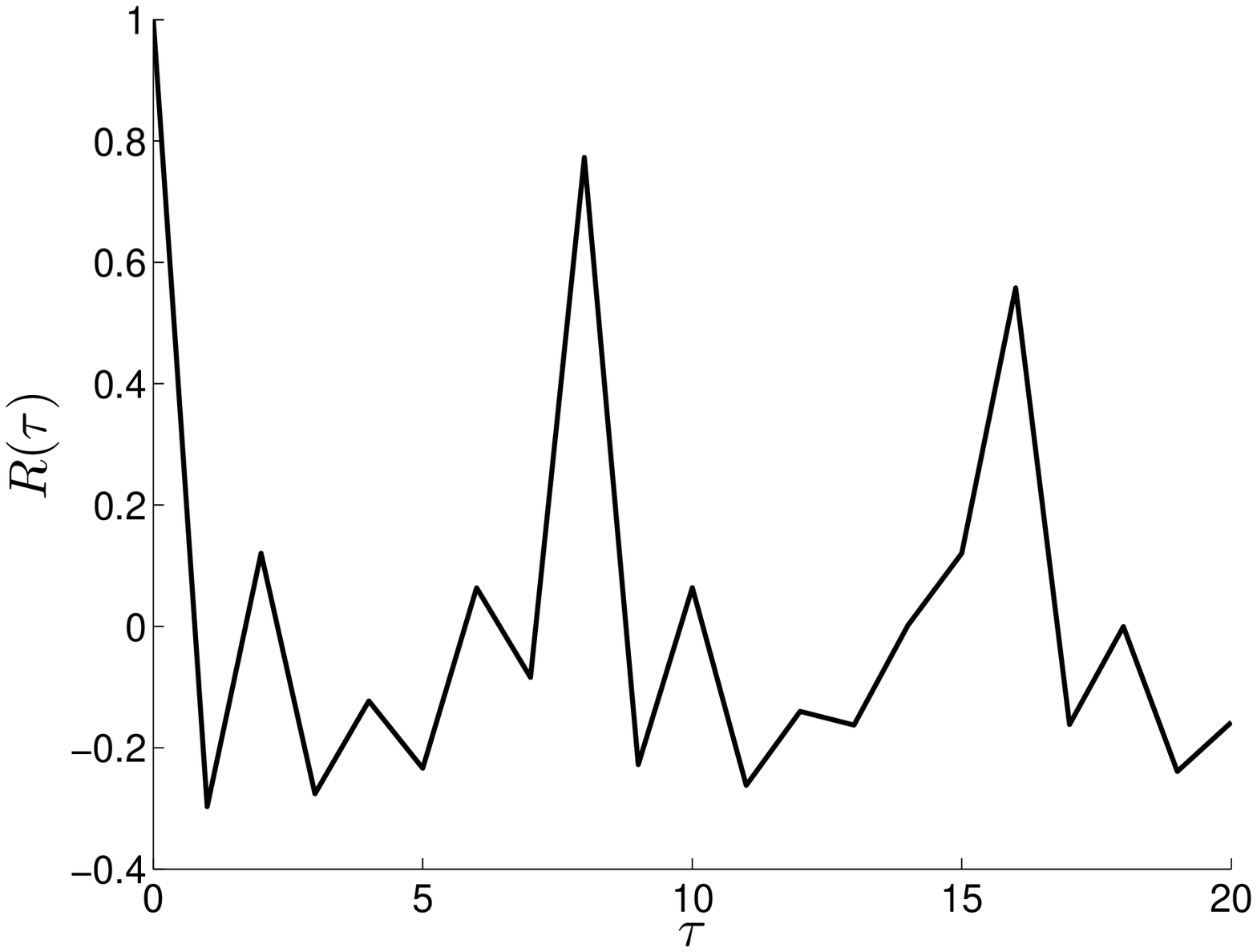} & \hspace{2mm}
\includegraphics[scale=.18]{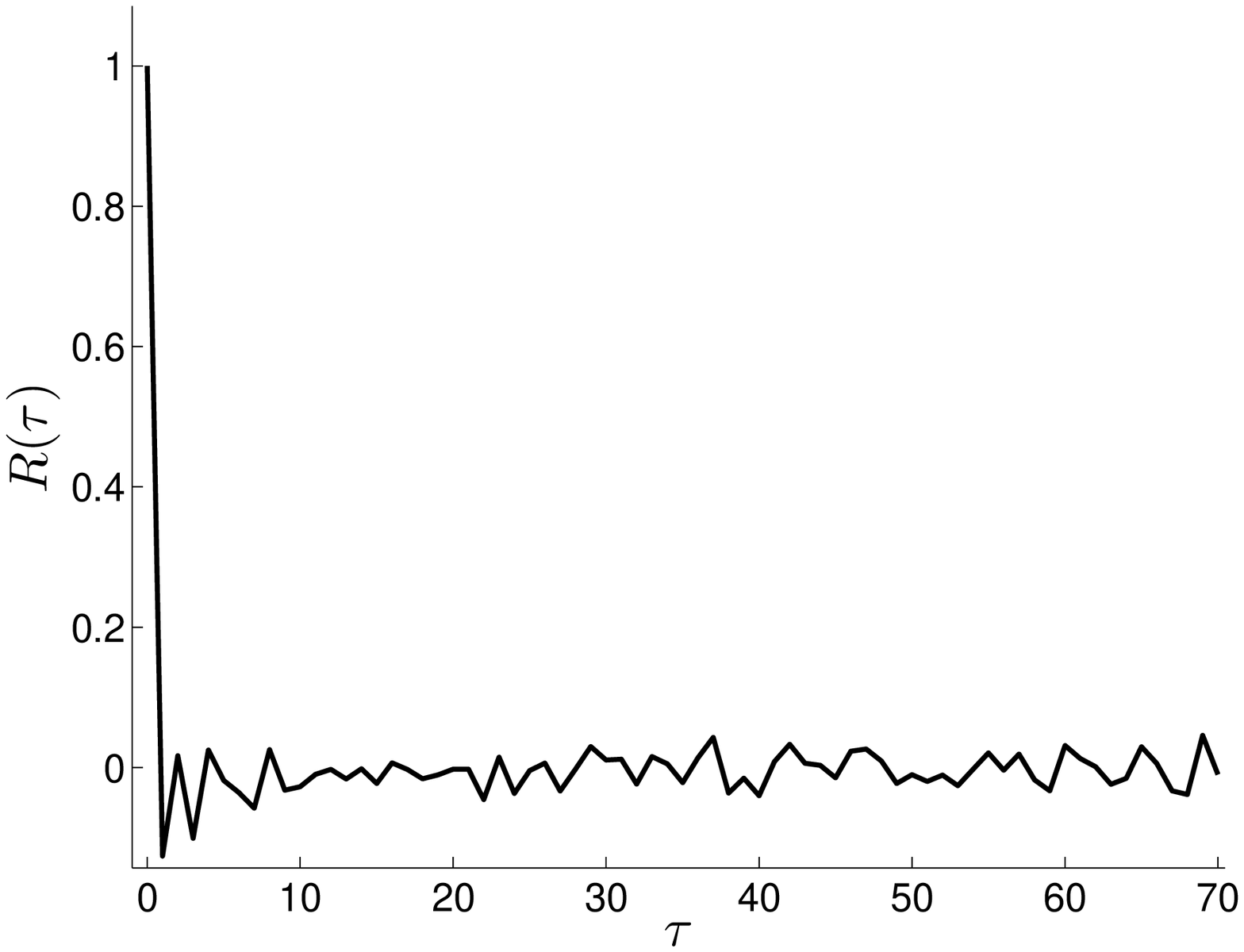}
\end{tabular}
\end{center}
\vspace*{8pt} \caption{\label{fig:R3_linear}\em Autocorrelation function $R(\tau)$ for three
combinations of the population size $N$ and agent memory $m$: $N=401$, $m=1$ (left), $N=1601$,
$m=2$ (middle) and $N=1601$, $m=5$ (right). Simulations were done for $S=2$ and $g(x)=x$. The
highest values of $R$ are for $\tau = 2\cdot 2^m$, except for $\tau=0$, for all games fulfilling
the $NS\gg 2^P$ condition.}
%\end{figure}
%\begin{figure}[t]
\begin{center}
\begin{tabular}{ccc}
\includegraphics[scale=.20]{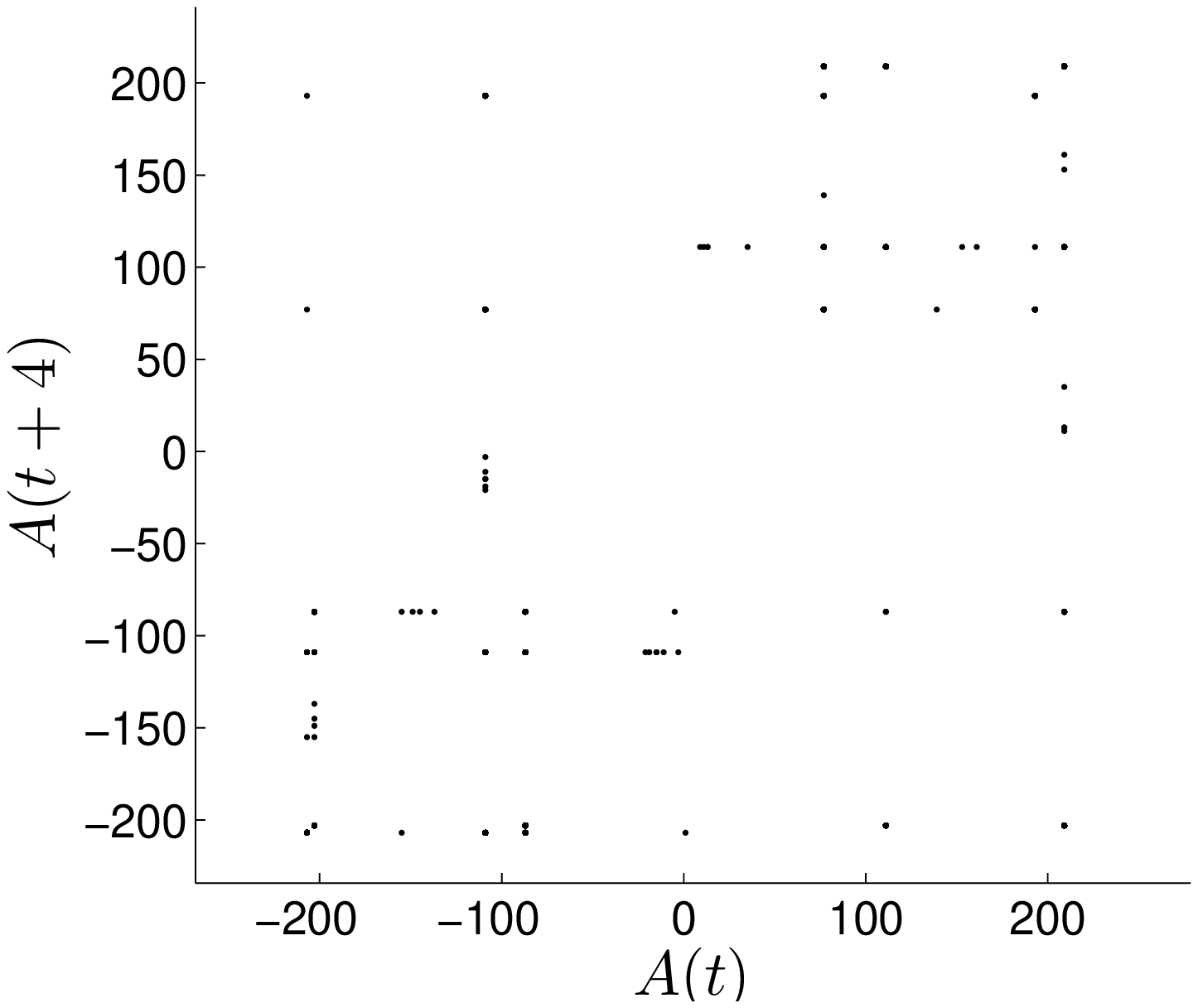} & \hspace{2mm}
\includegraphics[scale=.20]{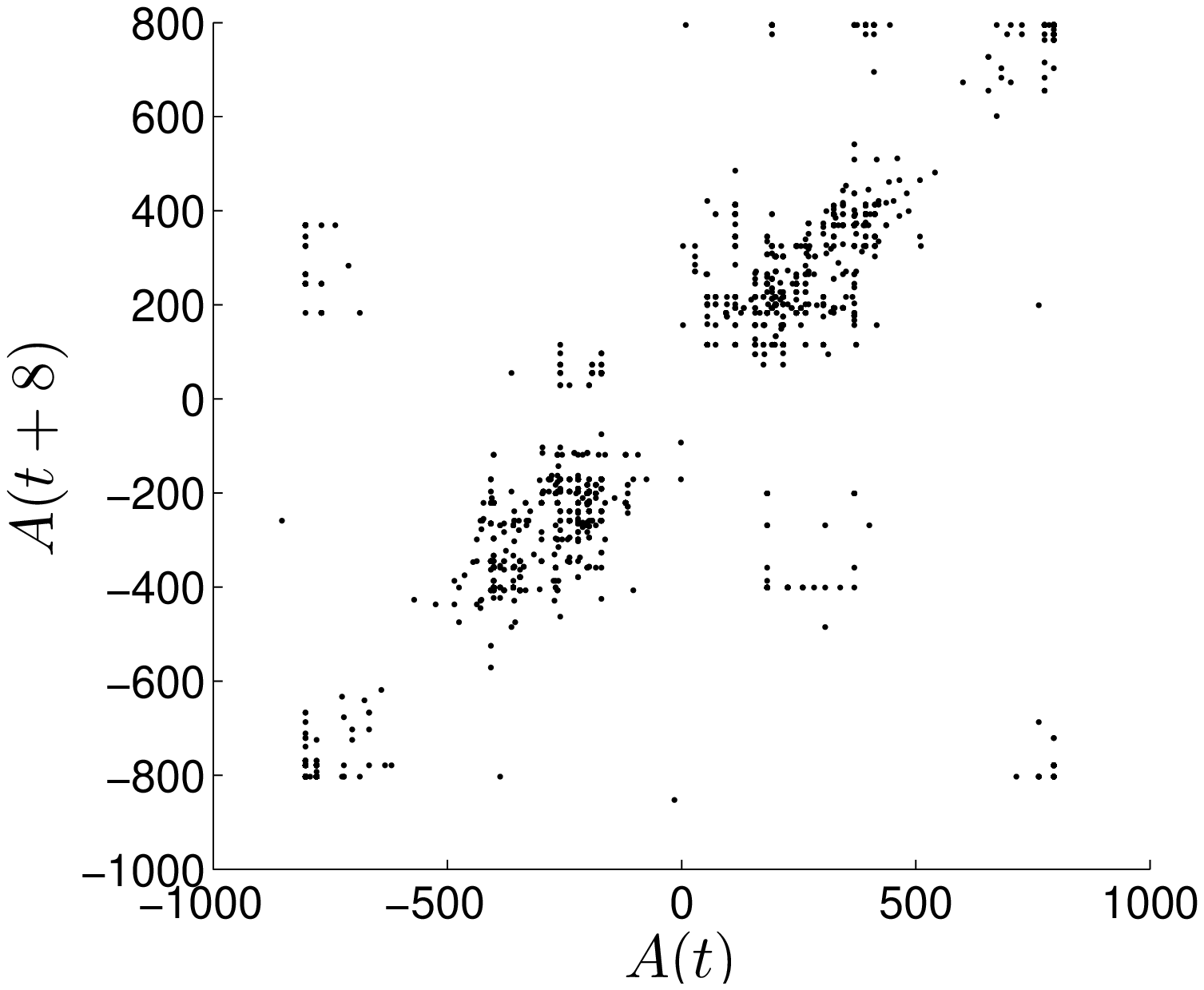} & \hspace{2mm}
\includegraphics[scale=.20]{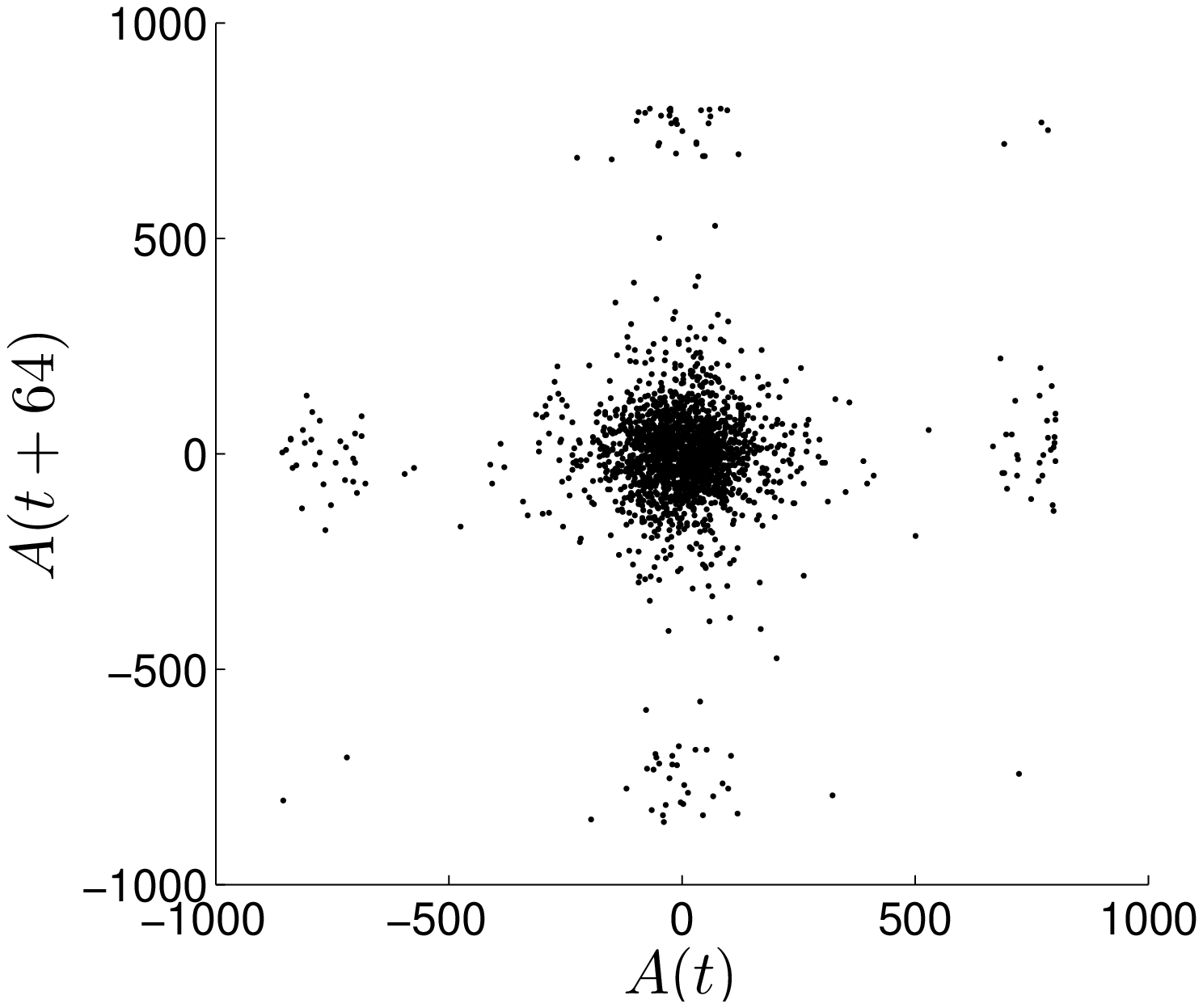}
%\hspace{-8pt} \includegraphics[scale=.27]{./fig36_a.eps} & \hspace{-8pt}
%\includegraphics[scale=.27]{./fig36_b.eps} & \hspace{-8pt}
%\includegraphics[scale=.27]{./fig36_c.eps}
\end{tabular}
\end{center}
\vspace*{8pt} \caption{\label{fig:AagainstA3_linear}\em Plots of the aggregated demand $A(t+2\cdot
2^m)$ vs. $A(t)$ for three combinations of the population size $N$ and agent memory $m$: $N=401$,
$m=1$ (left), $N=1601$, $m=2$ (middle) and $N=1601$, $m=5$ (right). Simulation was done for $S=2$
and $g(x)=x$. For $m=1$ and $m=2$ points tend to flock around diagonals, indicating positive
correlation, but clusterization of points is not much pronounced.}
\end{figure}

Figs \ref{fig:A3_signx}, \ref{fig:R3_signx} and \ref{fig:AagainstA3_signx} present results for the
steplike payoff function $g(x)=\mbox{sgn} (x)$: the time evolution of $A(t)$, the autocorrelation
function $R(\tau)$ and the scatter plots of $A(t+2\cdot 2^m)$ against $A(t)$, respectively. The
same results for the proportional payoff function $g(x)=x$ are given in Figs \ref{fig:A3_linear},
\ref{fig:R3_linear} and \ref{fig:AagainstA3_linear}.

Even a fleeting glance at Figs \ref{fig:A3_signx} and \ref{fig:A3_linear} reveals regularities in
$A(t)$ for both payoff functions but more regular and distinct for $g(x)=x$. In this case their
period increases with the memory length $m$ and their maximal values are equal to the half of the
population size $N/2$. This periodicity can be better seen using autocorrelation function $R(\tau)$
(cf. Figs \ref{fig:R3_signx} and \ref{fig:R3_linear}) where $\tau$ is the correlation time. The
autocorrelation $R$ exhibits statistically periodic peaks with

%\FloatBarrier
%
\FloatBarrier
\noindent periods $T=2\cdot 2^m$, as has been already observed in the efficient regime in
Refs.~\cite{zheng01PhysicaA301,jefferies01PhysRevE65}. The autocorrelation is much less pronounced
for games which do not meet the criterion $NS\gg 2^P$, as seen in Figs~\ref{fig:R3_signx} and
\ref{fig:A3_linear} (right). Relaxation of this criterion spoils periodicity of the aggregated
demand. Similar observations can be done inspecting the $A(t+2\cdot 2^m)$ {\it vs.} $A(t)$ scatter
plots in Figs \ref{fig:AagainstA3_signx} and \ref{fig:AagainstA3_linear} where points for games
fulfilling $NS\gg 2^P$ condition (left and middle panels in Figs \ref{fig:AagainstA3_signx} and
\ref{fig:AagainstA3_linear}) are stronger flocked around diagonals.

Another interesting feature of the aggregated demand, seen in the one-dimensional plots of $A(t)$,
and better in the two-dimensional plots $A(t+2\cdot 2^m)$ {\it vs.} $A(t)$, is an existence of
preferred values of $A$. These preferred values show up as specles in the two-dimensional plots.
The specles are better focused and more numerous for $g(x)=\mbox{sgn} (x)$ (Fig.
\ref{fig:AagainstA3_signx}) than for $g(x)=x$ (Fig. \ref{fig:AagainstA3_linear}).

\begin{figure}[h]
\begin{center}
\begin{tabular}{cc}
\includegraphics[scale=.38]{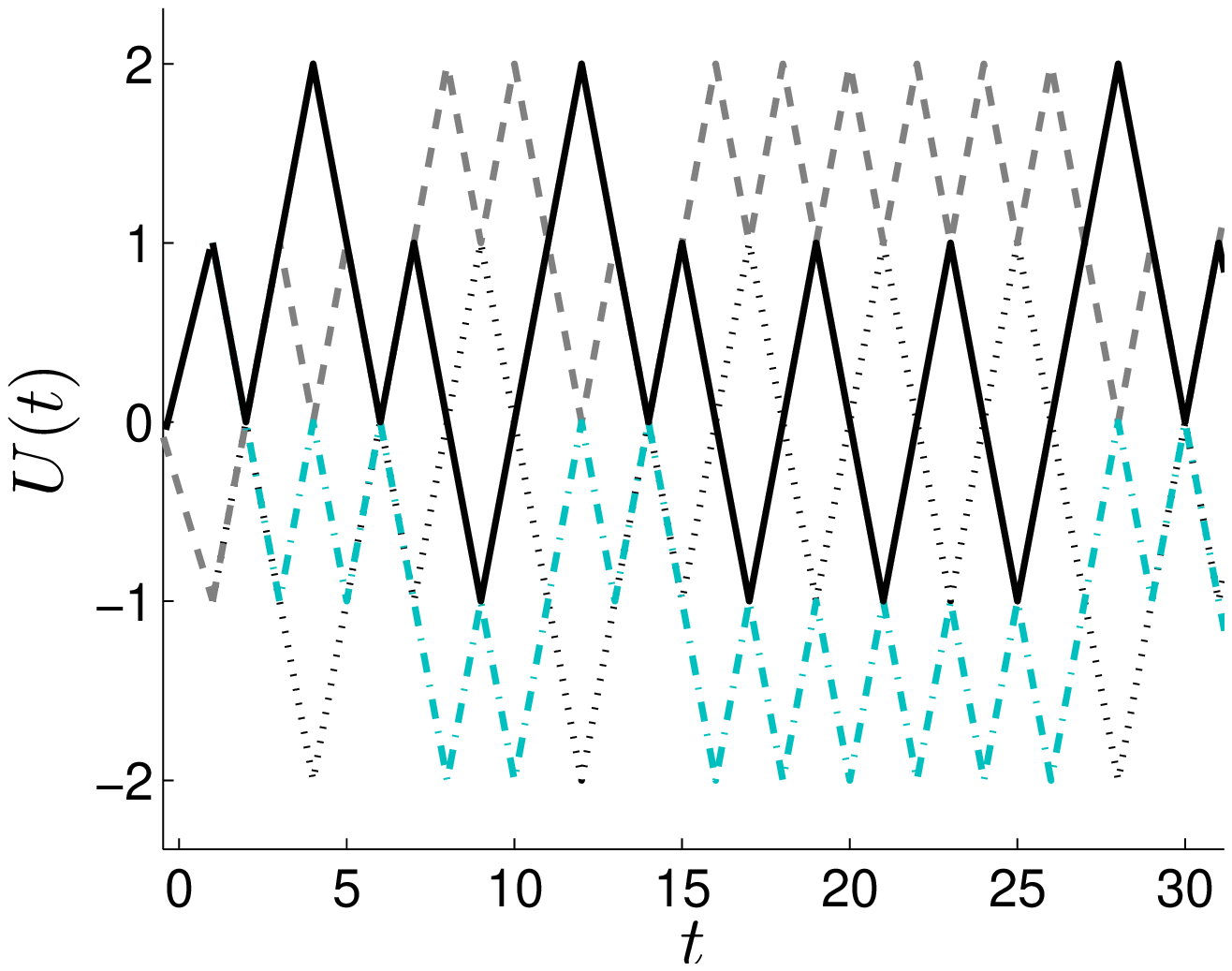} & \hspace{5mm}
\includegraphics[scale=.38]{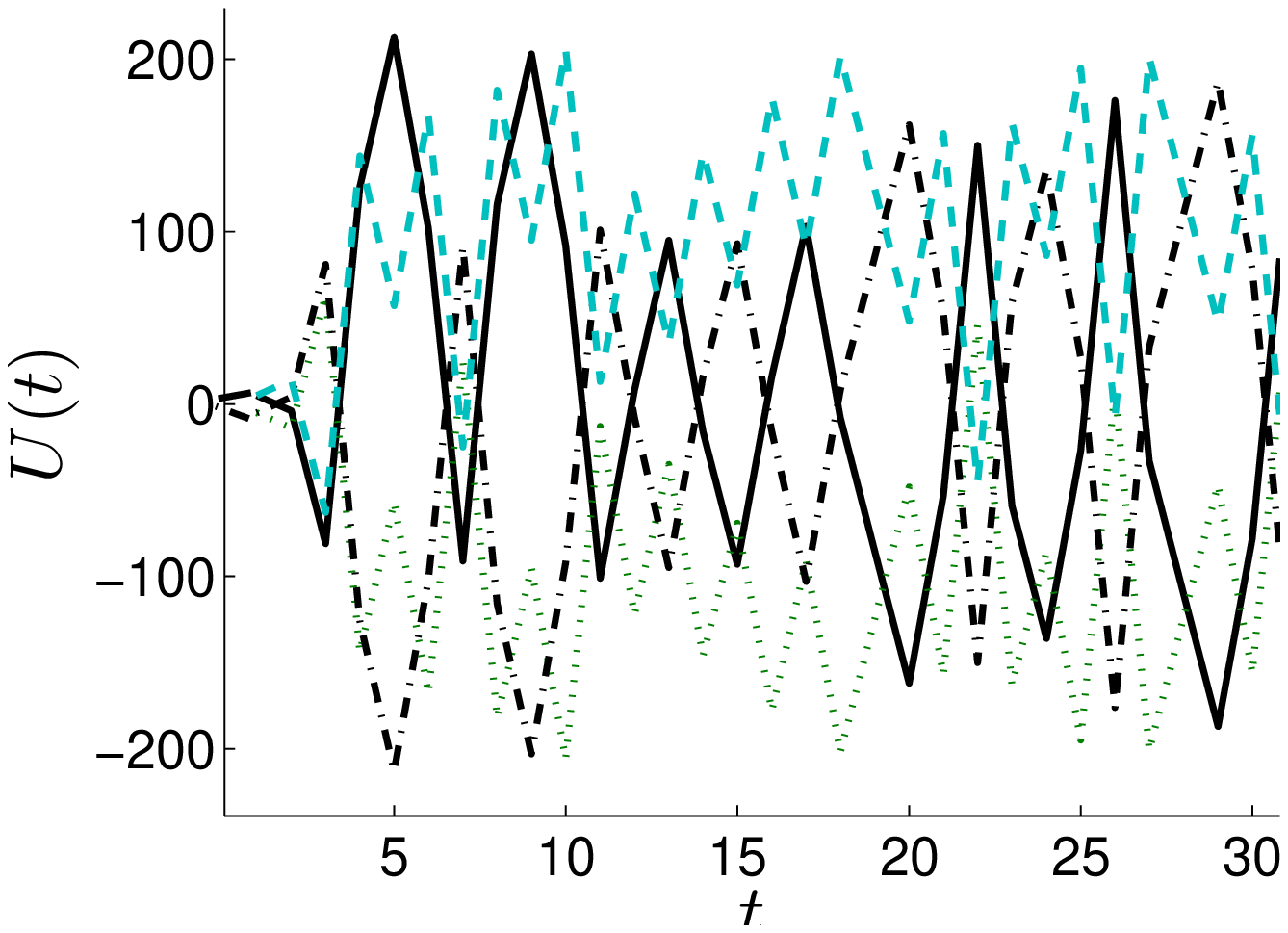}
\end{tabular}
\end{center}
\vspace*{8pt} \caption{\label{fig:UTrajectories}\em Trajectories of the utility function $U(t)$ for
all strategies of the MG with $S=2$ and $m=1$ and $N$ high enough to ensure the $NS\gg 2^P$ regime.
Two payoff functions are shown: the steplike $g(x)=\mbox{sgn}\,(x)$ (left) and the proportional
$g(x)=x$ (right). Lines correspond to all different strategies. Note difference of vertical scales
between panels.}
\end{figure}

Time evolution of the utility functions appears to be a strongly mean-reverting process,
independently of the payoff function, as seen e.g. in Figs \ref{fig:UTrajectories}. The more so,
for the steplike payoff $g(x)=\mbox{sgn} (x)$ the utility is bounded to rather narrow belt $-2^m\le
U(t)\le 2^m$, where here and in Fig.~\ref{fig:UTrajectories}, $U(t)$ stands for the utility for any
strategy. The formal proof of this statement is given in section \ref{sec:mesoscopic}. This feature
is observed for any $N$ and $S$, provided the criterion $NS\gg 2^P$ is met.

\section{The mesoscopic perspective}\label{sec:mesoscopic}
%\addcontentsline{toc}{chapter}{CHAPTER II \ \ Statistical
%properties of the model}\kappa
%\setcounter{equation}{0}  % reset counter
%\setcounter{chapter}{2}

In this section we present the effective description of MG by redefining MG as a Markov chain. The
general definition of state is found to be too complex for analytical treatments (cf. Sec.
\ref{sec:generalstate}). Fortunately, in the herd regime, many agents have identical sets of
strategies and their aggregation is possible. The set of individuals using the same strategies, called
fraction, is further treated as a single agent (cf. Sec. \ref{sec:fraction}).
All possible fractions exist provided the
game is large enough. Knowledge about utilities of pairwise different strategies and
history of past winning decisions are enough to predict the action of any fraction. This set
of parameters fully characterizes the system and is considered to specify its state.
This definition is strictly suitable only for a step-like payoff function and can slightly
vary for other payoffs (cf. Secs. \ref{sec:sgnPayoff} and \ref{sec:linearPayoff}).

Once the representation of the state is known, two methodologies are tried to explain the
observations. In the simplified case we assumed the quenched disorder~\cite{hart01EurPhysJB20}, i.e.
an initial random choice of the strategy set at the start of the game and its later fixation, and in addition
equality of fractions. However, not all observables are properly explained by
that means and an extension of these assumptions is needed.

Transition probabilities
can be calculated in two ways: before and after assignment of strategies to agents. We thus
distinguish between \emph{a priori} and \emph{a posteriori} probability distribution of the aggregate
demand.

Using this approach we manage to explain all observed phenomena. Finally, in Sec. \ref{sec:stability} we define and study
stability of this game in order to understand asymmetries observed in aggregate variables.

\subsection{Definitions}
\subsubsection{The general concept of state}\label{sec:generalstate}
Since the MG represents a system with many degrees of freedom, dimensionality of states is expected
to be large. In general, for each time step $t$, specification of state $x(t)$ consists of:
\begin{itemize}
\item[A.] The history of decisions $\mu(t)$,
\item[B.] The set of strategies of all agents $\{\alpha_n^s\}_{n=1,\ldots,N}^{s=1,\ldots,S}$,
\item[C.] The set of utilities for all strategies of all agents
$\{U_{\alpha_n^s}(t)\}_{n=1,\ldots,N}^{s=1,\ldots,S}$,
\item[D.] A function relating strategies to agents: $\rho(n):n\rightarrow\Delta_n$.
\end{itemize}
Although the history of decisions $\mu(t)$ partially stores information about the past of the
process, transition probabilities depend only on the present state and the process is Markovian.

Substantial reduction of the number of state parameters and simplification of state description are
possible if the game is large i.e. $NS \gg 2^P$ (cf. Secs. \ref{sec:sgnPayoff} and \ref{sec:linearPayoff}).

\subsubsection{Fraction -- definition and statistical properties}\label{sec:fraction}
All agents behaving in the same manner - the fraction - can, in a sense, be treated as a whole.
The fraction can be defined in two ways.

In the first
approach it is a set of agents possessing a given, all the same, set of $S$ strategies. The set
of pairwise different strategies \footnote{Two strategies are called different if the Hamming
distance between them is not equal to zero. The number of pairwise different strategies is equal to
$2^P$.} is denoted as $\{\beta_\kappa\}_{\kappa=1}^{2^P}$. The number of agents in the fraction
$\nu$, or the size of this fraction, is marked as $F_{\nu}$, where $\nu=\{1 \ldots G\}$ and $G$ is
the total number of different fractions. In large games, the system comprises agents of all
possible fractions what results in constant $G$. In general, if strategies are assigned to agents
randomly then $F_{\nu}$ are random variables. The strategy space consists of $2^P$ possible
strategies and $G$ is represented by the number of $S$-combinations with repetition: $G =
\binom{2^P+S-1}{S}$.

However, such definition of $G$ makes the expected values of the fraction
sizes, ${\mathbb E}[F_{\nu}]$, not equal for different fractions, provided that strategies are
randomly chosen from the uniform distribution. For example, assuming $S=2$, the fraction with two
the same strategies, e.g. $\{\beta_1, \beta_1 \}$, is two times smaller than fraction with
different strategies $\beta_1$ and $\beta_2$, where the ordering of strategies matters: $\{\beta_1,
\beta_2\}$ or $\{\beta_2, \beta_1\}$. Therefore in the sequel we use another definition: the
fraction is a set of agents using given sequence of strategies. The fraction size is now equal to
$G=2^{PS}$. In such definition the strategy index $s\in\{1,\ldots,S\}$ is dummy. Nevertheless we
use this approach because it radically simplifies the analysis without biasing the outcome,
assuming assigning agents to fractions with equal probabilities. For example, consider the case
$S=2$. Fractions' indexes are assigned to each pair of strategies arbitrarily, e.g. as presented in
Tab \ref{tab:fractions}.
\begin{table}[h]
{\center
\begin{tabular}{|c|c|c|c|c|c|c|c|c|}
  % after \\: \hline or \cline{col1-col2} \cline{col3-col4} ...
\hline
$F:$ & $F_1$ & $F_2$ & $F_3$ & $F_4$ & $F_5$ & $F_6$ & $F_7$ & $F_8$\\
$\beta:$ & ${\beta_1, \beta_1}$ & ${\beta_1, \beta_2}$ & ${\beta_1, \beta_3}$ & ${\beta_1,
\beta_4}$ &
${\beta_2, \beta_1}$ & ${\beta_2, \beta_2}$ & ${\beta_2, \beta_3}$ & ${\beta_2, \beta_4}$  \\
\hline
\end{tabular}

\begin{tabular}{|c|c|c|c|c|c|c|c|c|}
\hline $F:$ & $F_9$ & $F_{10}$ & $F_{11}$ & $F_{12}$ & $F_{13}$ & $F_{14}$ & $F_{15}$ & $F_{16}$\\
$\beta:$ & ${\beta_3, \beta_1}$ & ${\beta_3, \beta_2}$ & ${\beta_3, \beta_3}$ & ${\beta_3,
\beta_4}$ & ${\beta_4, \beta_1}$ & ${\beta_4, \beta_2}$ & ${\beta_4, \beta_3}$ & ${\beta_4, \beta_4}$ \\
\hline
\end{tabular}
\caption{\label{tab:fractions}\em One of possible assignments between fraction's indexes and pairs
of strategies}
%NIE usuwac tabelki - potrzebna do wyjasnienia example 2 w 'the case of equal-size fractions'
}
\end{table}

\noindent If at the beginning of the game strategies are drawn with equal probabilities, it corresponds to
assigning agents to a specific fraction with probabilities $\frac{1}{G}$. Assume
$W_{\nu}^n\in\{0,1\}$ is a random variable equal to 1 if agent $n$ belongs to fraction $\nu$. Then
$F_{\nu} = \sum_{n=1}^N W_{\nu}^n$ follows the binomial distribution and $Pr(F_{\nu} = f_{\nu}) =
\binom{N}{f_{\nu}} (\frac{1}{G})^{f_{\nu}} (1 - \frac{1}{G})^{N - f_{\nu}}$. Hence, ${\mathbb
E}[F_{\nu}] = N/G$ or, if normalized, ${\mathbb E}[F_{\nu}/N] = 1/G$.

For $N\rightarrow \infty$, we have ${\mathbb Var}[F_{\nu}] \rightarrow \infty$ and ${\mathbb
Var}[F_{\nu}/N] \rightarrow 0$ \footnote{After normalization the random variable $Z_{\nu}^n =
W_{\nu}^n/N\in \{0, \frac{1}{N}\}$ obeys the Bernoulli distribution with $Pr(Z_{\nu}^n=0) =
1-\frac{1}{G}$ and $Pr(Z_{\nu}^n=\frac{1}{N})= \frac{1}{G}$. Hence, ${\mathbb E}[Z_{\nu}^n] =
\frac{1}{GN}$ and ${\mathbb Var}[Z_{\nu}^n] = \frac{1}{GN^2} (1 - \frac{1}{G})$. Resultantly,
${\mathbb Var}[F_{\nu}/N = \sum_{n=1}^N Z_{\nu}^n] = \frac{1}{GN} (1 - \frac{1}{G})$.}. This means
that, asymptotically for large $N$, (i)  the absolute differences between sizes of fractions grow
indefinitely, and (ii) percentages of population assigned to any fraction are equal. Hence, the
larger the population, the larger expected difference between an actual size of a fraction
$F_{\nu}$ and its expected value ${\mathbb E[F_{\nu}]}$.

\subsubsection{Stability}\label{sec:stability} The game is considered stable if for any
strategy $\alpha_n$ the corresponding utility $U_{\alpha_n}(t)$ represents a mean-reverting
stochastic process, i.e. the time-average of its increments vanishes after sufficiently long time.
The MG has a build-in stabilization mechanism provided the game is large enough. The explanation is
as follows.

Imagine that a subset $Z$ of strategies ($Z \subset \{\beta_1,\ldots ,\beta_{2^{P}}\}$) gets on
average higher payoff than other subsets and the utilities in $Z$ grow up.
Then, there always exists the same number of
anticorrelated strategies with decreasing utility. The probability that an agent uses one of the
strategies with a high utility is $1 - (\#Z/2^P)^S$, compared to those who use strategies with a
low utility $(\#Z/2^P)^S$~\cite{wawrzyniak09ACSNo6} ($\#Z$ is the number of elements in $Z$).
Since the former probability is always higher, provided $S \geq 2$, then the
most of population uses better strategies and their utility decreases, i.e. the game stabilizes. As
long as fraction sizes are close to each other the above mechanism works and the game stays stable.
\subsection{The payoff \mbox{g(x) = sgn(x)}\label{sec:sgnPayoff}}
%\section{The payoff $\mathbf{g(x)=\mbox{sgn} (x)}$\label{sec:sgnPayoff}}
Here, the concept of the state for payoff $g(x)=sgn(x)$ is introduced. Applying it allows to
represent the game as a Markov process and constitutes a consistent basis for analytical
explanations of phenomena in the herd regime.
\subsubsection{The concept of the state}
Substantial reduction of the number of state parameters and simplification of state description are
possible in our case. Agents can use identical strategies. The expected number of identical strategies
in the whole population behaves asymptotically, for $N\rightarrow\infty$, like $NS/2^P$. The
condition $NS\gg 2^P$ assures that the game stays in that asymptotic regime and the number of
identical strategies is close to its asymptotic expected value. Identical strategies have the same
utilities over the whole game, provided the initial values of utilities are the same, e.g.
$U(0)=0$, for all strategies. It is thus enough to take into account only reduced set of pairwise
different strategies $\{\beta_{\kappa}\}_{\kappa=1}^{2^P}$ and utilities defined on them, and
therefore B and C from section~\ref{sec:generalstate} can be reduced:
\begin{itemize}
\item[B.] $\{\alpha_n^s\}_{n=1,\ldots,N}^{s=1,\ldots,S}\longrightarrow
\{\beta_{\kappa}\}_{\kappa=1}^{2^P}$, \item[C.]
$\{U_{\alpha_n^s}(t)\}_{n=1,\ldots,N}^{s=1,\ldots,S}\longrightarrow
\{U_{\beta_{\kappa}}(t)\}_{\kappa=1}^{2^P}$.
\end{itemize}
Concerning point D, it is sufficient to find probabilities for agents to have strategies from the
set of pairwise different strategies. The probability that given agent has any particular strategy
from this set is equal to $1-(1-1/2^P)^S$. For large $N$, the expected number of agents having this
strategy is equal to $N(1-(1-1/2^P)^S)$. Therefore point D, i.e. a function ascribing strategies to
agents, corresponding to the agent grouping tensor $\Omega$ of Ref.~\cite{jefferies01PhysRevE65},
can be dropped out entirely in this case. Note that this expected number in general differs from
the actual number, which has some consequences explained later.

Finally, we describe states using $\mu(t)$ and the set of utilities for the complete set of $2^P$
pairwise different strategies $\{\beta_{\kappa}\}_{\kappa=1}^{2^P}$:
\begin{eqnarray}
x(t)=[\,\mu(t),\,U_1(t),U_2(t),\ldots,U_{2^P}(t)\,]. \label{eq:state_sgnx}
\end{eqnarray}

Similar description of state was used in Ref.~\cite{jefferies01PhysRevE65} but there are two
important differences between these two: (i) the authors of
Ref.~\cite{jefferies01PhysRevE65} introduce a functional map giving time evolution of the system in
any regime, and (ii) they degenerate the game by following mean values of demand, thus making the
process deterministic and Markovian, and retaining possibility to randomize it perturbatively.
Contrary to them, we do not degenerate the game. We consider it as a stochastic Markov process and
eventually calculate the probability measure on states for the steplike payoff.

Utilities $\{U_{\beta_{\kappa}}(t)\}_{\kappa=1}^{2^P}$, considered as functions of time, are called
{\it trajectories}. In the majority of cases and provided the number of observed time steps is
large enough, strategies can be distinguished by their trajectories. The sufficient condition for
all $2^P$ trajectories $U_{\beta_{\kappa}}(t)$ $(0\le t\le t_0)$ to be distinguishable at $t_0$ is
that all $2^m$ possible histories $\mu$ appear until then in a row. On the other hand, appearance
of all histories $\mu$ until $t_0$, but not necessarily exclusively, represents a necessary
condition of distinguishability for trajectories. Examples of MGs in the regime $NS\gg 2^P$ are
shown in Figs~\ref{fig:UTrajectories} where trajectories are plotted for $m=1$ and $S=2$, and for
two payoff functions further studied in this paper: $g(x)=\mbox{sgn} (x)$ and $g(x)=x$.

\subsubsection{Finiteness of the number of states}
In this section we demonstrate that for any $t$ the
utility for any strategy is bounded from the bottom and top: $U_{min}\le U(t)\le U_{max}$, where $U_{min(max)}=-(+)\, 2^m$.
%$U_{min(max)}=-\mbox{\scriptsize (}+\mbox{\scriptsize )}2^m$.
At least two approaches are possible.
In the first approach one aggregates agents using strategies of a given utility value. Another one
is based on fractions. Here we elaborate in detail on the former one and only present the sketch of
proof of the latter.

Assume that at given time $t$ two different strategies have the same utilities. From
Eq.~(\ref{eq:U}) for the steplike payoff function it follows that after one time step these
utilities can either differ by two units or remain the same. If the initial values of the utilities
at $t=0$ are the same and after $\tau$ time steps at least one of them attains its extremal value,
$U_{min}$ or $U_{max}$, then the trajectories cover the set of $2^m+1$ values (cf.
Fig.~\ref{fig:UTrajectories}, left)
\begin{eqnarray}
U(\tau) & \in & \{u_l\}_{l=1}^{2^m+1} \nonumber \\
         & = & \{2^m, 2^m-2, \ldots, 2, 0, -2, \ldots, -2^m+2, -2^m\}.
%\label{eq:UScope}
\end{eqnarray}
Using this notation we have $u_1=U_{max}$ and $u_{2^m+1}=U_{min}$. The number of different
strategies characterized by the same $u_l$ is given by combinatorics as the number of trajectories
starting from 0 and ending at $u_l$ is
\begin{eqnarray}
\#\{\beta_{\kappa}: U_{\beta_{\kappa}}=u_l\}=\left (\begin{array}{c} U_{max} \\ l-1
\end{array}\right ), \quad\quad l=1,\ldots,2^m+1. \label{eq:nrOfTrajectories}
\end{eqnarray}
The probability that the active strategy of the $n$-th agent $\alpha_n^\prime$ has utility $u_l$ is
equal to
\begin{eqnarray}
Pr\big [U_{\alpha_n^\prime}(t)=u_l\big ]=\left\{\begin{array}{lr} 1-Pr\big
[U_{\alpha_n^\prime}(t)<u_l\big ], & \quad l=1 \\ Pr\big [U_{\alpha_n^\prime}(t)<u_{l-1}\big
]-Pr\big [U_{\alpha_n^\prime}(t)<u_l\big ], & \quad l>1 \end{array} \right .
\label{eq:PrOfActiveStrEqUl}
\end{eqnarray}
Using argumentation similar to that of Ref.~\cite{hart01EurPhysJB20}, but extended to the full
strategy space, one finds that
\begin{eqnarray}
Pr\big [U_{\alpha_n^\prime}(t)<u_l\big ] & = & \prod_{s=1}^S \Big [1-Pr\big [U_{\alpha_n^s}(t)\ge u_l\big ]\Big ] \nonumber \\
                                                 & = & \Big [ 1-\frac{\#\{\beta_{\kappa}: U_{\beta_{\kappa}}\ge u_l\}}{2^P}\Big ]^S,
%\label{eq:PrOfActiveStrEqOrSmallerUl}
\end{eqnarray}
where, for $t=\tau$,
\begin{eqnarray}
\#\{\beta_{\kappa}: U_{\beta_{\kappa}}\ge u_l\}=\sum_{j\ge l} \left (\begin{array}{c} U_{max} \\
j-1
\end{array}\right ).
\label{eq:nrOfBeta}
\end{eqnarray}
Denoting $Pr_{max(min)}=Pr\big [U_{\alpha_n^\prime}(\tau)=U_{max(min)}\big ]$, one sees from
Eq.~(\ref{eq:PrOfActiveStrEqUl}) that $Pr_{max}>Pr_{min}$. For any utility $u_l$, different than
$U_{min}$ or $U_{max}$, the number of different strategies (\ref{eq:nrOfTrajectories}) is even.
Even more, a half of strategies corresponding to each level $U_{min}< u_l < U_{max}$ suggests the
opposite action than another half. According to Eq.~(\ref{eq:PrOfActiveStrEqUl}), if two (or more)
strategies have the same utility, then all have the same probability to be the best strategies for
the $n$-th agent. This means that, if one excludes the best and the worst strategies, a half of
remaining strategies recommends the same action as the best or the worst strategy. Hence the
probability that an agent plays according to the strategy suggesting the same action as the best
strategy is equal to
\begin{eqnarray}
Pr\big [a_{\alpha_n^\prime}(\tau)=a_{\alpha^B}(\tau)\big ] & = & Pr_{max}+\frac{1}{2}\big (1-Pr_{max}-Pr_{min}\big ) \nonumber \\
 & = & \frac{1}{2}\big (1+Pr_{max}-Pr_{min}\big ),
\label{eq:PrAgentPlaysAsBestStr}
\end{eqnarray}
where $\alpha^B(t)$ is the best strategy from the whole set of strategies in the game, i.e.
$U_{\alpha^B(t)}=u_1$, and $1-Pr_{max}-Pr_{min}$ refers to the probability that the agent's best
strategy is neither the worst nor the best of all strategies. The factor $\frac{1}{2}$ reflects
that a half of strategies with non-extremal utilities suggests the same action as the best one. As
$Pr_{max}>Pr_{min}$, from Eq.~(\ref{eq:PrAgentPlaysAsBestStr}) it follows that if one of strategies
has the utility $U_{max}$, then more than half of the population plays according to the best
strategy. Subsequently, this subpopulation loose and gets the negative payoff. The rest are the
winners and get the positive payoff. This mechanism bounds the utility to stay between $U_{min}$
and $U_{max}$. In addition, we know the formula for the fraction of agents playing the same action.
For example, if $S=2$ and $m=1$ then $Pr_{max} = \frac{7}{16}$ and $Pr_{min} = \frac{1}{16}$.
Hence, ${\mathbb E}[A] = \frac{3}{8}N$.

The analogical results are achieved when the concept of fraction is used. The number of
different strategies characterized by the levels $u_l$ follows Eq.~(\ref{eq:nrOfTrajectories}).
Additionally, for all intermediate levels $U_{min}< u_l < U_{max}$ there exists the same number of
strategies that suggest $+1$ and $-1$. Hence, all fractions that use one of these intermediate
strategies compensate on average their mutual decisions. The last point is to find the number of
fractions that use the best and the worst strategy, which are equal to $2^{PS} - (2^P-1)^S$ and $1$,
respectively. For example, for $S=2$ and $m=1$ there are $G=2^{PS} = 16$ fractions: seven using
the best strategy and one using the worst one. Hence, ${\mathbb E}[A] = \frac{7}{16}N -
\frac{1}{16}N = \frac{3}{8}N$, in compliance with the previous example.

\subsubsection{The Markov process representation}\label{sec:MPsgnx}
The MG can be described in terms of the Markov process with the finite number of states. The
$\mbox{sgn}\,A(x_i)$ fully defines the utility and $\mu$ values of the next state and takes $\pm 1$. But in
some specific states $A(x_i)$ is always positive or negative and only one value of
$\mbox{sgn}\,A(x_i)$ appears. Hence, the transition may be either stochastic or deterministic and
the transition probability is equal to
\begin{eqnarray}
Pr(x_j|x_i) = \frac{1}{2}\big({\mathbb E}[\mbox{sgn}\,A(x_i)]+1\big). \label{eq: TransitionPr}
\end{eqnarray}
The probability (\ref{eq: TransitionPr}) depends only on the shape of the distribution \footnote{The
lack of explicit dependence of $Pr(x_j|x_i)$ on $x_j$ in Eq.~(\ref{eq: TransitionPr}) does not mean
that both transition probabilities are the same for stochastic transition. They can be different
for asymmetric distribution of $A(x_i)$ (cf. discussion in sec. 4.4.1 below).} of $A(x_i)$. Using
the concept of fractions, we redefine $A(x_i)$ as follows:
\begin{eqnarray}
A(x_i) = \sum_{\nu} C_{\nu}(x_i) F_{\nu} \label{eq: AasFractions},
\end{eqnarray}
where $C_{\nu}(x_i)\in [-1,1]$ is a common action of all members of the fraction $\nu$ in the state
$x_i$
\begin{eqnarray}
C_{\nu}(x_i) = \frac{1}{F_{\nu}}\sum_{n=1}^{F_{\nu}} a_{\alpha_n^{'}(x_i)}.
\end{eqnarray}
In other words, $C_{\nu}(x_i)$ represents the aggregated demand {\it per capita} within
fraction~$\nu$. The $C_{\nu}(x_i)$ depends on the action suggested in the state $x_i$ by the best
strategy, or strategies, of the $\nu\,$-th fraction.

There are the following groups of fractions:
\begin{itemize}
\item Fractions with only one best strategy in the state $x_i$. All agents in the fraction react
according to this strategy.
\item Fractions with many best strategies where all best strategies in a given fraction suggest the same action in the
state $x_i$. Although agents use different strategies, they all react identically.
\item Fractions with many best strategies, where for each fraction some of the best strategies suggest the opposite action than another ones in the state $x_i$.
Actions of agents are thus inhomogeneous and an overall action of such fraction is a random variable $C_{\nu}(x_i)$,
taking values $c_{\nu}^{\varphi} = \frac{-F_{\nu}+2\varphi}{F_{\nu}}$ for
$\varphi=\{0,\ldots,F_{\nu}\}$, where $\varphi$ represents the possible numbers of agents acting $-1$ in the fraction
$\nu$. This distribution depends on a proportion between best strategies suggesting
opposite actions. Assuming there is $p^{+(-)}$ strategies suggesting the positive (negative)
action, the $C_{\nu}(x_i)$ obeys the binomial distribution
\begin{eqnarray}
 Pr\big(C_{\nu}(x_i) = c_{\nu}^{\varphi}\big) = \binom{F_{\nu}}{\varphi} \Big(\frac{p^+}{p^+ +
p^-}\Big)^{\varphi} \Big(\frac{p^-}{p^+ + p^-}\Big)^{F_{\nu} - \varphi},\label{eq: PrC}
\end{eqnarray}
where ${\mathbb E}[C_{\nu}(x_i)] = -1 + 2\Big(\frac{p^+}{p^+ + p^-}\Big)$ and ${\mathbb
Var}[C_{\nu}(x_i)] = \frac{4}{F_{\nu}}\Big(\frac{p^+}{p^+ + p^-}\Big) \Big(\frac{p^-}{p^+ +
p^-}\Big)$.
\end{itemize}

Fractions from the first two groups and suggesting $+1$ are marked with $d$, suggesting
$-1$ are marked with $q$, and those belonging to the third group are indexed with $w$. Hence, Eq.
(\ref{eq: AasFractions}) transforms into:
\begin{eqnarray}
 A(x_i) & = & \sum_{F_d:C_{d}(x_i)=1} F_d - \sum_{F_q:C_{q}(x_i)=-1} F_q \nonumber\\
& & + \sum_{F_w:C_{w}(x_i)\in [-1,+1]} C_{w}(x_i) F_w, \label{eq: ADecomposed}
\end{eqnarray}
where
\begin{eqnarray}
{\mathbb Var}[A(x_i)] = 4 \sum_{F_w:C_{w}(x_i)\in [-1,+1]} F_w \Big(\frac{p^+}{p^+ + p^-}\Big)
\Big(\frac{p^-}{p^+ + p^-}\Big). \label{eq: VarAglobal}
\end{eqnarray}
Further analysis is relatively easy when fractions are of equal sizes and it complicates if their
sizes are random.
%\subsubsubsection{The case of equal-size fractions}
\newline
\newline
\noindent{\it The case of equal-size fractions}

\noindent The system with the same numbers of agents per fraction we call the \emph{reference
system} and the corresponding MP -- the \emph{reference MP}. The $A(x_i)$ is a random variable
which can be expressed as:
\begin{eqnarray}
A(x_i) = \frac{N}{G} \bigl( D - Q + \sum_{F_w:C_{w}(x_i)\in [-1,+1]}C_{w}(x_i) \bigr), \label{eq:
ADecomposedEF}
\end{eqnarray}
where $D$ and $Q$ refer to the total numbers of fractions from the first two groups suggesting $+1$
and $-1$, respectively. If the state is deterministic then the components with opposite signs do
not compensate and
\begin{eqnarray}
 |D-Q|>\mbox{max}\,\big(\sum_{F_w:C_{w}(x_i)\in [-1,+1]}C_{w}(x_i)\big). \label{eq: deterministicIneq}
\end{eqnarray}
In the limit $NS\rightarrow\infty$, inequality (\ref{eq: deterministicIneq}) is satisfied always
when the negative and positive components are unbalanced, i.e. $D \neq Q$. This can be proved at
least in two ways.

The general proof uses the strong law of large numbers where the sample average $C_{w}(x_i)$
converges almost surely to the expected value, i.e.
\begin{eqnarray}
Pr(\lim_{N \rightarrow \infty} C_{w}(x_i) = {\mathbb E}\,\big[ C_{w}(x_i) \big]) = 1.
\end{eqnarray}
Each ${\mathbb E}\,\big[ C_{w}(x_i) \big]$ is equal to zero. Therefore the sum over
$F_w:C_{w}(x_i)\in [-1,+1]$ is equal to zero as well.

Another approach is applicable not only in the limit and requires separate analyses per state, as
given in \emph{Example 2}. For stochastic transitions there is always $D = Q$. For such states,
$A(x_i)$ has distribution symmetric around zero, ensuring that also distribution of
$\mbox{sgn}\,A(x_i)$ is symmetric and ${\mathbb E}[\mbox{sgn}\,A(x_i)]=0$. Thus, transitions to two
following states are equally probable.

Knowing how to distinguish the deterministic and stochastic
states, the algorithm of defining the MP is the following:
\begin{enumerate}
\item Consider all $2^m$ initial states. Such states are characterized by $U_{\beta_{\kappa}} = 0$
for all strategies $\kappa$ and different histories $\mu$. Due to equality of all strategies, two
minority decisions are equally possible for each of initial states and the transition is
stochastic. These minority decisions determine strategies that get positive or negative payoff. The
updated $U$ and $\mu$ values determine $2^{m+1}$ next states.
\item If, in the next state, there are many best pairwise different strategies suggesting opposite
actions, then $D=Q$ and, again, two minority decisions and two successive states are possible, and
the transition is stochastic. Hence, two next states have to be determined.
\item If, in the next state, there are many best pairwise different strategies suggesting the same
action, then $D\neq Q$ and the minority decision is determined by this action, and transition is
deterministic.
\item If there is only one strategy characterized by the highest value of the utility, then $D\neq
Q$ and the minority decision is determined by the best strategy, and transition is deterministic.
\end{enumerate}
Here we illustrate how one can find subsequent states and their transition probabilities using the
algorithm presented above (\emph{Example 1}). Next, in the \emph{Example 2} we show how to check
step-by-step that the transition is stochastic/deterministic assuming finite number of agents.
\newline

{\it \noindent Example 1: transition scenarios for $m=1$ case}

\begin{figure}[h]
\begin{tabular}{cc}
\includegraphics[scale=.24]{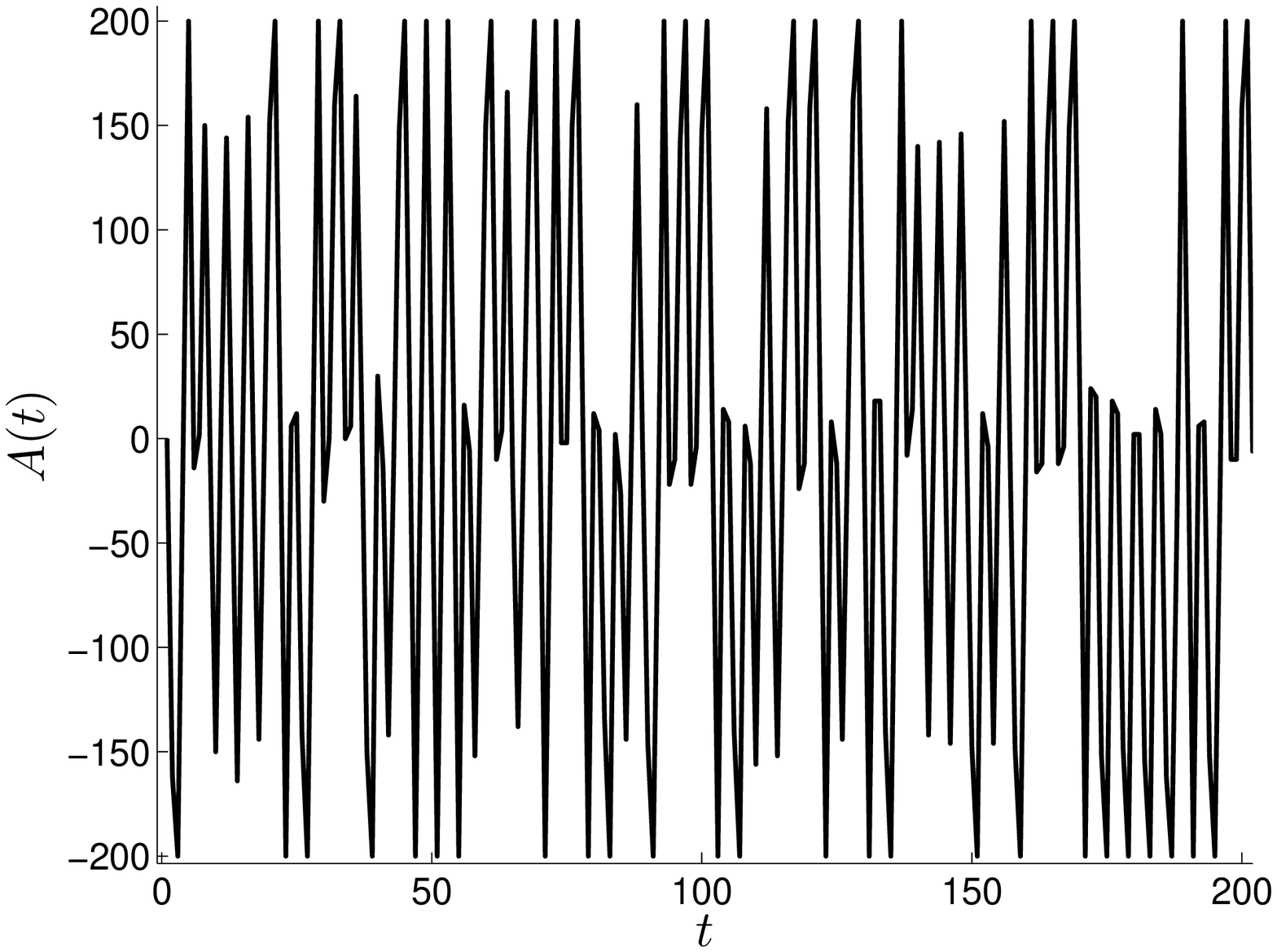} & \hspace{5mm}
\includegraphics[scale=.24]{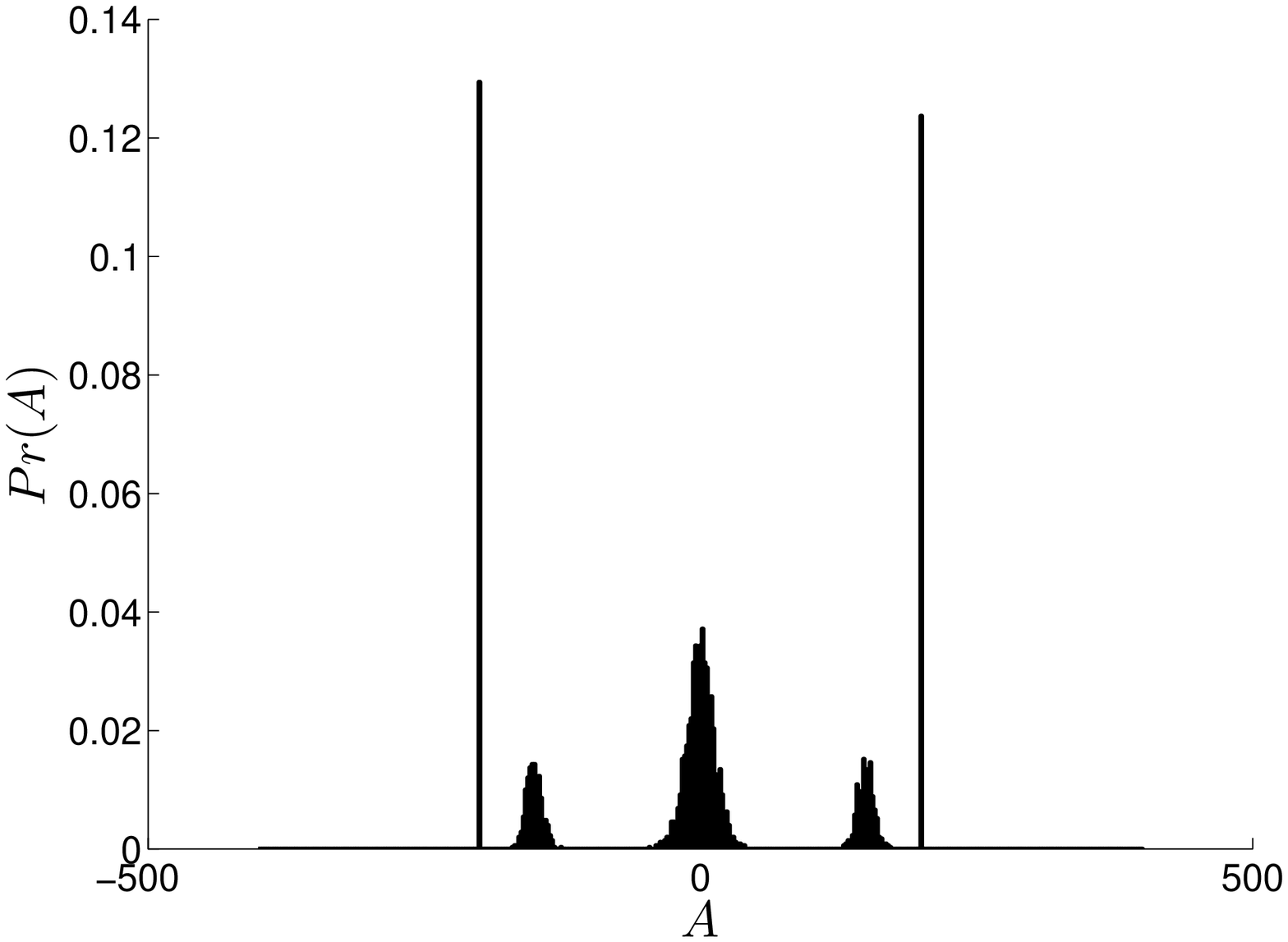}\\
\includegraphics[scale=.24]{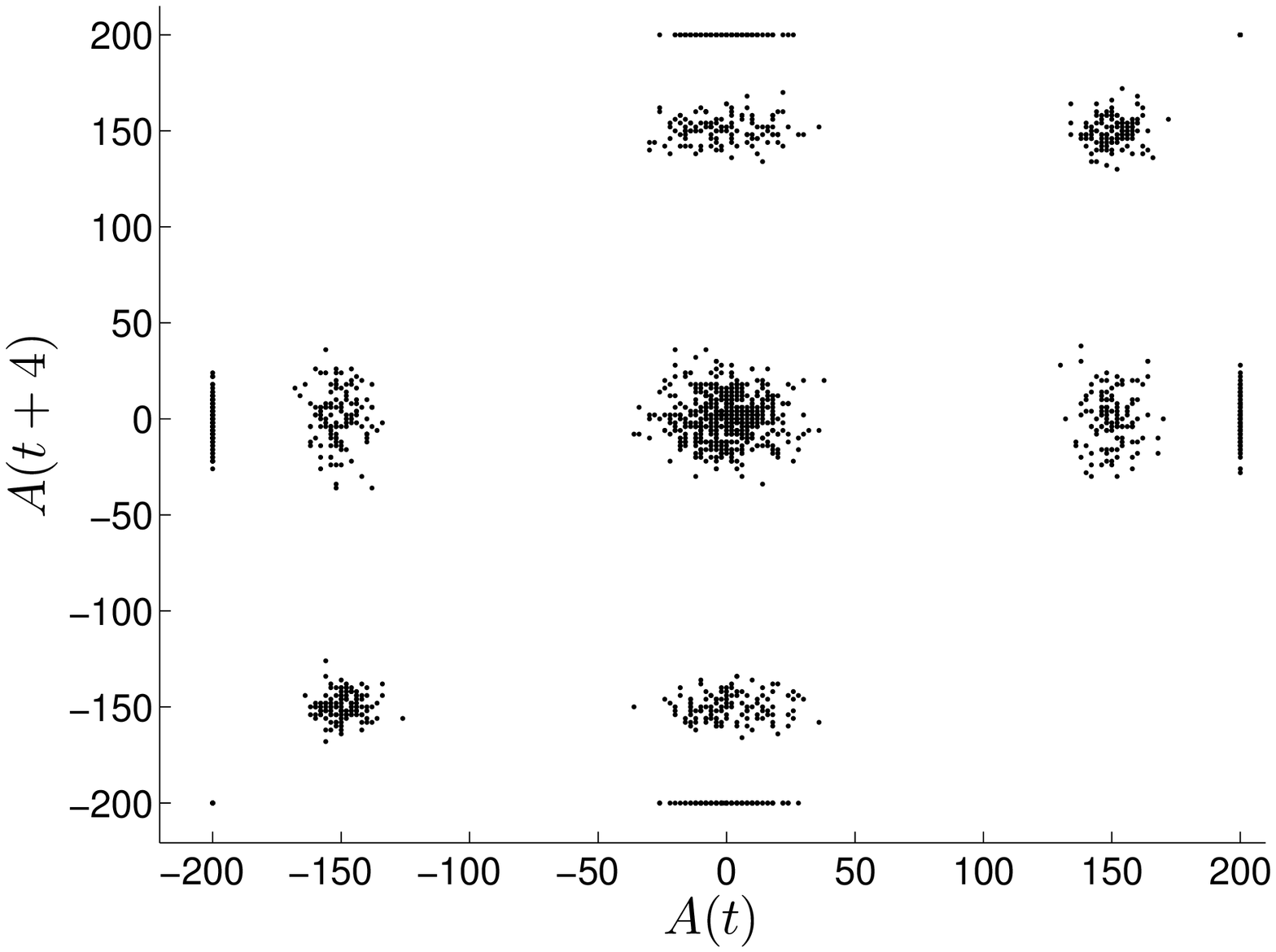} & \hspace{5mm}
\includegraphics[scale=.24]{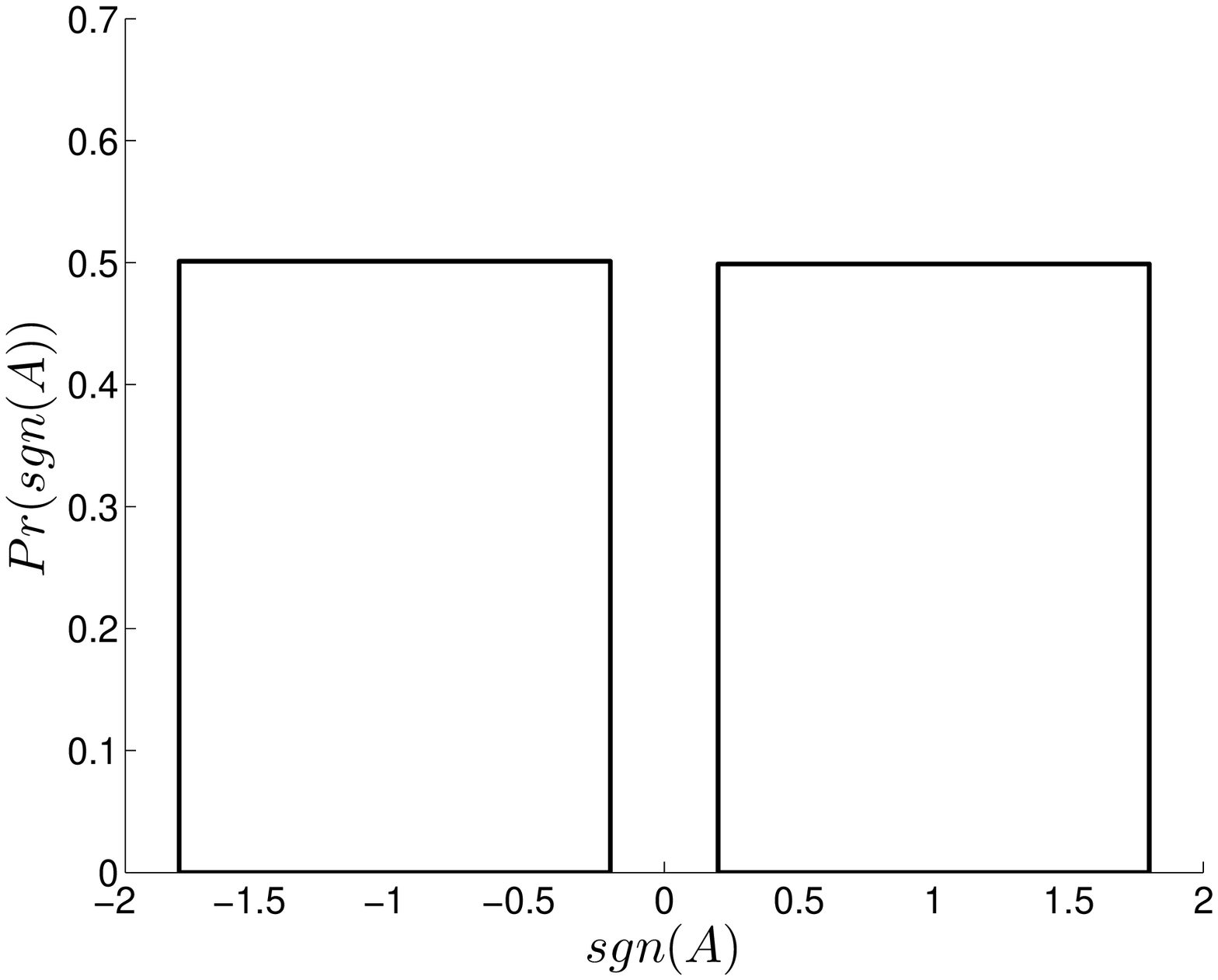}
\end{tabular}
\caption{\label{fig:A_sgnxRegular}\em Time evolution of the aggregated demand $A(t)$ (upper left),
Plots of the aggregated demand $A(t+2\cdot 2^m)$ vs. $A(t)$ (lower left), Estimated $Pr(A)$ (upper
right) and $Pr(\mbox{sgn}(A))$ (lower right) for the population size $N = 400$ and agent memory $m
= 1$, $S=2$ strategies per agent and identical sizes of fractions.}
\end{figure}

\noindent An example realization of the $A(t)$ for the reference MP is given in Fig.
\ref{fig:A_sgnxRegular} (upper left). The estimated $A$-distribution is symmetric (upper right) and
the distribution of $\mbox{sgn}(A)$ is symmetric likewise (lower right)\footnote{Small asymmetries
visible in Fig. \ref{fig:A_sgnxRegular} are due to finite number of samples used for estimation.}.
The scatter plot of $A(t+\tau)$ as a function of $A(t)$, where $\tau = 2^{m+1}$, indicates
periodicity and existence of preferred values of $A$ (lower left).
In this case the complete specification of states and calculation of the transition matrix are
relatively easy. All strategies are listed in Tab. \ref{tab:strategies}.
\begin{table}[h]
\begin{center}
\begin{tabular}{|r|rrrr|} \hline
$\mu$ & $\beta_1$ & $\beta_2$ & $\beta_3$ & $\beta_4$ \\ \cline{1-5} \hline
 -1 & -1 & -1 &  1 & 1 \\
  1 & -1 &  1 & -1 & 1 \\ \hline
\end{tabular}
\end{center}
\caption{\label{tab:strategies}\em Strategies for $m=1$}
\end{table}
\begin{figure}[h]
\begin{center}
\includegraphics[scale=.32]{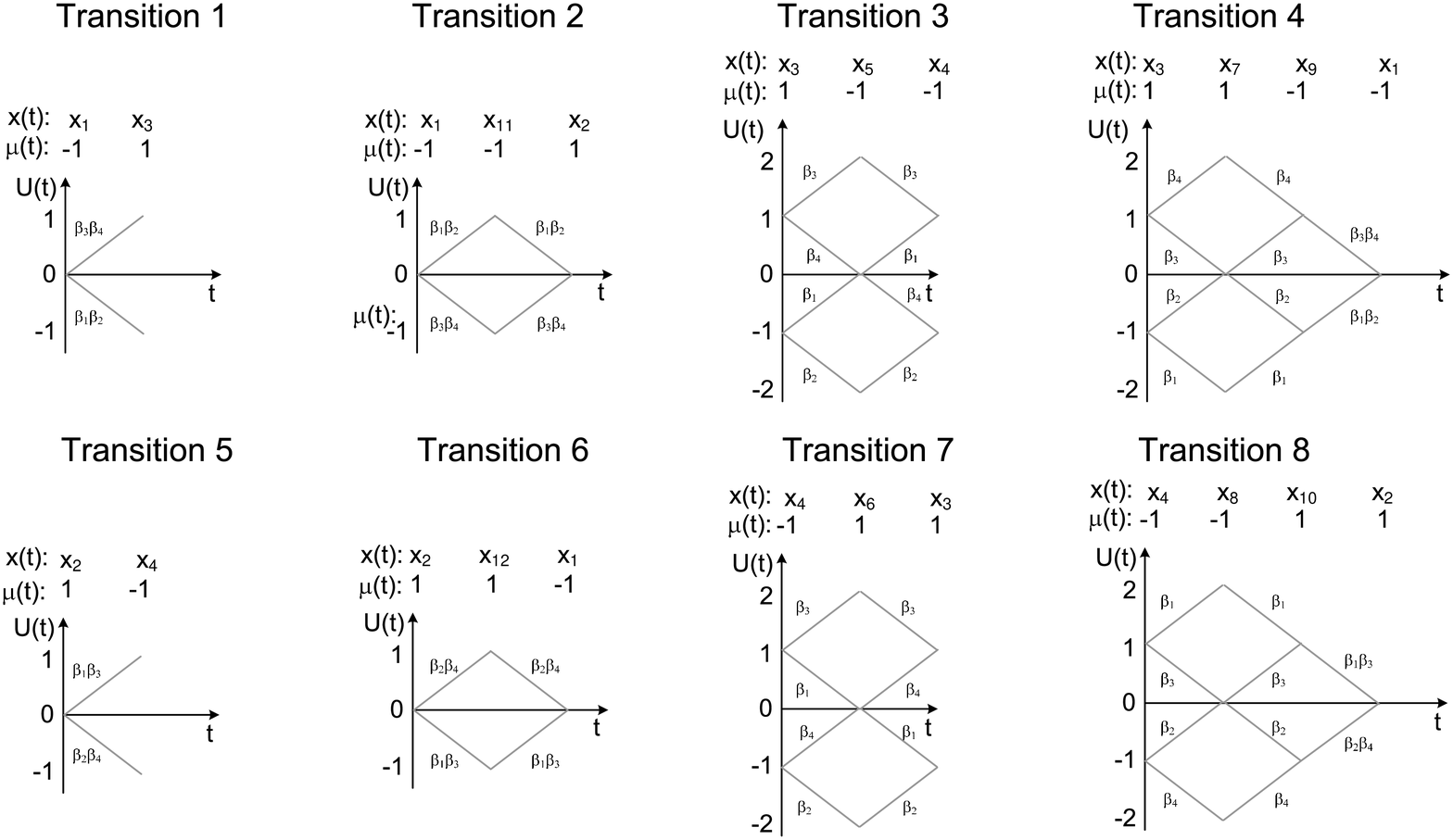}
\end{center}
\vspace*{8pt} \caption{\label{fig:AnalyticalTrajectories}\em Trajectories of utilities for $m=1$.}
\end{figure}
Possible transition scenarios for the $m=1$ MG, represented as the Markov chain, are illustrated in
Figs \ref{fig:AnalyticalTrajectories}.

\begin{figure*}[h]
\begin{tabular*}{0.5\textwidth}{ccc}
\includegraphics[scale=0.55]{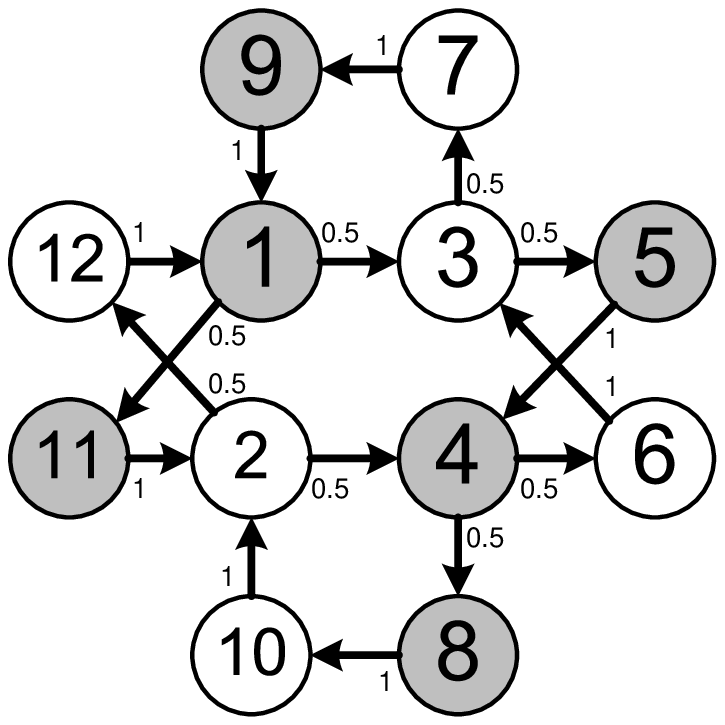} & \hspace{30mm}
\includegraphics[scale=0.55]{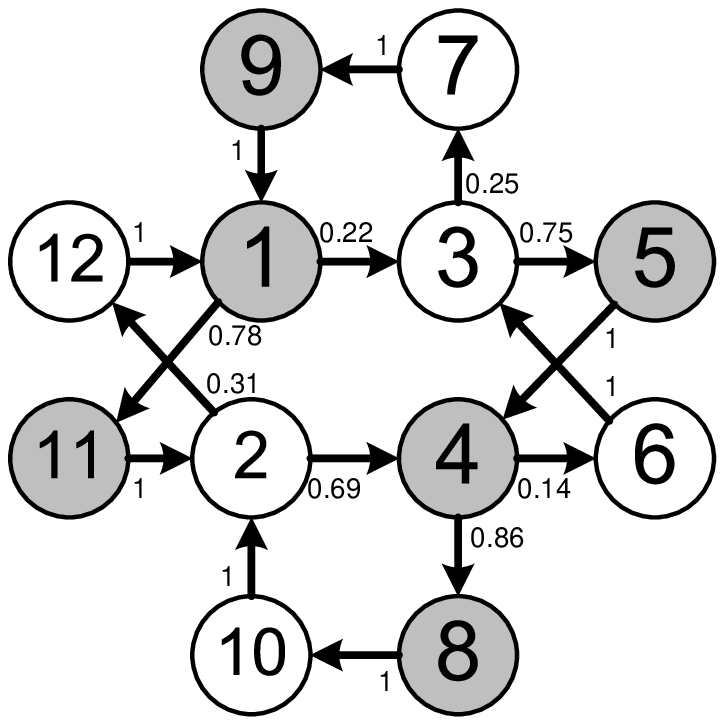}
\end{tabular*}
\caption{\label{fig:stages}\em Diagrams of the Markov chain representation of the MG in the
efficient regime for $m=1$. Numbers in circles represent the following states: $x_1 = [-1, 0, 0, 0,
0]$, $x_2=[1,0,0,0,0]$, $x_3=[1,-1,-1,1,1]$, $x_4=[-1,1,-1,1,-1]$, $x_5=[-1,0,-2,2,0]$,
$x_6=[1,0,-2,2,0]$, $x_7=[1,-2,0,0,2]$, $x_8=[-1,2,0,0,-2]$, $x_9=[-1,-1,-1,1,1]$, $x_{10} =
[1,1,-1,1,-1]$, $x_{11}=[-1,1,1,-1,-1]$, $x_{12}=[1,-1,1,-1,1]$. States marked as grey incorporate
$\mu=-1$ while the white ones $\mu=+1$. Values assigned to arrows reflect transition probabilities.
Two cases are shown: equal fractions (left) and unequal ones where agents draw strategies with
uniform probability (right). In the case of equal fractions, if transitions to two states are
possible from a given state, both transition probabilities are the same.}
\end{figure*}

At the beginning of the game all utilities are equal to
zero. Depending on the history $\mu$, only two initial states can exist: $x_1=[-1,0,0,0]$ and
$x_2=[1,0,0,0]$. For each of these two states two further scenarios are equally possible, because
the utilities of corresponding strategies are the same. The choice depends on the ratio between
numbers of agents in two groups: one with $a=1$ and another one with $a=-1$. These scenarios are as
follows.
\begin{itemize}
\item[\underline{Transition 1}]
~\\
Being in the state $x_1$, the majority of agents use strategies suggesting $a=-1$. Then

\begin{itemize}
\item the minority action in the next step is $a^\ast=1$, \item strategies $\beta_1$ or $\beta_2$
give negative payoff, \item strategies $\beta_3$ and $\beta_4$ give positive payoff.
\end{itemize}
The system goes to the state $x_3=[1,-1,-1,1,1]$ (cf. Fig.~\ref{fig:AnalyticalTrajectories},
Transition~1) where $U_{\beta_3}=U_{\beta_4}=1$ and these strategies suggest different actions on
the last history $\mu=1$. Similarly, there are two strategies with the utilities
$U_{\beta_1}=U_{\beta_2}=-1$ suggesting different actions on $\mu=1$. Hence, there are two
equiprobable scenarios, further described as Transitions 3 and 4. \item[\underline{Transition 2}]
~\\
Being in the state $x_1$, the majority of agents use strategies suggesting $a=1$. Then
\begin{itemize}
\item the minority action in the next step is $a^\ast=-1$, \item strategies $\beta_3$ or $\beta_4$
give negative payoff, \item strategies $\beta_1$ and $\beta_2$ give positive payoff.
\end{itemize}
The system goes to the state $x_{11}=[-1,1,1,-1,-1]$ (cf. Fig.~\ref{fig:AnalyticalTrajectories},
Transition~2) where $U_{\beta_1}=U_{\beta_2}=1$ and give the same actions on the last history
$\mu=-1$. Most of agents use these strategies (e.g. $3/4$ of the population, provided $S=2$) and
the sole possibility is that the system goes to the state $x_2$. \item[\underline{Transition 3}]
~\\
Being in the state $x_3$, the majority of agents use strategies suggesting $a=1$ and the system
passes to $x_5$. In this state $U_{\beta_3}=U_{max}$ and $U_{\beta_2}=U_{min}$ (cf.
Fig.~\ref{fig:AnalyticalTrajectories}, Transition~3). According to the reasoning from section 5.1,
if one utility attains its maximal or minimal value, most agents use strategies suggesting the same
action as the best strategy. Consequently, there is only one scenario possible in $x_5$: the best
strategy, and all strategies giving the same output as the best one, loose and the system goes to
the state $x_4$. \item[\underline{Transition 4}]
~\\
Another possibility in $x_3$ is that most of agents decide $a=-1$ and the system goes to $x_7$. In
this state $U_{\beta_4}=U_{max}$ and $U_{\beta_1}=U_{min}$ (cf.
Fig.~\ref{fig:AnalyticalTrajectories}, Transition~4). Subsequently, the best strategy, and all
strategies giving the same output as the best one, loose and the system goes to the state $x_9$. In
$x_9$ both best strategies suggest the same for the last history $\mu=-1$. The majority of the
population uses one of these best strategies and the system moves to $x_1$.
\item[\underline{Transition 5--8}]
~\\
These transitions are analogical to Transitions 1--4, but the initial state is $x_2$.
\end{itemize}
\renewcommand{\arraystretch}{1.3}
\begin{table}[h]
\begin{center}
\begin{tabular}{|r|rrrrr|r|r|r|} \hline
 & $\mu$ & $U_1$ & $U_2$ & $U_3$ & $U_4$ & $Pr(x_i)$ & ${\mathbb E}\,[A(x_i)]$ & ${\mathbb Var}\,[A(x_i)]$ \\ \cline{1-8} \hline
 $x_1$ &    -1 &  0 &  0 &  0 &  0 & $\frac{1}{8}$ & 0 & $\frac{N}{2}$\\
 $x_2$ &     1 &  0 &  0 &  0 &  0 & $\frac{1}{8}$ & 0 & $\frac{N}{2}$\\
 $x_3$ &     1 & -1 & -1 &  1 &  1 & $\frac{1}{8}$ & 0 & $\frac{N}{4}$\\
 $x_4$ &    -1 &  1 & -1 &  1 & -1 & $\frac{1}{8}$ & 0 & $\frac{N}{4}$\\
 $x_5$ &    -1 &  0 & -2 &  2 &  0 & $\frac{1}{16}$ & $\frac{3}{8}N$ & $\frac{N}{8}$\\
 $x_6$ &     1 &  0 & -2 &  2 &  0 & $\frac{1}{16}$ & $-\frac{3}{8}N$ & $\frac{N}{8}$\\
 $x_7$ &     1 & -2 &  0 &  0 &  2 & $\frac{1}{16}$ & $\frac{3}{8}N$ & $\frac{N}{8}$\\
 $x_8$ &    -1 &  2 &  0 &  0 & -2 & $\frac{1}{16}$ & $-\frac{3}{8}N$ & $\frac{N}{8}$\\
 $x_9$ &    -1 & -1 & -1 &  1 &  1 & $\frac{1}{16}$ & $\frac{1}{2}N$ & $0$\\
 $x_{10}$ &  1 &  1 & -1 &  1 & -1 & $\frac{1}{16}$ & $\frac{1}{2}N$ & $0$\\
 $x_{11}$ & -1 &  1 &  1 & -1 & -1 & $\frac{1}{16}$ & $-\frac{1}{2}N$ & $0$\\
 $x_{12}$ &  1 & -1 &  1 & -1 &  1 & $\frac{1}{16}$ & $-\frac{1}{2}N$ & $0$\\ \hline
\end{tabular}
\end{center}
\caption{\label{tab:states}\em States $x_i$ $(i=1,\ldots,12)$, their probabilities $Pr(x_i)$ and
demands for $m=1$. The ${\mathbb E}\,[A(x_i)]$ stands for the expected value of $A$ for the state
$x_i$.}
\end{table}
The states are listed in Tab. \ref{tab:states}. These states and transitions are sufficient to
define a memoryless representation of the MG with a transition graph displayed in Fig.
\ref{fig:stages}.
Some of its states have the same expected demand ${\mathbb E}\,[A]$ over realizations of the game,
e.g. ${\mathbb E}\,[A(x_i)]=0$ ($i=1,...,4$), since the same numbers of agents play according to
strategies recommending opposite actions. Using formulas
(\ref{eq:PrOfActiveStrEqUl}-\ref{eq:nrOfBeta}) we can find ${\mathbb E}\,[A]$ for all states (cf.
Tab. \ref{tab:states}), consistently with observations in Fig. \ref{fig:A_sgnxRegular}, where five
clusters on the diagonal are found around values from Tab.~\ref{tab:states}. Our process is a
stationary Markov chain for which the stationary Master Equation can be solved with respect to the
state probabilities. Their values are given in Tab.~\ref{tab:states}, in the column marked
$Pr(x_i)$ $(i=1,\ldots,12)$. The state probabilities from Tab.~\ref{tab:states} can be also used to
find statistical periods of the demand
\begin{eqnarray}
Pr\big[A(t)=A(t+\tau)\big] & = & \sum_{ij}\,\delta\big[A\big(x_j(t+\tau)\big),A\big(x_i(t)\big)\big] \nonumber \\
& \cdot & Pr\big[x_j(t+\tau)\,|\,x_i(t)\big]\cdot Pr\big[x_i(t)\big],
%Pr\big[A(t)=A(t+\tau)\big]=\sum_{ij}Pr\big[A\big(x_j(t+\tau)\big)=A\big(x_j(t)\,|\,x_i(t)\big)\big]\cdot Pr\big[x_i(t)\big].
\label{eq61}
\end{eqnarray}
where $\delta(x,y)$ stands for the Kronecker symbol. The maximal value of $7/16$ is found for
$\tau=4$ and this explains why the largest correlation is found also for $\tau =4$ (cf. Figs
\ref{fig:R3_signx} and \ref{fig:R3_linear}).
\newline

\noindent {\it Example 2: Deterministic transitions}

\noindent Here, we show an example how to prove that the transition from a given state is
deterministic provided that the system is a reference one and the game is in herd regime but not
necessarily in the limit $NS\rightarrow \infty$. Additionally, we present that the transition can
change if agents are assigned to fractions randomly.

Let us consider an arbitrarily chosen state for $S=2$ and $m=1$ where the transition is
deterministic, e.g. $x_5$ defined as $x_5 = [-1, 0, -2, 2, 0]$. Assume that fractions' indexes are
assigned to each pair of strategies according to Tab. \ref{tab:fractions}. Analyzing each fraction
one finds that:
\begin{itemize}
\item For fractions $F_{11},F_{12},F_{15},F_{16}$ both strategies suggest $+1$. Hence $C_{\nu}(x_i)
= +1$, for $\nu\in\{11,12,15,16\}$.
\item For fractions $F_{3},F_{7},F_{8},F_{9},F_{10},F_{14}$ strategy with higher $U$ suggests $+1$.
As a result for these strategies $C_{\nu}(x_i) = +1$, for $\nu \in\{3,7,8,9,10,14\}$.
\item In fractions $F_1,F_2,F_5,F_6$ both strategies suggest $-1$. Thus $C_{\nu}(x_i) = -1$, for
$\nu\in\{1,2,5,6\}$.
\item Finally, fractions $F_4,F_{13}$ have two strategies with equal probabilities but suggesting
opposite actions. Hence, $C_{\nu}(x_i)$, for $\nu\in\{4,13\}$, follows binomial distribution
 (\ref{eq: PrC}).
\end{itemize}
For the reference system (equal fractions) one can calculate ${\mathbb E} [A(x_5)] =
\frac{5}{16}N$. The uncertainty is introduced by agents belonging to fractions $F_4,F_{13}$ because
they choose $-1$ or $+1$ with the same probability. It means that $A(x_5)\in\{ \frac{3}{16}N \ldots
\frac{5}{16}N \}$ and $\mathbb{V}ar [A(x_5)]  = N/8$. Hence, $A(x_5)$ is always positive and
$a^*(x_5) = -1$, thus the successor state is determined unambiguously. Such analysis can be
performed for arbitrary state which makes easy calculation of variance of the aggregate demand (cf.
Tab. \ref{tab:states}).

Any MG with $m>1$ in the efficient regime can be represented as a Markov process with a finite
number of states. The same method as for $m=1$, but more demanding computationally, can be used to
calculate state probabilities. The reasoning presented is strictly true only in the ideal case
where subpopulations of agents in different fractions are equal, or if the system is considered
\emph{a priori}, i.e. before strategies are assigned to agents at the beginning of the game. In
\emph{a posteriori} analysis we consider the game where strategies are already assigned. In most
cases such game is characterized by an inequality between sizes of fractions due to the initial
randomness in the strategies' generation process (quenched disorder). In \emph{Example 2},
considering system \emph{a priori}, the expected value ${\mathbb E} [A(x_5)]$ remains the same but
the variance changes distinctly enough to allow for appearance of negative samples. Considered
\emph{a posteriori}, also ${\mathbb E} [\widetilde{A}(x_5)]$ is most likely biased compared to
${\mathbb E} [A(x_5)]$. We show that some interesting phenomena, among them the sensitivity of the
predictability $H_A$ to the payoff, appear only when the quenched disorder is taken into account
i.e. imbalance between fractions exists.
%\subsubsubsection{The case of unequal-size fractions}
\newline
\newline
\noindent{\it The case of unequal-size fractions}

\begin{figure}[h]
\begin{tabular}{cc}
\includegraphics[scale=.24]{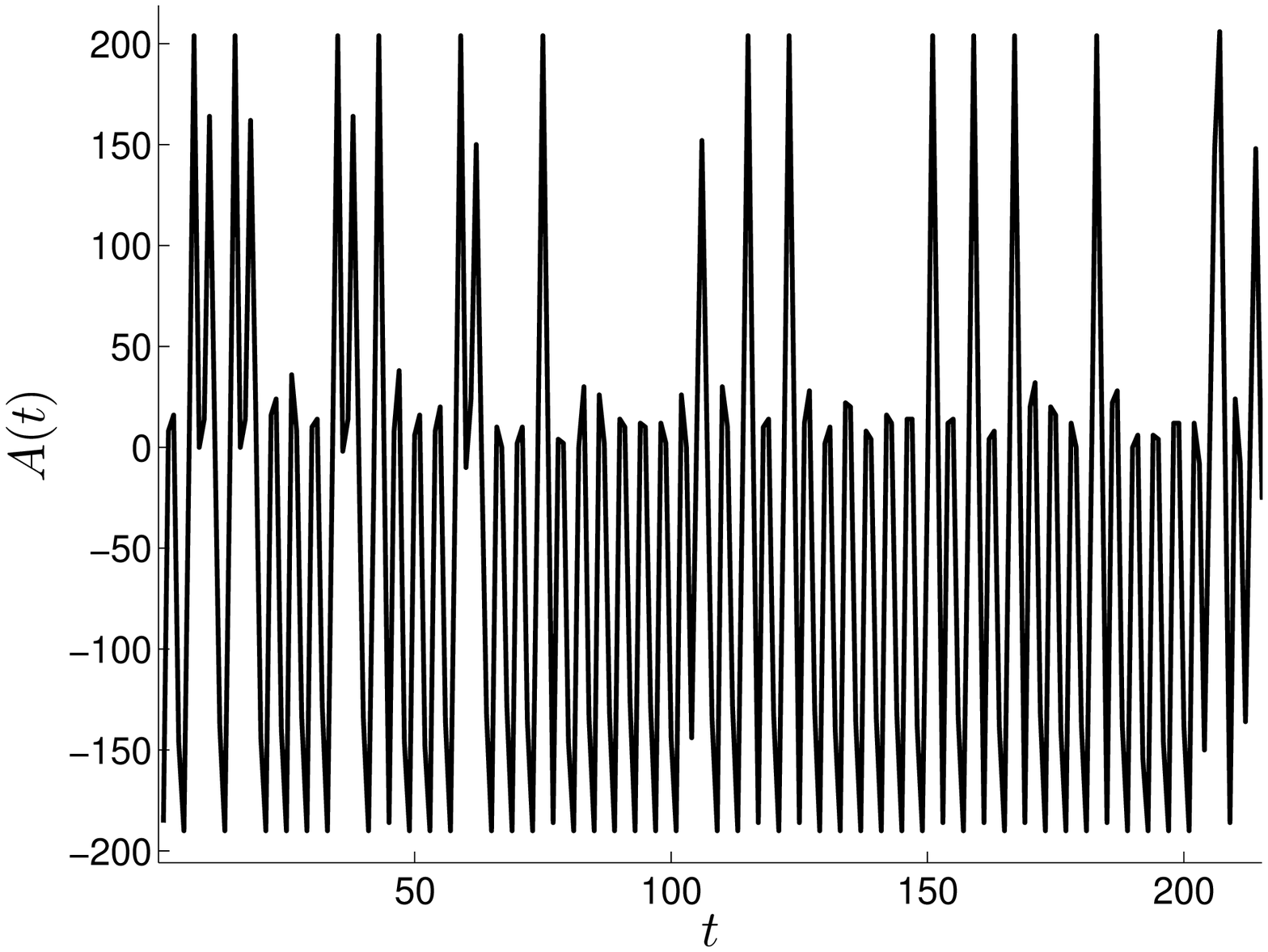} & \hspace{5mm}
\includegraphics[scale=.24]{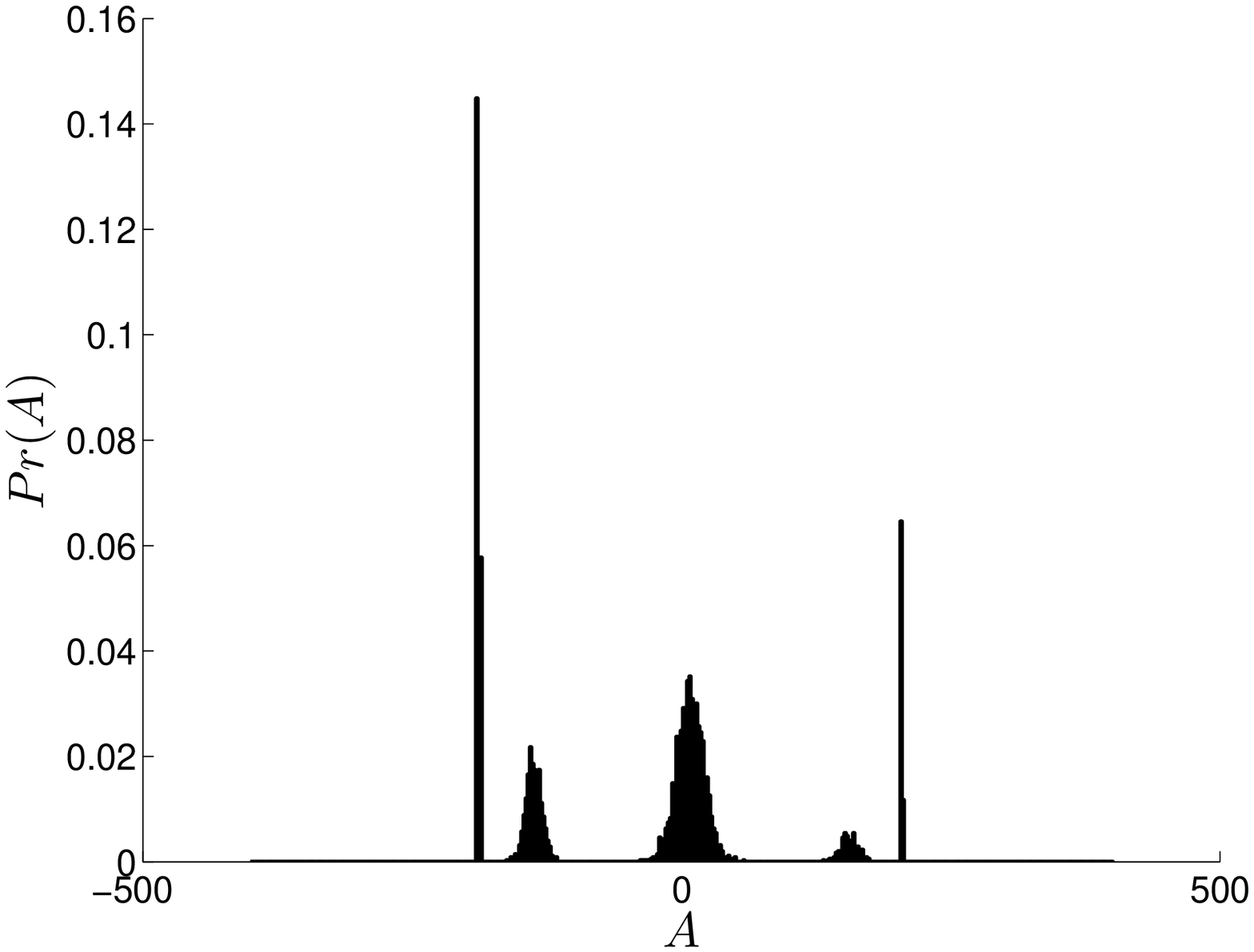}\\
\includegraphics[scale=.24]{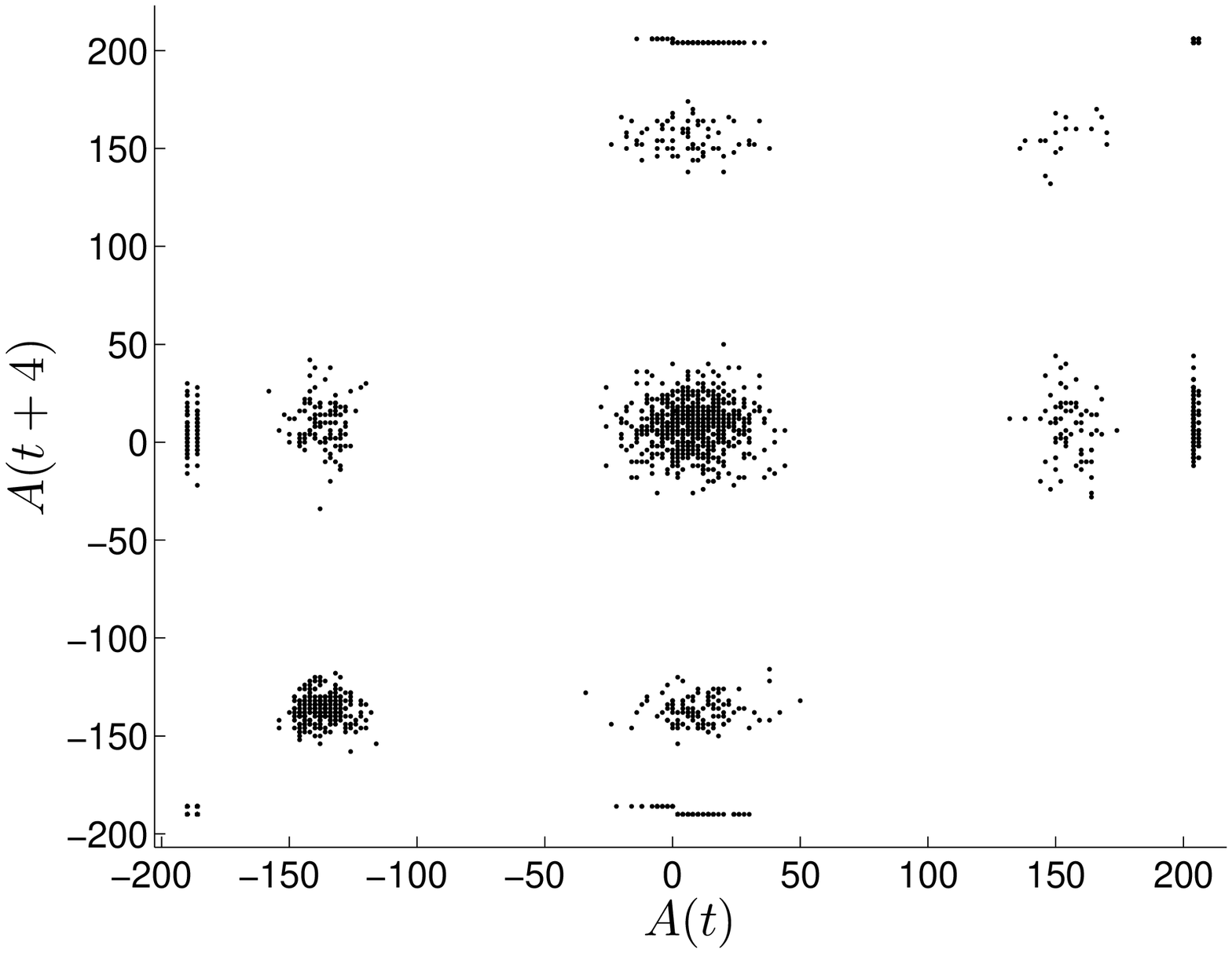} & \hspace{5mm}
\includegraphics[scale=.24]{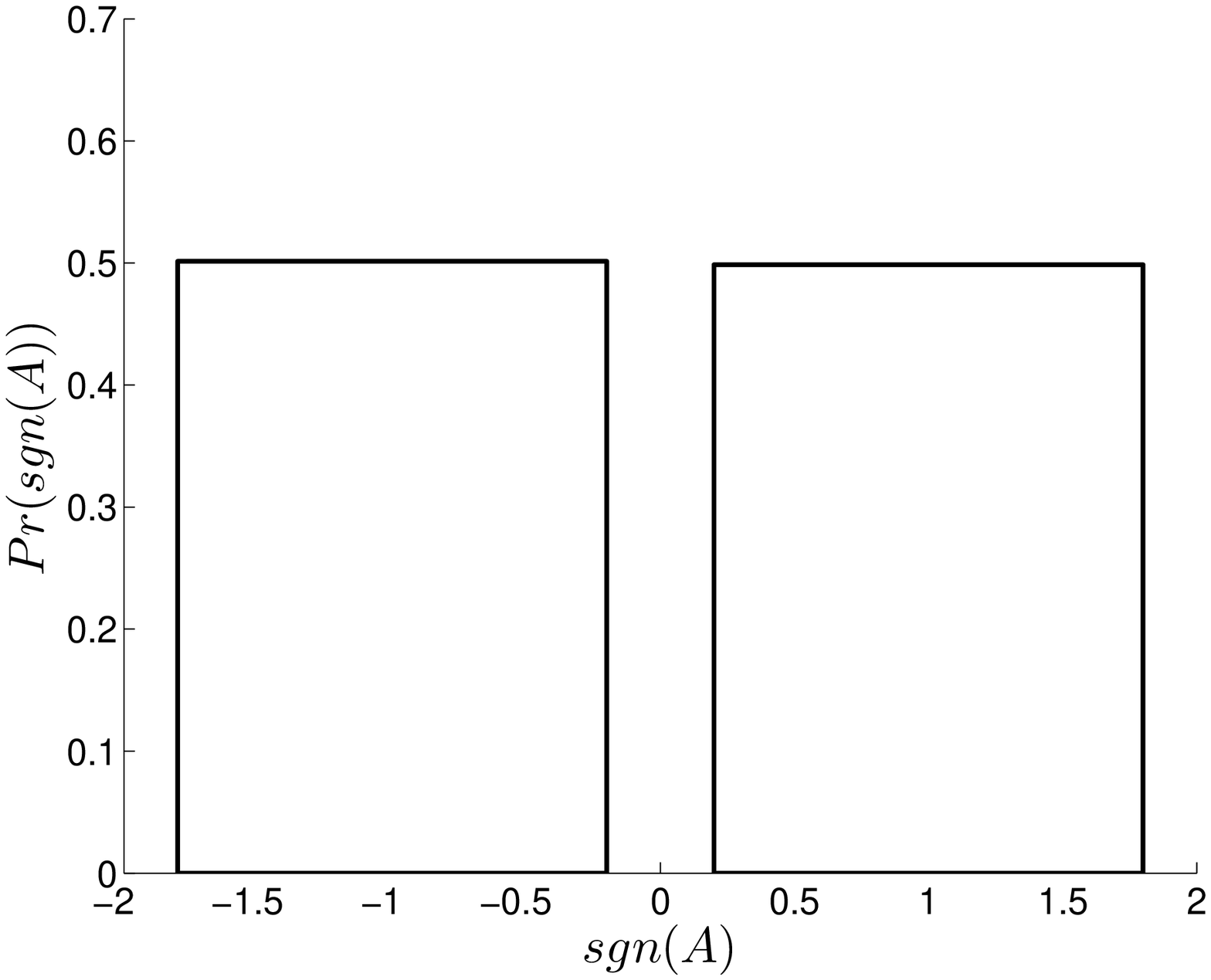}
\end{tabular}
\caption{\label{fig:A_sgnxIrregular}\em Time evolution of the aggregated demand ${A}(t)$ (upper
left), Plots of the aggregated demand ${A}(t+2\cdot 2^m)$ vs. ${A}(t)$ (lower left), Estimated
$Pr({A})$ (upper right) and $Pr(\mbox{sgn}({A}))$ (lower right) for the population size $N = 400$
and agent memory $m = 1$, $S=2$ strategies per agent and unequal sizes of fractions.}
\end{figure}

\noindent If strategies are assigned randomly to agents then fraction sizes are likely to be
unequal. Let us consider one of the simplest cases where strategies are drawn with equal
probabilities, which corresponds to assigning an agent to any fraction with the probability
$\frac{1}{G}$. Interestingly, numerical experiments show that in this case the reconstructed MP
usually follows the sequence of states of the reference MP but the values of transition probabilities are
not reproduced. This bias does not disappear even if the game is enlarged (see Figs
\ref{fig:A_sgnxRegular} and \ref{fig:A_sgnxIrregular}).
The explanation is as follows.

States in the reference MP, where stochastic transition appears, are characterized by the same
number of positive and negative components in formula (\ref{eq: ADecomposed}). Calculating
the transition probability we considered two cases: before and after assignment of strategies
to agents, i.e. the \emph{a priori} and \emph{a posteriori} one.

Calculating \emph{a priori} expected value
of $\mbox{sgn}\, A$ we do not know yet the specific number of agents in the $\nu$-th fraction and we just operate on
random variables:
\begin{eqnarray}
 {\mathbb E}[\mbox{sgn}\,A(x_i)] & =&  {\mathbb E}\Big[ \mbox{sgn}\sum_{\nu} C_{\nu}(x_i) F_{\nu}\Big] \nonumber \\
 & = & {\mathbb E}\Big[ \mbox{sgn}\big(\sum_{F_d:C_{d}(x_i)=1} F_d - \sum_{F_q:C_{q}(x_i)=-1} F_q \nonumber\\
 & & + \sum_{F_w:C_{w}(x_i)\in [-1,+1]}
 C_{w}(x_i) F_w \big)\Big] \label{eq: apriori}
\end{eqnarray}
Each fraction size $F_\nu$ obeys the same binomial distribution. Since we consider stochastic
transitions in the reference system, then there is the same number of elements in the first and
second sum of Eq.~(\ref{eq: apriori}). The distribution of the third sum is symmetric around zero
because it contains pairwise symmetric components. Thus, the distribution of $A(x_i)$ is also
symmetric, as well as the distribution of $\mbox{sgn}\,A(x_i)$. By that means ${\mathbb
E}[\mbox{sgn}\,A(x_i)] = 0$.

When strategies are assigned to agents, then the numbers of agents in fractions, $f_{\nu}$, are
known and the system is considered as {\it a posteriori}. The ${\mathbb E}[\mbox{sgn}\,{A}(x_i)]$
can be decomposed:
\begin{eqnarray}
{\mathbb E}[\mbox{sgn}\,{A}(x_i)] & = & {\mathbb E}\Big[ \mbox{sgn}\,\big(\sum_{f_d:C_{d}(x_i)=1}
f_d -
\sum_{f_q:C_{q}(x_i)=-1} f_q \nonumber\\
& & + \sum_{f_w:C_{w}(x_i)\in [-1,+1]} C_{w}(x_i) f_w\big)\Big]. \label{eq: aposteriori}
\end{eqnarray}
Provided $S=2$, the last sum in Eq.~(\ref{eq: aposteriori}) is symmetric around zero due to $C_{w}$
symmetry but the first two sums introduce a bias, shifting distribution of $A(x_i)$. If $S>2$ then
also the third term may be biased. Since ${\mathbb Var}[F]\rightarrow \infty$ for
$N\rightarrow\infty$, then considering only first two components one gets
\begin{eqnarray} {\mathbb Var}[\sum_{F_d:C_{d}(x_i)=1} F_d - \sum_{F_q:C_{q}(x_i)=-1} F_q] \rightarrow \infty,
\end{eqnarray}
which means that the probability of a large bias grows indefinitely with $N$.

If the \emph{a
posteriori} distribution of $A(x_i)$ is shifted then the \emph{a posteriori} distribution of
$\mbox{sgn}\,A(x_i)$ is asymmetric, regardless of a symmetry of the last term. Consequently, most
likely ${\mathbb E}[\mbox{sgn}\,{A}(x_i)] \neq 0$. The equality between ${\mathbb
E}[\mbox{sgn}\,{A}(x_i)]$ calculated using \emph{a priori} distribution and ${\mathbb
E}[\mbox{sgn}\,A(x_i)]$ calculated using distribution \emph{a posteriori} occurs only if numbers
of agents per fraction are equal for all fractions. In other cases the expected absolute bias of
distribution increases with $N$ and probabilities in stochastic transitions are most likely
unequal. In some experiments we found that for specific states the bias can shift the distribution
so heavily that it is always positive or negative. Therefore the state, being
\emph{a priori} stochastic, may become deterministic when analyzed \emph{a posteriori}.

Finally, consider the states with deterministic transitions in the reference system. If now $F$ is
a random variable, then with some, usually very small, probability the transition becomes
stochastic due to the specific realization of $F$. The analysis of one specific state is given in
\emph{Example 2} of the present section.
%in Sec. \ref{sec:MPsgnx}.

\subsubsection{Stochasticity of the game depends on initial conditions} We assumed that
$U_{\alpha_n^s}(t=0)=0$ for all $\alpha_n^s$. This assumption seems natural as reflecting no {\it a
priori} preference for any strategy. However, it appears to be critical for the MG dynamics for
$g(x)=\mbox{sgn} (x)$. Stochastic transitions show up for the degenerate state, i.e. with more than
one strategy with the same utility. Removing this ambiguity suppresses stochasticity and the game
becomes deterministic. In such a case, our simplified description of the state fails because
strategies have unique utilities and cannot be aggregated. Consequently, the Markovian treatment,
as presented in Sec. \ref{sec:MPsgnx}, is no longer useful but its description in terms of the
Markov process, defined as for proportional payoff $g(x)=x$, becomes interesting. In particular,
the game follows the Eulerian path on de Bruijn graph and is deterministic (cf. Sec.
\ref{sec:linearPayoff}).

\subsubsection{Stability of the game and behaviour of the predictability $H$} Disproportions in
fractions affect transition probabilities. If the absolute disproportions are very large then some
transitions, which exist in the reference system, can disappear and the graph is reduced to its
subgraph. The game remains stable because each subgraph is characterized by sequence of states
assuring that $+1$ and $-1$ appear after given $\mu$ with the same frequency (cf. white and grey
circles, respectively, in Fig. \ref{fig:stages}). Equality of frequencies of the opposite minority
decisions after any $\mu$, is both the necessary and sufficient condition to assure stability,
provided $g(x) = \mbox{sgn}(x)$. Hence, the stability entails the same frequencies, resulting with
$\langle a^*|\mu \rangle = 0$. No matter whether the system is the reference one or not -- the
$H_a$ is always equal to zero, provided the game is stable. The above mechanism works as long as
the game is deep in herd regime, i.e. $NS \gg 2^P$, and if strategies are drawn from the uniform
distribution or the one close to it. If game moves to the cooperation mode, or strategies are drawn
from an asymmetric distribution, then the methodology of MP breaks down because relative
disproportions between fractions are large. This distorts stability and additional states appear.

The stability mechanism requires balance between frequencies of the negative and positive signs of
$A$ after any $\mu$, regardless of the value of $A$. The $\langle A|\mu\rangle$ in formula
(\ref{eq:HA}) can be redefined as follows:
\begin{eqnarray}
\langle A|\mu\rangle \simeq \sum_{i=1}^{\#X^\mu} {\mathbb E}\,\big[A(x_i^\mu)\big] Pr(x_i^\mu),
\label{eq:Amu}
\end{eqnarray}
where $X^{\mu}$ is the set of all states $x_i^\mu$ including history $\mu$. Approximation
(\ref{eq:Amu}) is based on replacing each partial sum of random variable in state $x_i^\mu$,
$A(x_i^\mu)$, by its expected value in this state, ${\mathbb E}\,[A(x_i^\mu)]$. Eq. (\ref{eq:Amu})
is strict in the limit of infinite time, $T\rightarrow\infty$.

Analyzing the system \emph{a posteriori}, $\sum_{x_i^\mu\in X^{\mu}}{\mathbb
E}\,[{A}(x_i^\mu)]Pr(x_i^\mu) = 0$ only in the case of equal fractions, because there always exists
a pair of states with the same $\mu$, the same probabilities and symmetric distributions around
zero. The larger the game, the larger possible disproportions of ${\mathbb E}\,[{A}(x_i^\mu)]$
between the reference and the real system, provided that in the real system strategies are drawn
from flat distribution. As a result, $\sum_{i=1}^{\# X^\mu} {\mathbb E}\,[{A}(x_i^\mu)]
Pr(x_i^\mu)$ grows with the population size. Hence, $H_A$ as a function of the control parameter $n
= N/P$ is larger than zero in the herd regime, if the system is different than the reference one.

\subsubsection{Variance per capita $\sigma^2/N$} For simplicity, we consider here only the case of
equal fractions and do not distinguish between \emph{a priori} and \emph{a posteriori} games. The
variance per capita (\ref{eq:sigma}) is defined using the sum over the set of all states $X$. If
game is large enough, then a suitable approximation based on the MP representation is given by
\begin{eqnarray}
\sigma^2(A) & \simeq & \sum_{i=1}^{\#X} Pr(x_i) {\mathbb E}\big[A(x_i)\big]^2 \label{eq: VarPA} \\
            &    =   & \sum_{i=1}^{\#X} Pr(x_i) \Bigg(\frac{N}{G} {\mathbb E}\Big[\sum_{\nu=1}^{G} C_{\nu}(x_i)\Big]\Bigg)^2.
\label{eq: VarPAExt}
\end{eqnarray}
In derivation of Eq.~(\ref{eq: VarPA}) from Eq.~(\ref{eq:sigma}) we use expansion of $\sigma^2(A)$
into the sum of partial sums over states and the fact that variation of ${\mathbb E}\,[A(x_i)]$
from state to state is significantly larger than the width of distribution of $A$ in any state (cf.
Figs \ref{fig:A_sgnxRegular} and \ref{fig:A_sgnxIrregular}, upper right). More detailed explanation
is as follows.

In Eq.~(\ref{eq:sigma}), each value of demand $A(t)$ is generated in one of $K$ possible states.
Assuming ergodicity, the sum over time steps $t$ in Eq.~(\ref{eq:sigma}) $(t=0,\ldots,T)$ can be
represented as a sum over all $T$ visits in states $x_k$ $(k=1,\ldots,K$ \mbox{and} $K=\# X)$.
Since each state is visited many times, the sum over visits in states can be decomposed into
partial sums over states
\begin{eqnarray}
\sigma(A)^2 & = & \frac{1}{T}\sum_{t=0}^T A(t)^2 \nonumber \\
            & = & \sum_{k=1}^K \frac{1}{I_k}\sum_{i_k=1}^{I_k}A(x_{i_k})^2 \nonumber \\
            & \simeq & \sum_{k=1}^K {\mathbb E}\,[A(x_k)^2],
\end{eqnarray}
where $i_k$ runs over subsequent moments when the system is in the $k$-th state and $I_k$ stands for
the number of visits in this state. For any state $x_k$ the random variable $A(x_k)$ can be
represented as a sum
\begin{eqnarray}
A(x_k)={\mathbb E}\,[A(x_k)]+\eta(x_k),
\end{eqnarray}
where $\eta(x_k)$ is a random variable and ${\mathbb E}\,[\eta(x_k)] = 0$. Hence
\begin{eqnarray}
{\mathbb E}\,[A(x_k)^2]={\mathbb E}\,[A(x_k)]^2+{\mathbb E}\,[\eta(x_k)^2]. \label{a:EA}
\end{eqnarray}
Since, depending on the state (see Tab.~\ref{tab:states}),
\begin{eqnarray}
{\mathbb E}\,[A(x_k)]^2 & \sim & 0 \quad \mbox{or} \quad N^2, \nonumber \\
{\mathbb E}\,[\eta(x_k)^2] & \sim & N \quad \mbox{or} \quad 0, \\
\end{eqnarray}
the second term in Eq.~(\ref{a:EA}) may be neglected for large $N$ and one arrives at Eq.~(\ref{eq:
VarPA}).

In order to guide intuition, let us consider example from Fig.~\ref{fig:A_sgnxRegular} (upper
right). This joint distribution of $A$ is a sum of distributions for twelve states. Five distinct
peaks correspond to distributions of $A$ in groups of states. States corresponding to peaks, as
well as expected values of $A^2$ and $\eta^2$, are given in Tab.~\ref{tab:var}.
\begin{table}[*h]
\begin{center}
\begin{tabular}{|c|c|c|c|} \hline
Peak (from left) & States $x_k, k=1,..,12$ & ${\mathbb E}\,[A(x_{k})]^2$ & ${\mathbb
E}\,[\eta(x_k)^2]$,\;$k=1,..,12$ \\ \hline
1 & $x_{11,12}$ & $(-N/2)^2$    & 0 \\
2 & $x_{8,6}$   & $(-3N/8)^2$   & $\sim N$ \\
3 & $x_{1,2,3,4}$ & 0           & $\sim N$ \\
4 & $x_{5,7}$   & $(3N/8)^2$    & $\sim N$ \\
5 & $x_{9,10}$  & $(N/2)^2$     & 0 \\ \hline
\end{tabular}
\end{center}
\caption{\label{tab:var}\em Squared expected values of $A$ and expected values of $\eta^2$ for five
peaks seen in Fig.~\ref{fig:A_sgnxRegular} (upper right).}
\end{table}

The number of fractions where all strategies suggest the same action after given $\mu$ is always
$(2^{P-1})^S$, where $2^{P-1}$ represents the half of the strategy space where all strategies
suggest the same action. Hence, at least, $2(2^{P-1})^S$ terms in $C_{\nu}$ in the sum (\ref{eq:
VarPA}) compensate mutually. By that means there is $2^{PS} - 2^{(P-1)S+1}$ terms which in the worst
case are not compensated. Indeed, one can find states where all actions of these fractions are
equal to $+1$ or $-1$, but also states where contributions of all fractions compensate to $0$.
Hence
\begin{eqnarray}
0 \leq |\sum_{j=1}^G C_{\nu}(x_i)| \leq 2^{PS} - 2^{(P-1)S+1},
\end{eqnarray}
where the upper boundary can be factorized into $2^{PS}(1-2^{1-S})$, and only the number of
different fractions $G=2^{PS}$ depends on $P$. In particular, for $S=2,3,4,5$ this factor is equal
to $\frac{G}{2}, \frac{3}{4}G, \frac{7}{8}G, \frac{15}{16}G$, respectively.

Generally:
\begin{eqnarray}
{\mathbb E}\,[A(x_i)] \sim  N.
\end{eqnarray}
As a result, in Eq.~(\ref{eq: VarPA}), $\sigma^2 \sim N^2$ and $\sigma^2/N \sim N$, in agreement
with numerical simulations~\cite{savit99PhysRevLetters82} and theoretical
results~\cite{hart01EurPhysJB20,johnson99PhysA269,hart01PhysicaA298}.

The variance is no longer proportional to $N^2$ if game leaves the herd regime. In the random mode,
there is less agents than fractions and therefore:
\begin{eqnarray}
\sigma^2(x_i) = \sum_{n=1}^N {\mathbb Var}\,[a_{\alpha_n^\prime}(x_i)].
\end{eqnarray}
Considering further the case $S=2$, on average the half of agents do not have choice because they
have two strategies suggesting the same action. Decisions in this half of the population compensate
mutually and do not influence $A$. There are states where the rest of the population has a choice
and thus $\sigma^2(x_i) = \sum_{n=1}^{N/2} {\mathbb Var}\,[a_{\alpha_n^\prime}(x_i)]$. Hence,
$\sigma^2 \sim \frac{N}{2}$.

In the cooperation regime, most of fractions are in game but fluctuations of $F$ are still
relatively large. Thus, there are fractions more and less populated. Strategies that are in less
populated fractions win more frequently. The impact of these fractions is compensated by larger
fractions and therefore the variance is minimal. It reflects the balance between the crowd and
anticrowd in the so called crowd-anticrowd approach~\cite{hart01EurPhysJB20}.

\subsection{The payoff \mbox{g(x) = x} \label{sec:linearPayoff}} The linear payoff $g(x) = x$ requires different methods
of analysis than the steplike one. For $g(x)=\mbox{sgn} (x)$, in each state there are strategies
suggesting different actions with the same utility. If an agent has two or more best strategies
with the same utility then it chooses one of them randomly. As a result, some transitions are
stochastic. The more so, the utility is bounded from the bottom and top: $U_{min}\le U(t)\le U_{max}$,
where $U_{min(max)}=-(+)\, 2^m$.
%$U_{min(max)}=-\mbox{\scriptsize (}+\mbox{\scriptsize )}2^m$.
The number of values of utility is relatively small.
% and the definition of the state can be effectively based on these values.
For $g(x) = x$, the probability that the pairwise different strategies have the same
utility is small, compared to the case of $g(x)=\mbox{sgn} (x)$, and the range of possible $U$ is
much wider, from $-N/2$ to $N/2$, provided that the system is the reference one. Resultantly,
stochasticity of transitions disappears almost completely but the game is still periodic. A
persuasive explanation of periodicity is proposed by the authors of
Ref.~\cite{jefferies01PhysRevE65} using de Bruijn representation of the memory sequences $\mu$.
Here we extend their analysis and explain the dynamics of $A(t)$ by introducing a novel definition
of the state.

\subsubsection{The initial phase} All steps with more then one strategy with the same utility are
called initial. If $U(t=0) = \mbox{const}$ for all strategies, then some initial steps are
necessary to split all utilities of pairwise different strategies. Now we show that the minimal
number of such steps is $2^m$ and the maximal is $2^{m+1}$.

Identical utilities of two different strategies at time $t$ can either differ by $2A(t)$ or remain
the same at $t+1$. They differentiate when corresponding pairwise different strategies suggest
opposite actions after $\mu(t)$. Therefore the shortest time to split utilities of all $2^P$
strategies is $2^m$. Such scenario requires appearance of all possible histories $\mu$ without any
repetitions.

\begin{figure}[h]
\begin{center}
\includegraphics[height=55mm,width=0.70\linewidth]{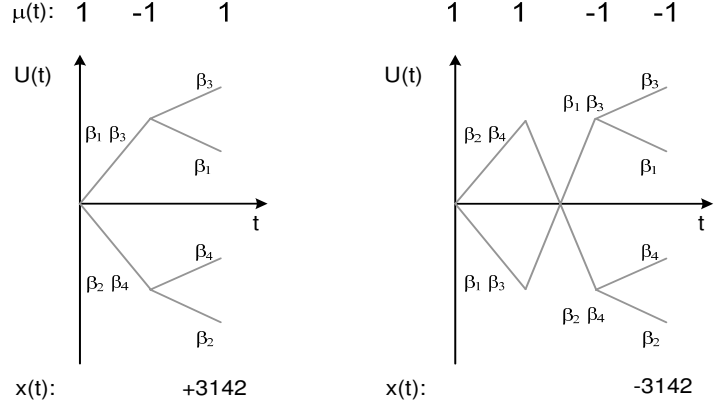}
\caption{\em The shortest (left) and longest (right) scenario of the initial phase for $m=1$ and
$\mu(0)=1$.} \label{fig:fig7}
\end{center}
\end{figure}

If strategies react in such way that their utilities do not split from step $t$ to $t+1$, then it
means that the same $\mu$ appears twice. Resultantly, the strategies that won in step $t$ have to
lose in step $t+1$, due to the positive change of the utility and being preferable to the majority
of the population at time $t+1$. Thereby the sign of $A(t+1)$ changes, compared to the sign of
$A(t)$, and different $\mu$ has to appear. There is only one $\mu$ for which given half of
different strategies reacts identically and for any other $\mu$ they have to split. Example of both
scenarios is presented in Fig. \ref{fig:fig7}, where strategies are defined as in Tab.
\ref{tab:strategies}.
The first scenario is relatively easy to follow and we focus on the second one. The initial value
is $\mu(0) = 1$ and all strategies have the same utility $U=0$. Each agent at $t=0$ draws one
strategy randomly. Let us assume that most of them decide to use the strategy suggesting $a(0)=-1$.
As a result (i) $\beta_2$ and $\beta_4$ get positive payoff and, (ii) the next history is $\mu=1$
(cf. Tab.~1). After $\mu=1$ both winning strategies suggest the same action and lose. So the next
history is $\mu=-1$. Since the history changed, the glued strategies have to react differently
because two different $\mu$'s cannot cause the same reaction of all strategies. Thus, the longest
time to split all $U$ trajectories is $2^{m+1}$ and requires every possible history to appear
twice.

\subsubsection{The concept of the state} At any step of the game one can rank all pairwise
different strategies as the best, second best, third best, etc. Sizes of fractions corresponding to
these strategies are known~\cite{hart01EurPhysJB20,wawrzyniak09ACSNo6}. An ordered list of indexes
of different strategies, complemented by $\mu$ value, is sufficient to fully describe the game at a
given moment and can be used as a characteristics of the state. Formally, assume
$\{\beta_{\kappa}\}_{\kappa=1}^{2^P}$ is the set of pairwise different strategies indexed
arbitrarily. There exists the sorting operator $\omega(\kappa)\rightarrow l$, ordering strategies
according to their utilities, such that $l_{\beta_k}$ stands for the position of the strategy
$\beta_{\kappa}$ in the ordered list. Then the state is as follows
\begin{eqnarray}
x(t)=[\,\mu(t),l_{\beta_1}(t),\ldots,l_{\beta_{2^P}}(t)\,]. \label{eq:state_x}
\end{eqnarray}
The total number of states is equal to $P\,\prod_{\kappa=0}^{2^P/2-1}(2^P-2\kappa)$ and accounts
for all possible orders of $P$ strategies, provided each strategy has its
anti-strategy \footnote{First arbitrarily chosen strategy from the set of $2^P$ strategies can be
placed on one of $2^P$ positions in the ordered list. When the position of the given strategy is
chosen, then the position for its anti-strategy is chosen automatically. Next, the strategy from
the reduced set of $2^P-2$ strategies is placed in one of $2^P-2$ positions, and so forth. Each
level occurs with different $\mu$ and there are $P$ different $\mu$'s.}, where the pair consisting
of the strategy and its anti-strategy is characterized by the normalized Hamming distance equal to
one.

As prevalent number of strategies have unique utility, the probability (\ref{eq:PrOfActiveStrEqUl})
for the active strategy $\alpha_n^\prime$ can be simplified (cf. also
Ref.~\cite{hart01EurPhysJB20})
\begin{eqnarray}
Pr \big [U_{\alpha_n^\prime}(t)=u_l\big ]=
\Big(1-\frac{l-1}{2^{P}}\Big)^S-\Big(1-\frac{l}{2^{P}}\Big)^S, \quad\quad l\ge 1.
\label{eq:PrOfActiveStrEqUlGx}
\end{eqnarray}
As a result, in the limit $N\rightarrow \infty$ about $N \; Pr\big [U_{\alpha_n^\prime}(t)=u_l\big
]$ agents use the $l$-th best strategy. Subsequently, analysis of actions of strategies provides
values of the aggregate demand in each state. Consider, for example, the case when $A$ is the
largest possible. Since $\{u_l\}$ is a sorted list of utilities, this is possible if the first
$l/2$ strategies in this list suggest actions opposite to the last $l/2$. Then the probability of
an action suggested by the best strategy is equal to
\begin{eqnarray}
Pr \big[a_{\alpha_n^\prime(t)}=a_{\alpha^B(t)}\big] & = & \sum_{l=1}^{2^{P-1}} Pr\big[U_{\alpha_n^\prime(t)}=u_l\big] \nonumber \\
                                                   & = & 1-\frac{1}{2^{S}}.
\label{eq:TheLargestProof}
\end{eqnarray}
This means that for large $NS$ for about $N(1-\frac{1}{2^{S}})$ agents their active strategy is the
same as the best strategy and the absolute value of the aggregated demand is equal to
\begin{eqnarray}
|A|=N\Big(1-\frac{1}{2^{S-1}}\Big). \label{eq:ATheLargest}
\end{eqnarray}
In particular, if $S=2$ then $|A|=N/2$.

\subsubsection{De Bruijn representation}
\begin{figure}[h]
\begin{center}
\begin{tabular}{cc}
\includegraphics[scale=.35]{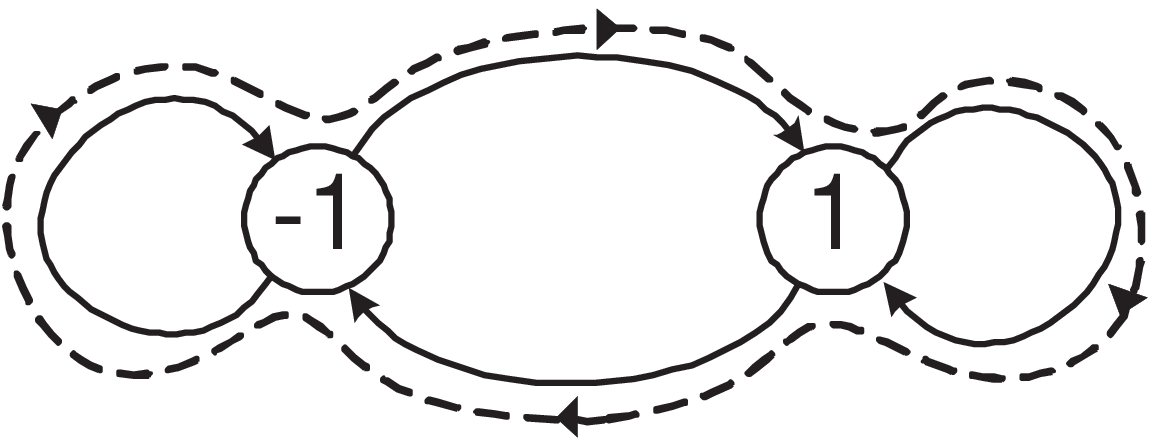} & \hspace{1cm} \includegraphics[scale=.35]{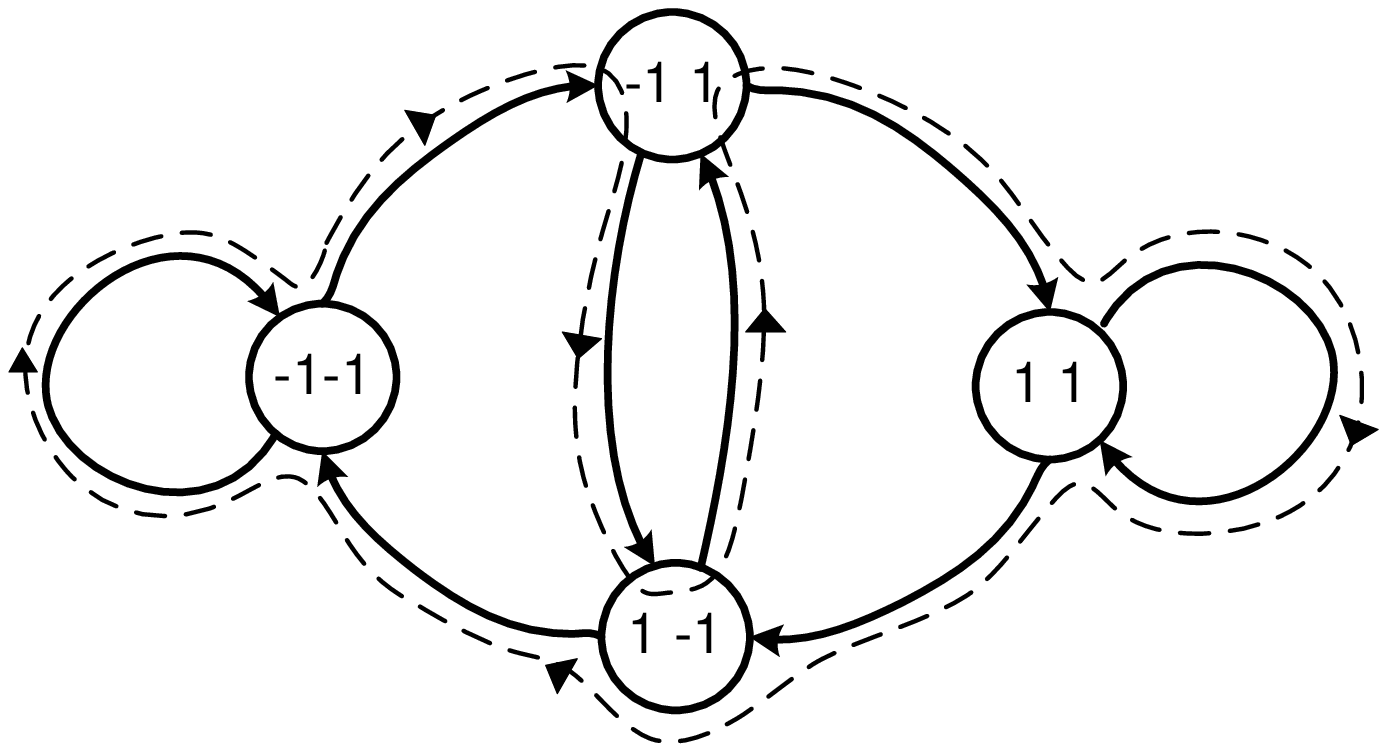}
\end{tabular}
\end{center}
\vspace*{8pt} \caption{\label{fig:DeBruijn_m12}\em De Bruijn graphs of orders $m=1$ (left) and
$m=2$ (right). Dashed lines represent examples of the Euler trails on the graph: one trail for
$m=1$ (left) and one of two possible Euler trails for $m=2$ (right)}
\end{figure}

We know that $U$ trajectories represent mean-reverting processes. Thus, the state space
(\ref{eq:state_x}) is projected onto the subspace $x(t)=\,\mu(t)$ and the dynamics of the MG can be
efficiently studied using de Bruijn graphs, as shown in Ref.~\cite{challet00PhysRevE62}. The
decision history $\mu(t)$ is a sequence of $m$ minority actions
\begin{eqnarray}
\mu(t)=\big[a^\ast(t-m),a^\ast(t-m+1),\ldots,a^\ast(t-1)\big]. \label{eq:deBruijn}
\end{eqnarray}
The $\mu(t+1)$ is obtained by adding $a^\ast(t)$ to the right and deleting $a^\ast(t-m)$ from the
left of the vector (\ref{eq:deBruijn}), such that there are two possible successors $\mu(t+1)$ of
$\mu(t)$. If one history can be obtained from another one using this procedure, then the latter has
a directed edge to the former one. Histories may be represented by labelled edges. These rules
define de Bruijn graph of the order $m$. Examples for $m=1$ and $m=2$ are given in
Figs~\ref{fig:DeBruijn_m12}.

Histories in MGs are not equiprobable~\cite{challet00PhysRevE62}. Among all paths on the de~Bruijn
graph of the game, Euler paths define the shortest sequence of histories where each strategy loses
and wins equally likely. In the non-Eulerian paths some histories are more frequent and therefore
some strategies are more profitable. We show in the following that in the efficient mode the
non-Eulerian paths are rare compared to the Eulerian ones.

\subsubsection{Algorithm generating strong demand fluctuations}\label{sec:StrongFluctuations} 
\begin{figure}[t]
\begin{center}
\includegraphics[scale=.40]{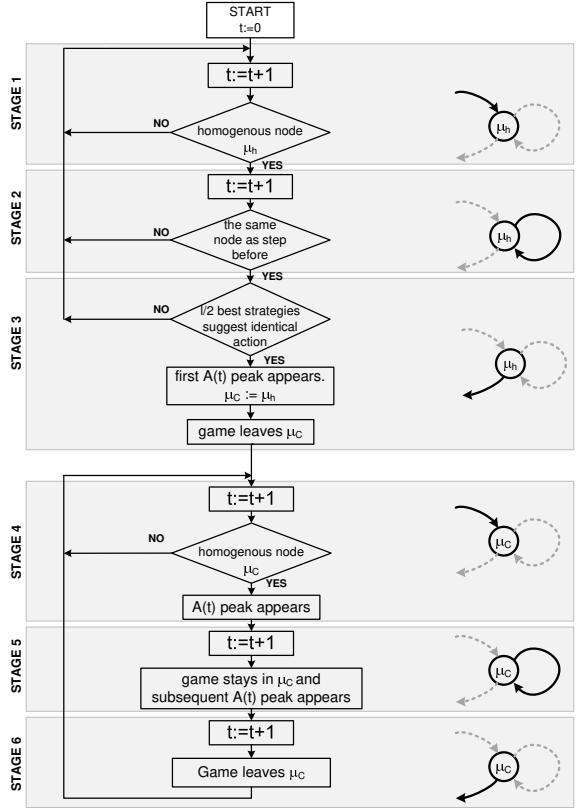}
\end{center}
%\vspace*{8pt}
\caption{\label{fig:pick_generation}\em The flow chart of the MG evolution algorithm,
illustrating appearance of distinct peaks of demand.}
\end{figure}

We
noticed that large fluctuation of $A$ is only possible if the game is in one of two de Bruijn nodes
called {\it homogeneous}, i.e. consisting of identical symbols:
$\mu_{h1(2)}=\big[-(+)1,\ldots,-(+)1\big]$. Interesting enough, peaks are observable only after one
of the homogenous histories, but not after both. In Fig.~\ref{fig:pick_generation} we present the
flow chart illustrating appearance of strong fluctuations of $A(t)$.
Below we describe the algorithm step by step. First three stages lead to the first peak. Next steps
explain why the subsequent peaks follow each other and why they have opposite signs.
\begin{itemize}
\item[\underline{Stage 1}]
~\\
If $A(t_1)$ stands for the first peak of demand then three prior conditions have to be fulfilled.
The first is that $\mu(t_1 - 1)=\mu_{h1(2)}$, where $\mu_{h1(2)}=[-(+)1,\ldots,-(+)1]$ is a
homogeneous node. \item[\underline{Stage 2}]
~\\
It is also required that at $t_1-1$ majority of agents decides to change the node. If this is
fulfilled then the minority action is
\begin{eqnarray}
a^\ast(t_1-1)=\left\{ \begin{array}{rr} -1, & \quad \mu(t_1-1)=\mu_{h1} \\
1, & \quad \mu(t_1-1)=\mu_{h2} \end{array} \right .. \label{eq:MinorityPeakGeneration}
\end{eqnarray}
Hence $\mu(t_1) = \mu(t_1-1)$, the minority action is to stay in the same node and gives the
positive payoff to the winning strategy
\begin{eqnarray}
R_{\alpha_n^s}(t_1-1)=-a_{\alpha_n^s} A(t_1-1). \label{eq:PayoffPeakGeneration}
\end{eqnarray}
~\\
\item[\underline{Stage 3}]
~\\
There is a non-zero probability that strategies corresponding to the first $l/2$ utilities in
$\{u_l\}$ have won in the last step. Such circumstance is possible provided stages 1 and 2 are
realized. If this third condition is fulfilled then we mark such history $\mu_C$. Then all first
$l/2$ strategies suggest the same reaction after $\mu_C$. Hence the majority decision at $t_1$ is
to stay in the node and the maximal demand (cf. Eq. (\ref{eq:TheLargestProof})) is generated. All
strategies with high utility get the penalty and the low-utility ones are rewarded by the same
amount. The game follows the minority decision and escapes from the de Bruijn node $\mu_C$. When
the game leaves $\mu_C$, the strategy set is split into two groups of high and low utility, as
illustrated in Fig.~\ref{fig:UandA_Ns1600m2linearZoom}. In the next steps the game goes to
$\mu\neq\mu_C$.
\begin{figure}[h]
\begin{center}
\begin{tabular}{cc}
\includegraphics[scale=.40]{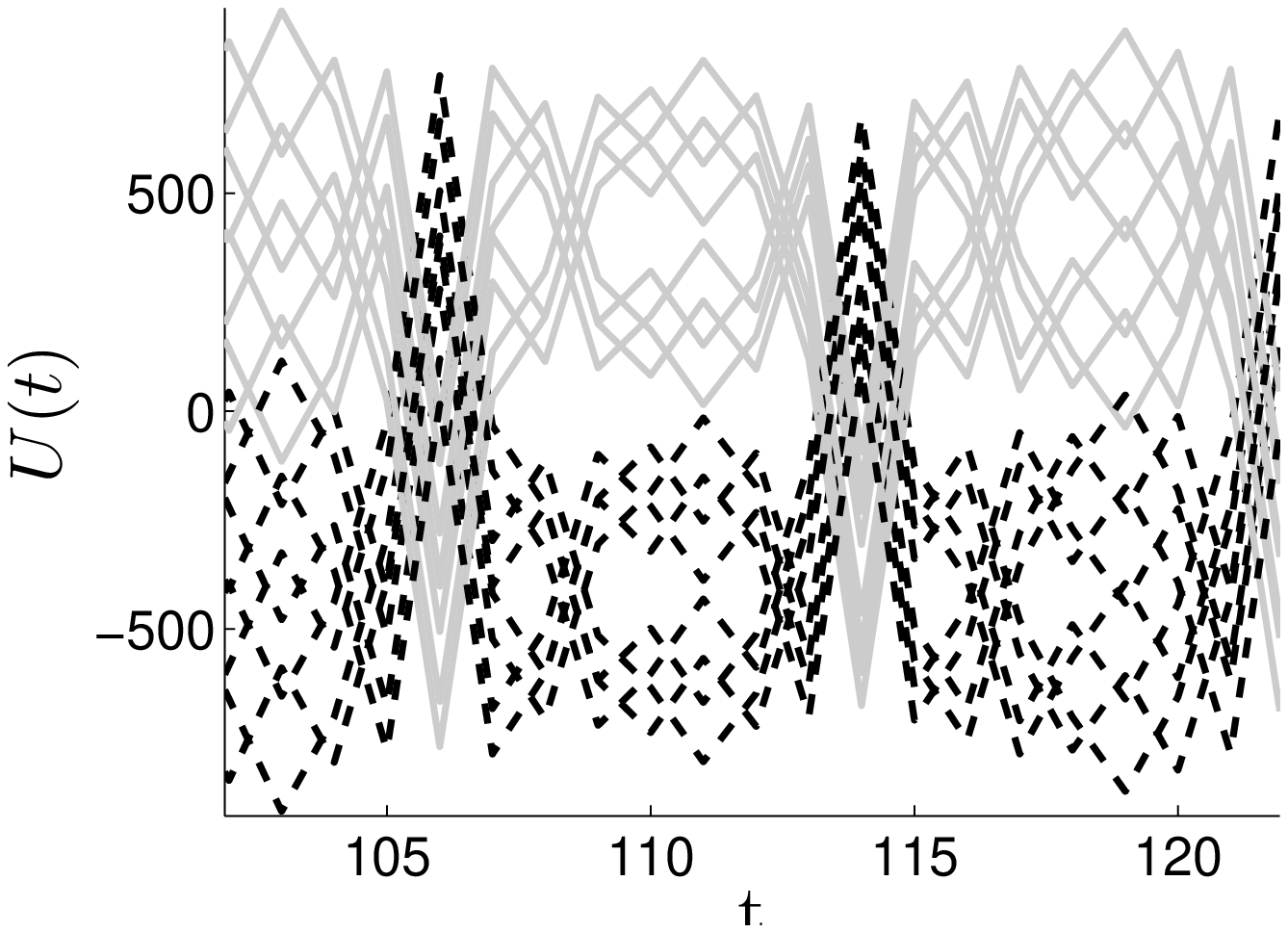} &
\includegraphics[scale=.28]{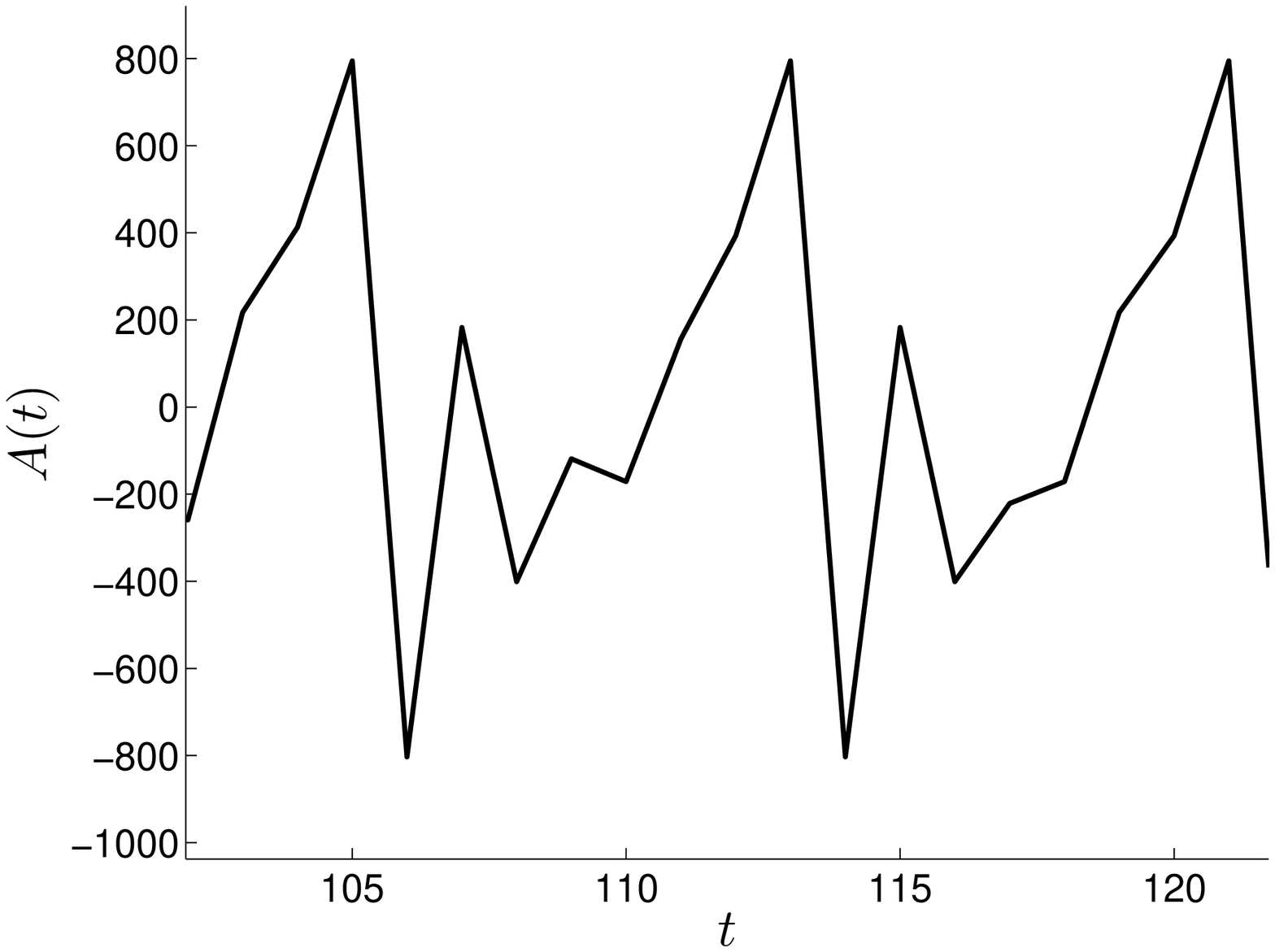}
\end{tabular}
\end{center}
\vspace*{8pt} \caption{\label{fig:UandA_Ns1600m2linearZoom}\em The time evolution of the utilities
(left) and the aggregated demand (right) for the MG with $N=1601$, $S=2$, $m=2$ and $g(x)=x$. Grey
solid trajectories represent strategies reacting in the same way after particular history $\mu_C$.
Black dashed trajectories represent anti-correlated strategies reacting in the opposite way after
$\mu_C$. The appearance of $\mu_C$ is in $t=105$ and $106$ and then after every $2^{m+1}$ steps.}
\end{figure}
\item[\underline{Stage 4}]
~\\
Next steps do not substantially affect utilities as long as the history $\mu_C$ does not reappear.
There is no history other than $\mu_C$ assuring that the first $l/2$ strategies in the $\{u_l\}$
list suggest a collective action resulting with the most spiky demand. Hence, after $t_1$, the
variations of $A$ do not affect the utility significantly until the $\mu_C$ reappears at $t_2>t_1$
and when the set of the best $l/2$ strategies is the same as at $t_1$. Then the $l/2$ best strategies
suggest the game to shift to another node characterized by history $\mu(t_2+ 1) \neq \mu_C$ and the
maximal demand $|A(t_2)| = N(1-\frac{1}{2^{S-1}})$ is generated. All the $l/2$ best strategies get
penalty proportional to the absolute value of the aggregated demand. Concurrently, the $l/2$
strategies with the lowest utility are rewarded with the same amount (cf.
Fig.~\ref{fig:UandA_Ns1600m2linearZoom}). \item[\underline{Stage 5}]
~\\
Next, the game follows the edge leading to the same node. Subsequently, the $l/2$ best strategies
suggest staying in the same vertex $\mu_C$. Again, high absolute value of demand is generated but
the sign of $A(t_2+1)$ is opposite to the sign of $A(t_2)$. Consequently, all strategies with high
$U(t_2 + 1)$ get penalty $N(1-\frac{1}{2^{S-1}})$ and, concurrently, strategies with low utility
get reward of the same size. \item[\underline{Stage 6}]
~\\
The game goes to the vertex $\mu_C(t_2 + 2) \neq \mu_C$ and stages 4--6 repeat.
\end{itemize}

Since high $A(t)$ appears only after the history $\mu_C$, we have just two transitions in the
Eulerian path starting from this history. From this it follows that the frequency of peaks is
equal to
\begin{eqnarray}
f & = & \frac{2}{2^{m+1}} \nonumber \\
  & = & \frac{1}{2^m},
\label{eq:peakFrequency}
\end{eqnarray}
in agreement with our simulations. The value $2^{m+1}$ is the length of the Euler path and it
corresponds to the period of $A$ observed in Figs.~\ref{fig:A3_linear}-\ref{fig:AagainstA3_linear}.

\subsubsection{The Markov process representation}
%\newline
%\newline
\noindent{\it The case of equal-size fractions}

\noindent As pointed out in Sec. 4.2, rewards and penalties have to compensate if the game is
stable. This requires specific order of states (cycle), such that every $\mu$ has to appear twice
over the cycle, in order to assure the same magnitudes of reward and penalty for any strategy. Such
cycles are considered as attractors because, as we will see, they tend to pull in other initial
states. The question is: how many attractors exist and how one can find them? At least two ways of
dealing with the problem are possible for equal-size fractions. The first is a brute force method
where for each state its successor is determined. But usefulness of this method is limited only to
small $m$. Another approach requires analysis of the Euler paths on de Bruijn graph and is
applicable for any $m$. We will show subsequently that the number of attractors is two times larger
than the number of Euler paths. Below are examples of both methods.

We present the brute force method for $m=1$ and strategies defined as in Tab.~\ref{tab:strategies}.
For simplicity, we use abridged notation for the state, e.g. $-3412$ stands for $[-1,3,4,1,2]$.
Each state has to be analyzed and its successor has to be found. Fig. \ref{fig:basin} presents
relations between states. There are two attractors:
\begin{figure}[h]
\begin{center}
\includegraphics[width=0.85\linewidth]{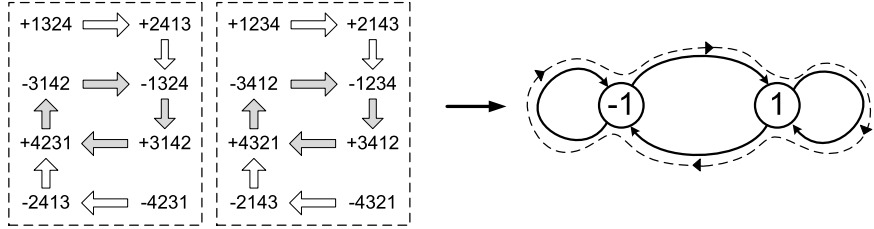}
\caption{\em Two basins of attraction for $m=1$ (left). Attractors are marked by grey arrows. Both
attractors are projected to the same Euler path in de Bruijn graph (right). For simplicity, we use
abridged notation for the state, e.g. $-3412$ stands for $[-1,3,4,1,2]$.} \label{fig:basin}
\end{center}
\end{figure}
\begin{eqnarray}
\mbox{Attractor 1} & = &  [+4231, -3142, -1324, +3142]   \\
\mbox{Attractor 2} & = &  [+4321, -3412, -1234, +3412]
\end{eqnarray}
Both attractors are equally possible, provided that $U(t=0) = \mbox{const}$ for all strategies.
Each attractor assures that every possible history appears twice. One appearance rewards half of
strategies and another one penalizes them. Each reward and subsequent penalty are of the same
magnitude. Moving along attractors assures that the game follows the Euler trail in the de Bruijn
graph, consistently with results of Refs.~\cite{jefferies01PhysRevE65,challet00PhysRevE62}. An
example of $U$ trajectories corresponding to these attractors is presented in Fig.
\ref{fig:attractors}.
\begin{figure}[h]
\begin{center}
\includegraphics[width=.85\linewidth]{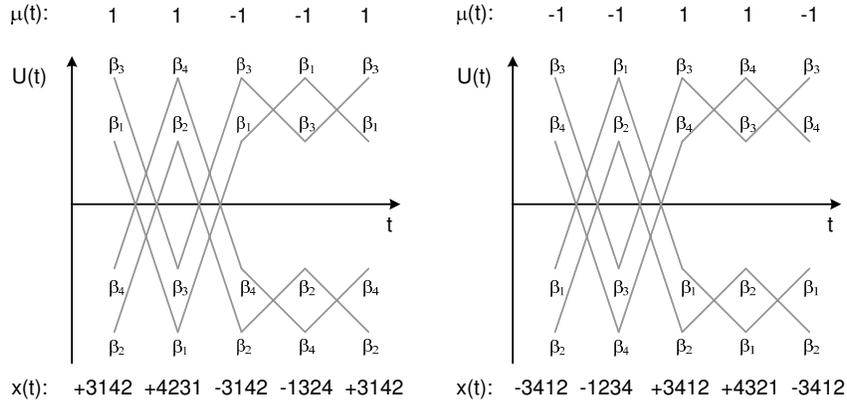}
\caption{\em Utility trajectories for two possible attractors for $m=1$.} \label{fig:attractors}
\end{center}
\end{figure}
The absolute changes of utilities in analyzed case $m=1$ are equal to one of two values: $N/2$ or
$N/4$, depending on state. In the former case, both best strategies suggest the same action. Thus
the $3/4$ of population acts according to these actions and an aggregate demand is equal to $|A| =
\frac{3}{4}N - \frac{1}{4}N = \frac{N}{2}$. In the latter case, the first and the third strategy
suggest the same action. Hence, the $\frac{10}{16}$ of the population chooses the same action and
consequently $|A| = \frac{10}{16}N - \frac{6}{16}N = \frac{N}{4}$. Exemplified realization for
$m=1$ is shown in Fig.~\ref{fig:A_linearRegular}.
\begin{figure*}[t]
\begin{tabular}{cc}
\includegraphics[scale=.27]{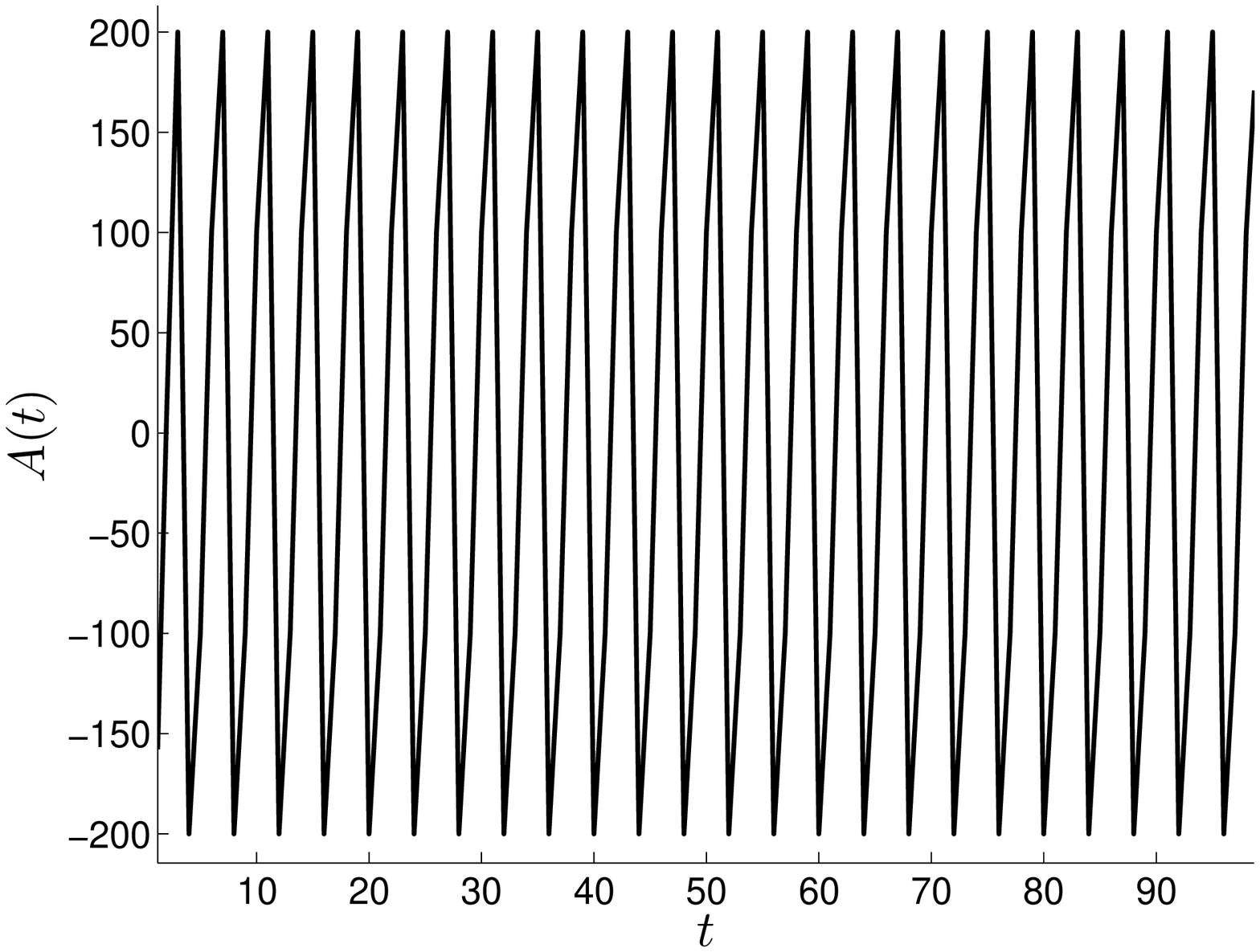} &
\includegraphics[scale=.27]{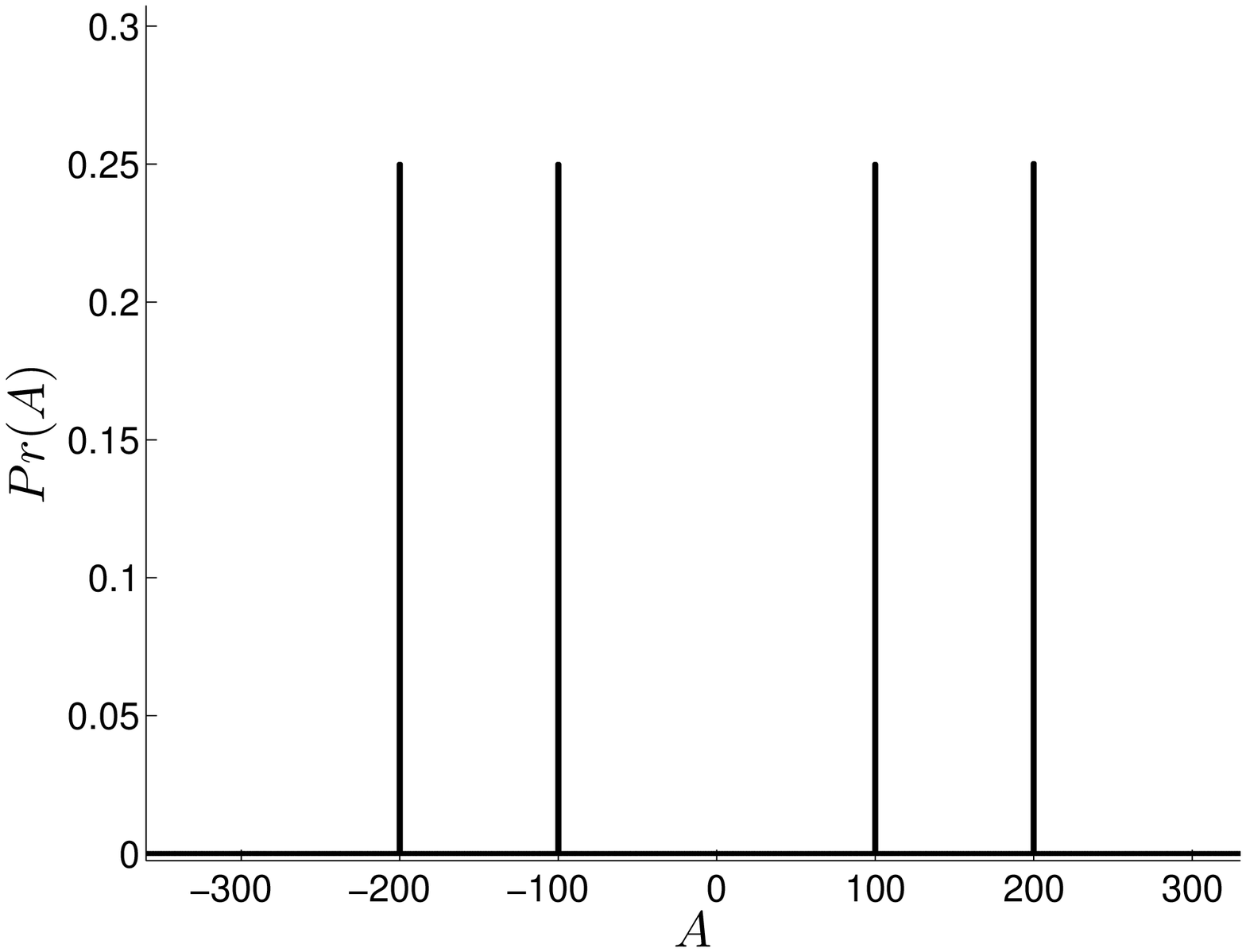}\\
\includegraphics[scale=.27]{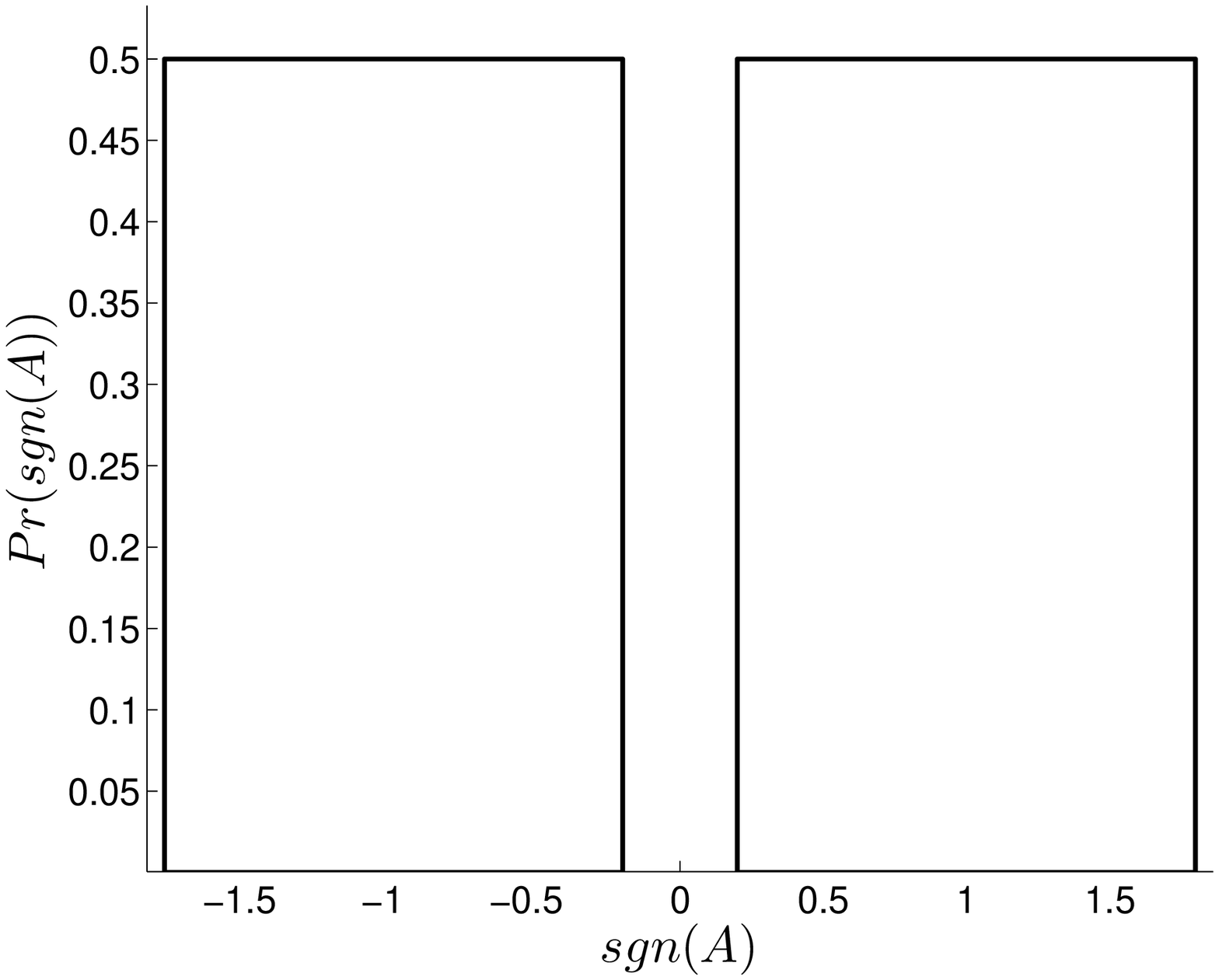}
\end{tabular}
\caption{\label{fig:A_linearRegular}\em Time evolution of the aggregated demand $A(t)$, estimated
$Pr(A)$ and $Pr(\mbox{sgn}(A))$ for the population size $N = 400$, agent memory $m = 1$, $S=2$
strategies per agent, identical sizes of fractions (reference system) and linear payoff $g(x) = x$}
\end{figure*}
\begin{figure*}[t]
\begin{tabular}{cc}
\includegraphics[scale=.27]{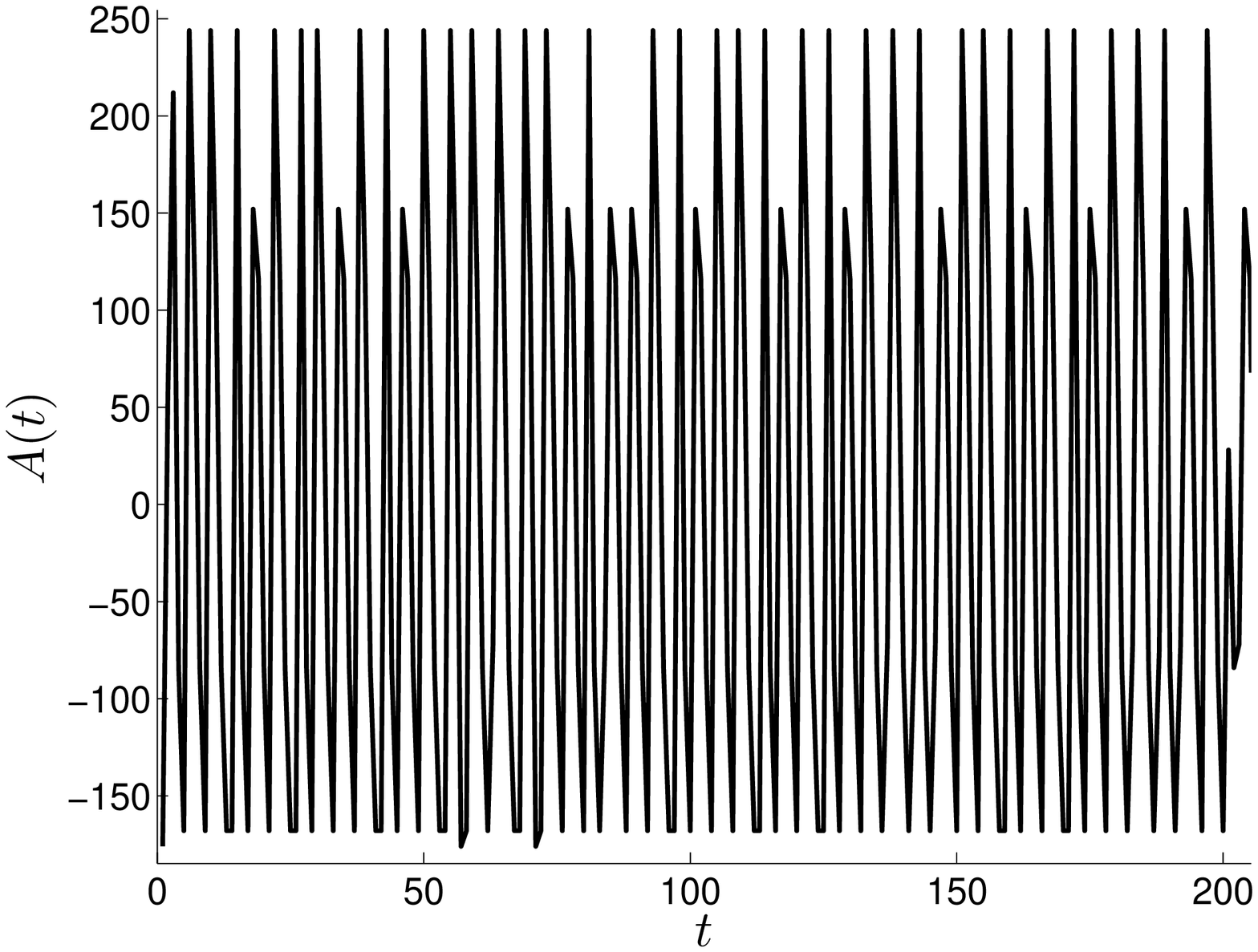} & \hspace{5mm}
\includegraphics[scale=.27]{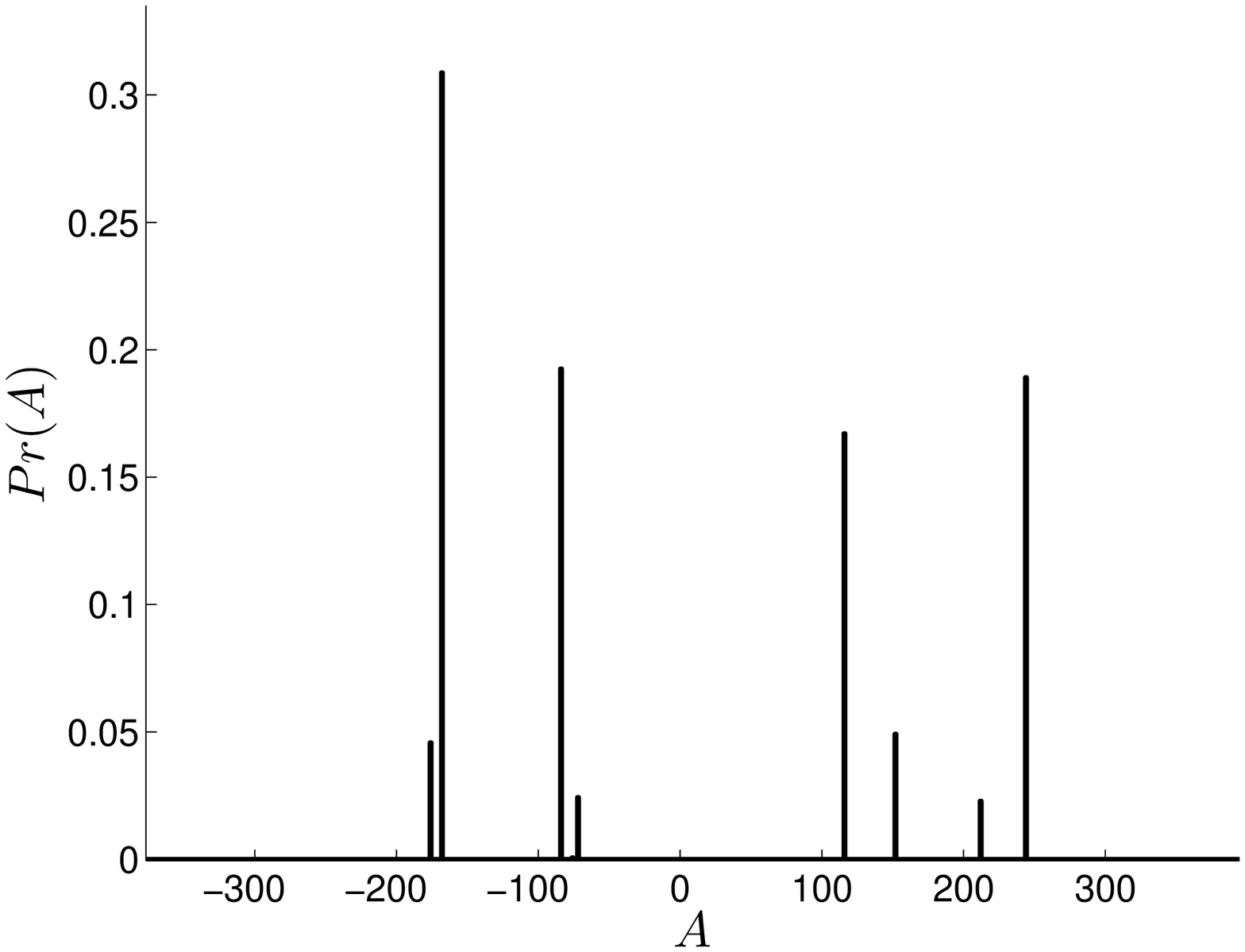}\\
\includegraphics[scale=.27]{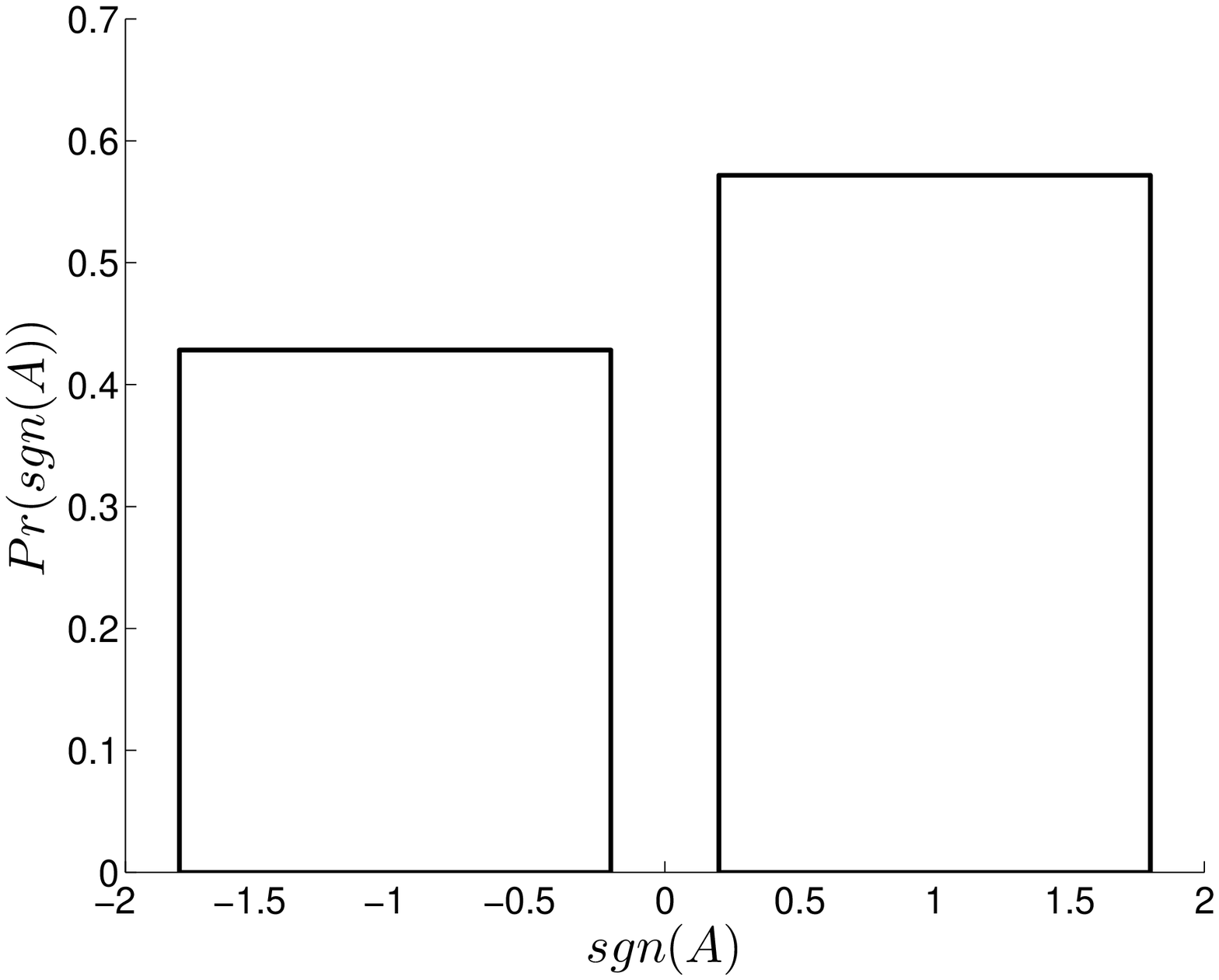}
\end{tabular}
\caption{\label{fig:A_linearIrregular}\em Time evolution of the aggregated demand ${A}(t)$, plots
of the estimated $Pr({A})$ and $Pr(\mbox{sgn}({A}))$ for the population size $N = 400$ and agent
memory $m = 1$, $S=2$ strategies per agent and unequal sizes of fractions.}
\end{figure*}
It is seen that both distributions, $A$ and $\mbox{sgn}(A)$, are symmetric. Since the game is fully
deterministic, each of four states $x_1 \ldots x_4$ is related to only one value of $A(x)$. Hence,
the $A$ distribution has four peaks.

More general way to determine the number of attractors is to count the number of Eulerian paths in
de Bruijn graph. Each attractor consists of the unique set of states that do not appear in other
attractors. We proved in Sec.~\ref{sec:StrongFluctuations} that each attractor comprises of exactly
one state characterized by the large oscillation \footnote{A large oscillation is explicitly
connected with a state characterized by half of best strategies suggesting the same action.} $|A| =
N(1-\frac{1}{2^{S-1}})$. This state has to incorporate the $\mu$ representing one of
the two possible homogenous nodes of the de Bruijn graph. As a consequence, there are two different
states belonging to two different attractors where both attractors are projected on the same
Eulerian path in de Bruijn graph. According to the theory of de Bruijn sequences, there is
$2^{2^{m}}/2^{m+1}$ Eulerian paths~\cite{aardenne51Circuits}. Hence, there is twice that many
attractors, $2^{2^m}/2^m$, e.g. there are $2,4$ and $32$ attractors for $m=1,2$ and $3$,
respectively.
%\vspace{3mm}
%\subsubsubsection{The case of unequal-size fractions}
\newline
\newline
\noindent{\it The case of unequal-size fractions}

\begin{figure}[h]
\begin{center}
\includegraphics[width=0.65\linewidth]{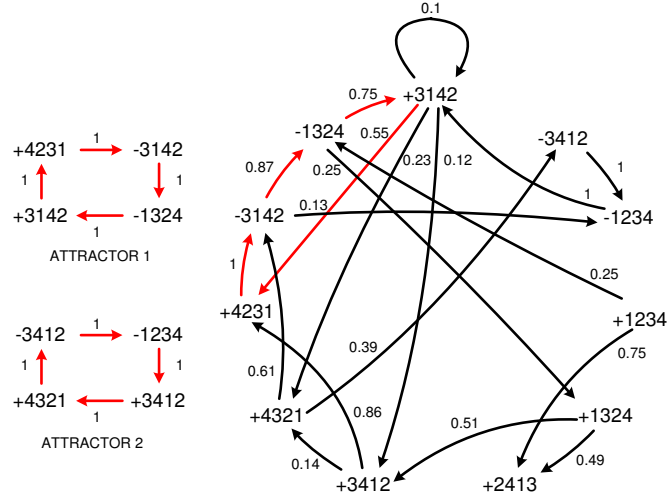}
\caption{\em Two possible attractors for $m=1$ (left) for game with equal sizes of fractions. The
transition graph for a real game where sizes are unequal, i.e. strategies are drawn from uniform
distribution (right).} \label{fig:stagesGfX}
\end{center}
\end{figure}

\noindent The size of different fractions most likely varies for strategies drawn randomly. This
shifts the \emph{a posteriori} ${A}$ distribution with respect to that of the reference system. The
mechanism is the same as for the steplike payoff. Consequently, in each state belonging to the
attractor, the values of $A$ are different than in the case of equal-size fractions. If the game
follows an attractor, the $A$ would not compensate to zero along the path and the utility values
would grow or shrink indefinitely. The minority mechanism stabilizes the game and prevents such
scenario by adding states to the attractor.
 Exemplified realization for the case, where strategies are drawn from uniform distribution, is shown in Fig. \ref{fig:A_linearIrregular}.
It is seen that both $Pr({A})$ and $Pr(\mbox{sgn}({A}))$ are asymmetric if distributions are
considered \emph{a posteriori}. The comparison of Markov chains, where sizes of fractions are equal
and different, is shown in Fig. \ref{fig:stagesGfX}.
It is seen that the game with unequal fractions mostly follows attractor~$1$ (red arrows) but in
three of four states transitions to other states can appear either. The probability of these
transitions is relatively small, indicating that sizes of fractions do not differ a lot. The MP
representation for unequal fractions is different for each realization.

\subsubsection{The variance per capita $\sigma^2/N$} We proved in Sec. \ref{sec:StrongFluctuations}
that large oscillations are periodic and equal to:
\begin{eqnarray}
|A| = N\Big(1-\frac{1}{2^{S-1}}\Big).
\end{eqnarray}
In particular, if $S=2$ then $\sigma^2  \sim \frac{N^2}{4}$, consistently with observations and
results of Ref.~\cite{hart01EurPhysJB20}.

The argumentation in Sec. \ref{sec:StrongFluctuations} becomes strict and
Eq.~(\ref{eq:TheLargestProof}) is exact in the efficient mode when $NS\gg 2^P$, ideally in the
limit $NS\rightarrow\infty$. But we also observe cyclic peaks of demand for $N=1601$ and $m=5$,
when the efficiency condition is not met (cf. Fig.~\ref{fig:A3_linear}, right). In fact, the
condition $NS\gg 2^P$ can be slacken off to the requirement that the population is numerous enough
that the game is in the herd mode. Games in that mode do not follow Eulerian paths because for
smaller $N$ the pool of strategies is too sparse and some histories occur more frequently.
Nevertheless, the mechanism of peak creation is approximately preserved, as long as $N$ is large
enough to cause the split of utilities into two groups.
\begin{figure}[h]
\begin{center}
\begin{tabular}{cc}
\includegraphics[scale=.40]{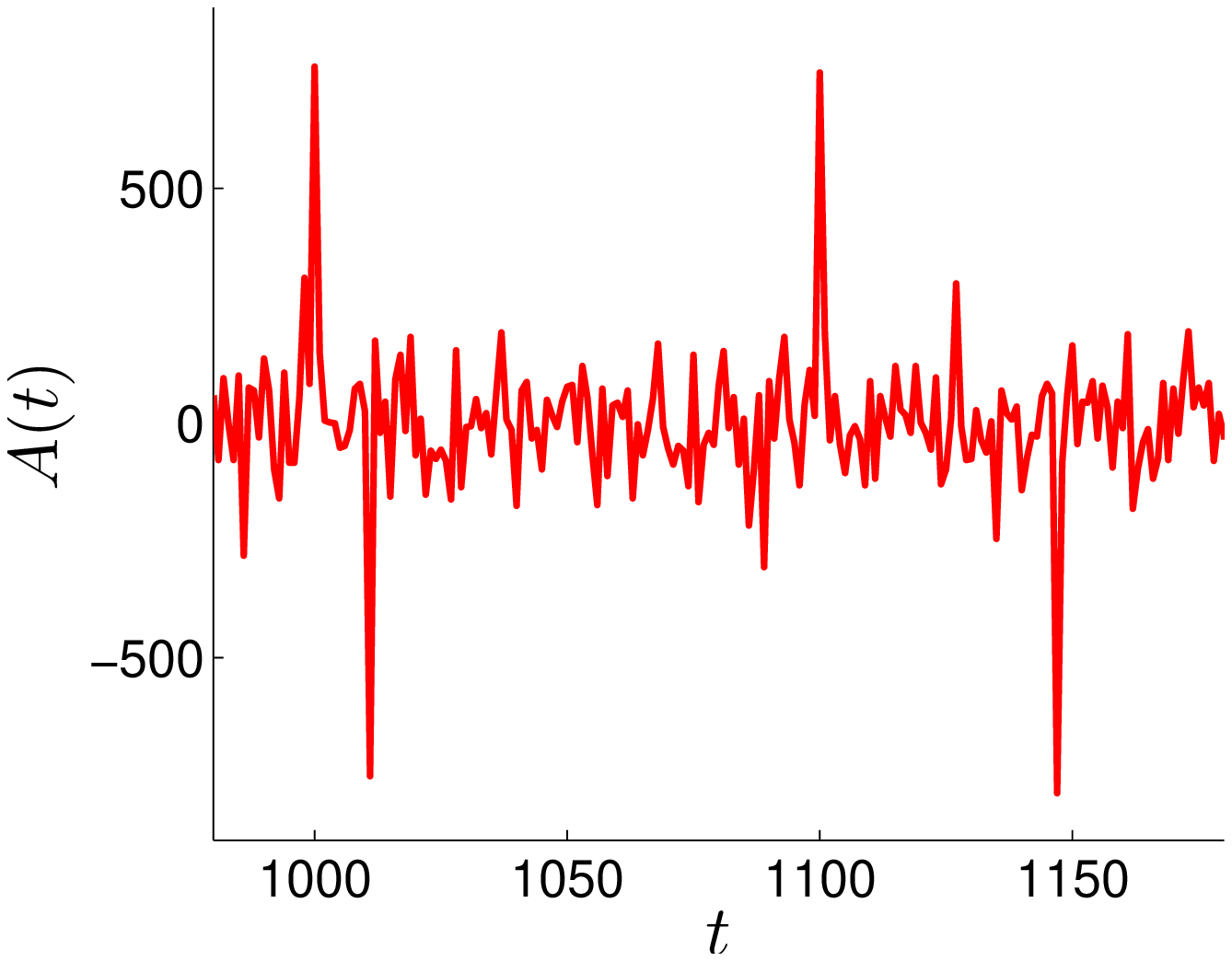} & \includegraphics[scale=.33]{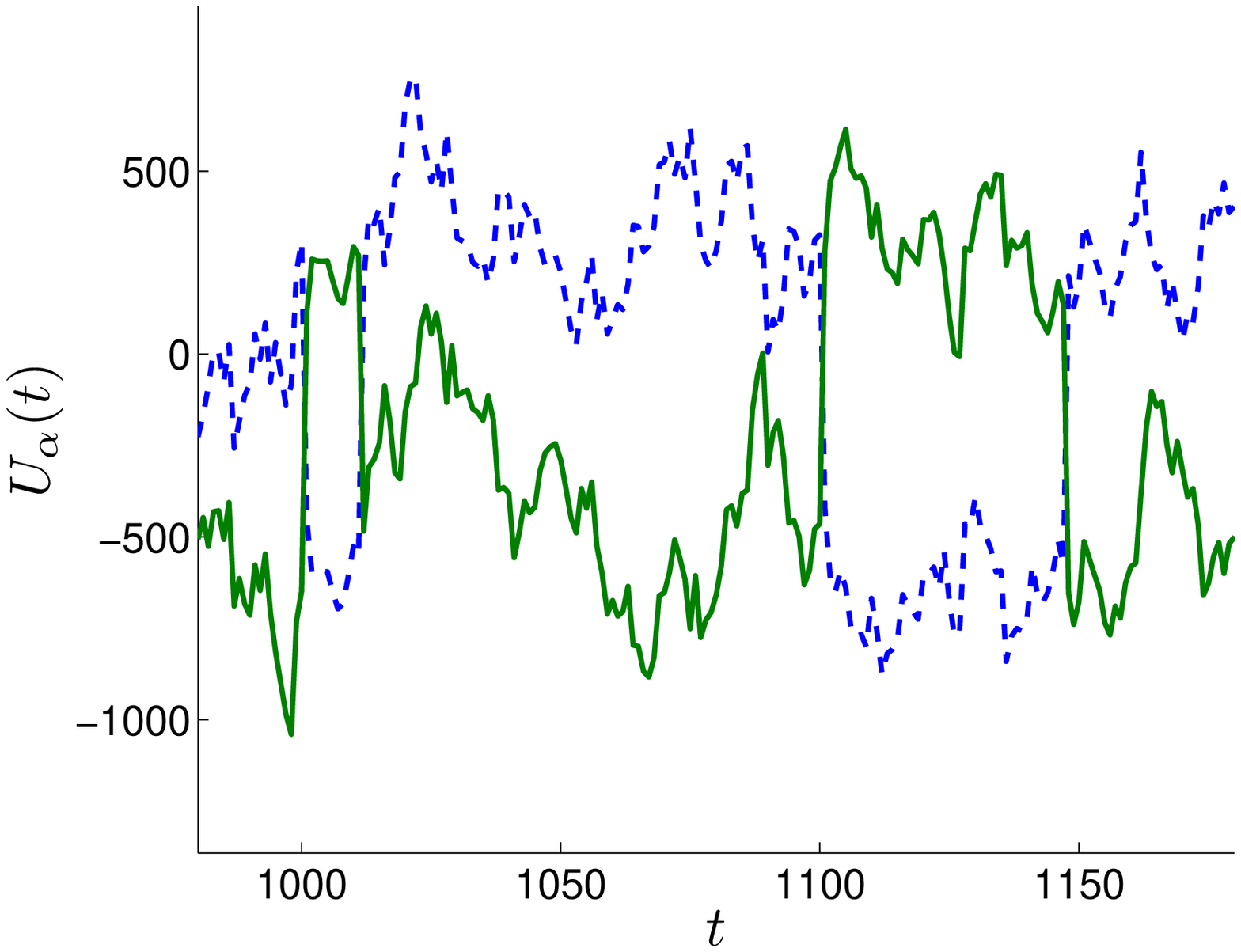} \\
\includegraphics[scale=.33]{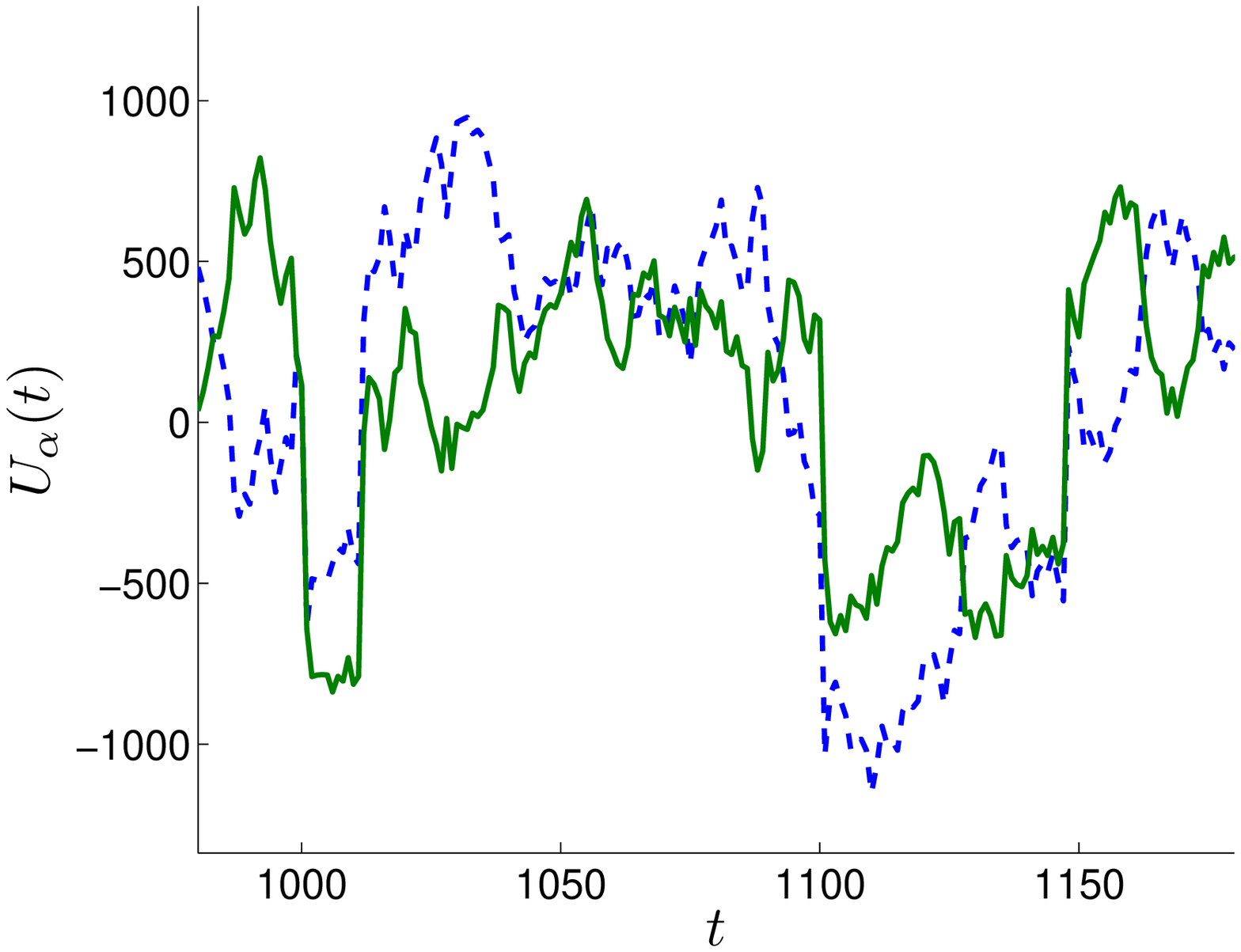} & \includegraphics[scale=.33]{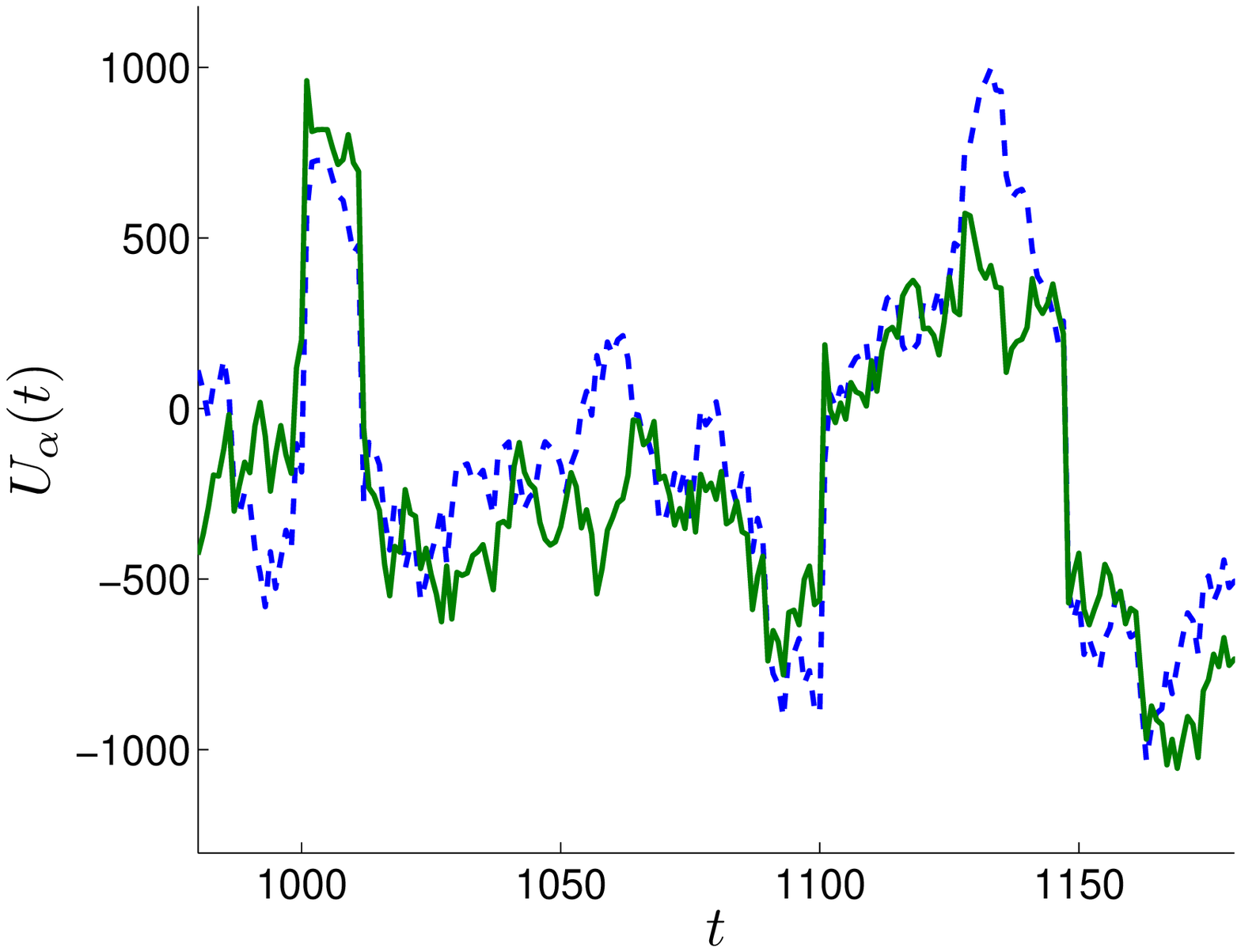}
\end{tabular}
\end{center}
\vspace*{8pt} \caption{\label{fig:tEvolutionOfAUSingleMarket}\em The time evolution of the
aggregated demand (upper left) and utilities for three cases: an agent with one high- and one
low-utility strategy (upper right), two low-utility strategies (lower left) and two high-utility
strategies (lower right) at $t=1000$. These three cases may be quantitatively distinguished using
the values of utilities at $t=1000$, corresponding to the location of the first maximum of $A(t)$
in the upper left panel. Simulation was performed for the MG with $N=1601$, $S=2$, $m=5$ and
$g(x)=x$.}
\end{figure}
At any time a somewhat simpler explanation of large oscillations may be given by dividing
strategies into two categories: the {\it good} with the positive payoff, and {\it bad} with
negative. Probability that an agent has no good strategies, or at least one good, is equal to
$\frac{1}{2^{S}}$ and $1-\frac{1}{2^{S}}$, respectively. Rapid fluctuations of demand are
transferred to similar fluctuations of the utility. The $A(t_1)$ fluctuates after the history
$\mu_C=\mu(t_1)$ when the strategies with higher utility indicate identical actions. If $A(t_1)$
strongly fluctuates, then at $t_1+1$ about $N(1-\frac{1}{2^{S}})$ agents have at least one strategy
with high utility and they choose it.
Strategies split into two groups: the first group consisting of high utilities and the second of low utilities, with
a gap between these two groups (cf. Fig.~\ref{fig:UandA_Ns1600m2linearZoom}).
Strategies with utilities belonging to the same group do not suggest the same actions, provided $\mu\ne\mu_C$, and therefore no
peak of $A$ is generated.
The $\mu_C$ has a non-vanishing probability to reappear at some
$t_2>t_1$. All agents belonging to the group with at least one high-utility strategy tend to react
identically and $A(t_2)$ fluctuates maximally, i.e. $A(t_2)=N(1-\frac{1}{2^{S-1}})$. This is
illustrated in Fig.~\ref{fig:tEvolutionOfAUSingleMarket} (upper left), where for $S=2$ we have
$A(t=1000)=\frac{N}{2}$. At $t_2$, all strategies with high $U(t_2)$ fail and get the penalty
$-A(t_2)$, whereas those with low $U(t_2)$ are rewarded with  $A(t_2)$. After $t_1$ agents are
divided into three groups, provided $S=2$: the group with two good strategies, with one good and
one bad, and with two bad. As seen in Fig.~\ref{fig:tEvolutionOfAUSingleMarket}, at $t=1000$ a
quarter of the population with two high-utility strategies evolves into two low-utility groups
(lower left), and {\it vice versa} for another quarter with two initially low-utility strategies
(lower right). Remaining half of the population just swaps utilities of their strategies (upper
right).

Results showing periodicity of $A(t)$ from simulations become closer to the theoretical results for
large $NS/2^P$ ratio. If this ratio is small, then the game hardly follows the Eulerian path and peaks of
$A(t)$ appear randomly.

\subsubsection{Stability of the game and behavior of the predictability $H$} The behavior of $H_A$
is driven by absolute disproportions between fractions' sizes. The payoff is an explicit function
of $A$ and, in order to stabilize the game, the negative and positive payoffs following the same
$\mu$ have to compensate mutually. Hence, for any $\mu$: $\langle A^-|\mu\rangle = \langle
A^+|\mu\rangle$ and $\langle H_A \rangle = 0$. For this kind of payoff the same frequency of the
negative and positive payoffs do not have to be preserved as it is required for $\mbox{sgn}(x)$
(see Fig. \ref{fig:A_linearIrregular}, bottom left).

The last point to understand is the plot of $H_a/N$ that seems to be equal to zero in the herd
regime. The $H_a$ is the sum of $\langle a^*|\mu\rangle$ over $P$ different $\mu$'s. Each of these
components is most likely nonzero and is bounded: $\langle a^*|\mu\rangle \in [-1, 1]$. Thus
$\mbox{max}(H_a) = \mbox{const} =P$ and in the limit $N \rightarrow \infty$ one has $H_a/N = 0$.

\subsection{The effect of imbalance between fractions} One can try to measure how the size of
disproportion between fractions affects transition probabilities in the Markov chain. To this end
we incorporate a measure of the distance between two arbitrary processes. Denoting the set of
reference processes by ${\cal R}$ and the set of examined ones by ${\cal E}$, this measure is
defined as
\begin{eqnarray}
\Upsilon=\sum_{i\in{\cal E \bigcup \cal R}} \sum_{j\in{\cal E \bigcup \cal R}}\big{|}Pr^{\cal
E}(x_i)Pr^{\cal E}(x_j|x_i) - Pr^{\cal R}(x_i)Pr^{\cal R}(x_j|x_i)\big{|}\label{eq:Distance}.
\end{eqnarray}
where $Pr^{\cal R}$ and $Pr^{\cal E}$ stand for the probabilities for the reference and examined
system. The $\Upsilon \in [0,2]$ is suitable to compare any MPs, comprising even such where
processes are based on different sets of states. If $\Upsilon=0$, then there are no differences
between processes. If $\Upsilon=2$, then processes are based on strongly disjunctive sets of
states.
\begin{figure*}[h]
%\begin{tabular*}{0.5\textwidth}{ccc}
\begin{tabular}{ccc}
\includegraphics[scale=0.27]{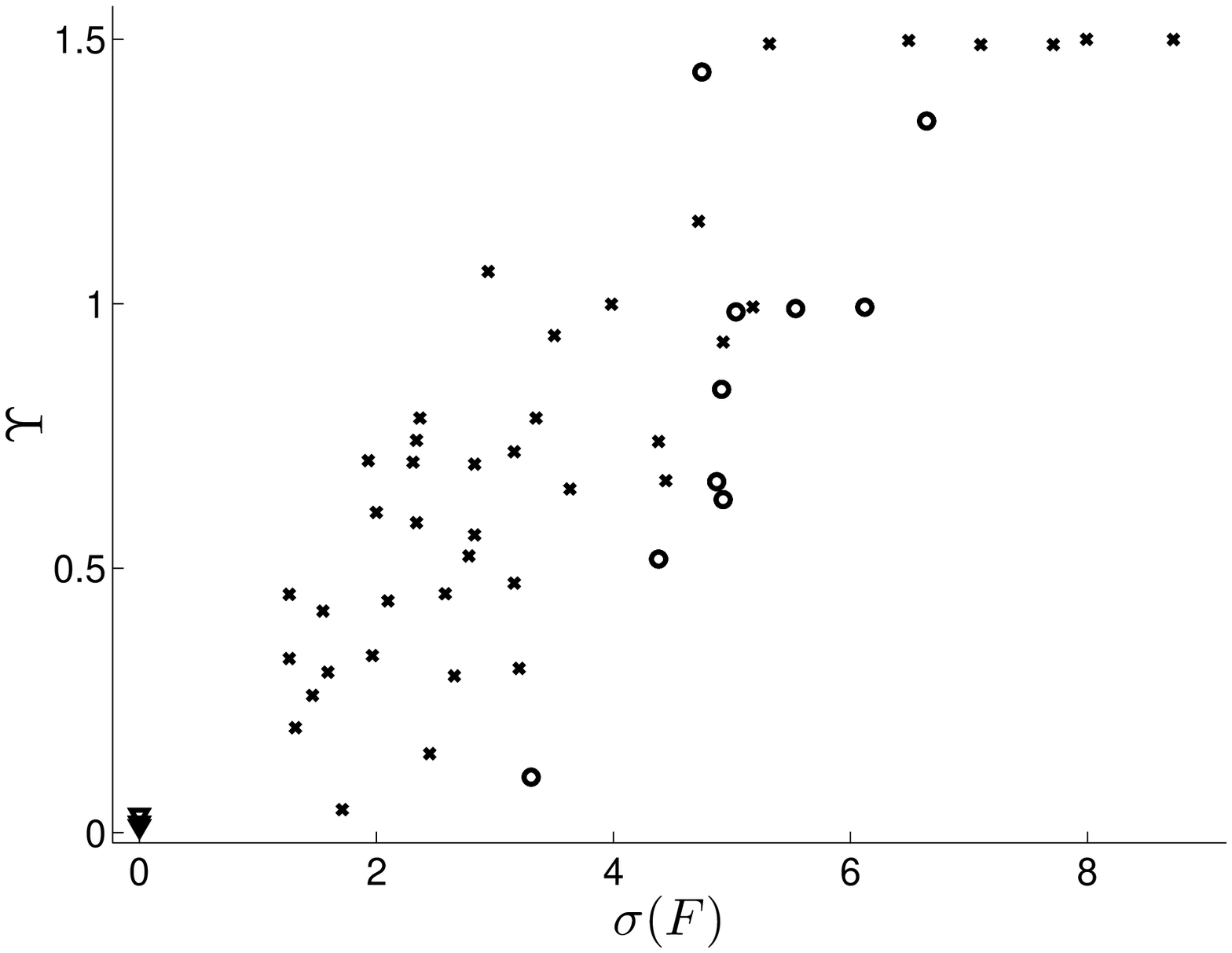} & \hspace{0mm}
\includegraphics[scale=0.27]{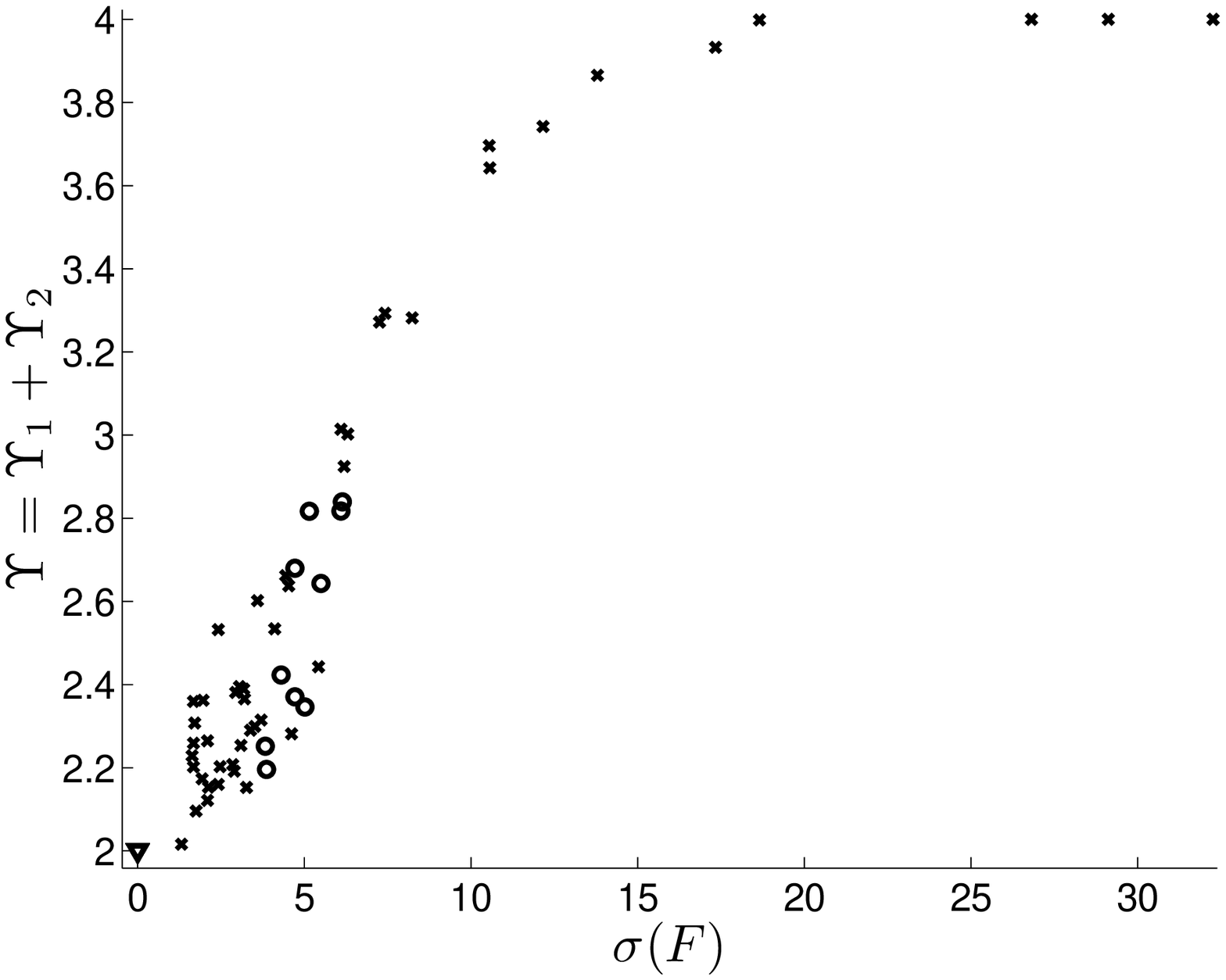}
\end{tabular}
\caption{\label{fig:DfStdF}\em Distance $\Upsilon$ as a function of $\sigma(F)$ for two different
payoffs $g(x) = \mbox{sgn}(x)$ (left) and $g(x) = x$ (right).}
\end{figure*}
The standard deviation $\sigma(F)$ is a measure of disproportion of fractions. The $Upsilon$ as a
function of $\sigma(F)$ is presented in Fig. \ref{fig:DfStdF}. The left panel presents
$\Upsilon$ measured between the reference MP (left-hand diagram in Fig.~\ref{fig:stages}) and 40
games where strategies are drawn from various distributions, provided $g(x)=\mbox{sgn}(x)$. In the
case $g(x)=x$, the function is more complicated because we do not have just one MP representing the
reference system but for $m=1$ there are two equiprobable attractors, corresponding to two MPs.
Therefore we use the sum of $\Upsilon_1+\Upsilon_2$ as a function of imbalance between fractions,
as presented in Fig. \ref{fig:DfStdF} (right). If the game follows attractor 1 then $\Upsilon_1 =
0$ and $\Upsilon_2 = 2$.

\section{Conclusions}
%\addcontentsline{toc}{chapter}{Conclusion}
%
In this paper we proposed a consistent, reductionist scheme explaining phenomenology of minority games in the efficient regime. In this mode the size of strategy space is much smaller than the number of strategies used by agents and the population as a whole can access complete information about the game.

Our discussion begun with the phenomenology. We considered a number of macroscopic random variables, or their moments, characterizing the game and being particularly important for applications, such as the aggregated demand, demand's variance \textit{per capita} and decision's or demand's predictabilities. We studied these variables as functions of the control parameters, e.g. the ratio of the total number of agents to the number of all possible winning histories, as well as their time evolution. Among interesting features we found that predictabilities may, or may not, be sensitive to the form of the payoff function, depending on how the predictabilities are actually defined: using winning decisions only or the overall demand.

Deeper insight into the mechanism of these behaviors was possible by performing coarse-graining and aggregation of some internal degrees of freedom of the game, thus defining an intermediate level of description, called mesoscopic.  At such mesoscopic level, fractions of agents using same strategies are treated as separate entities. Using this method, in the efficient regime when $NS\gg 2^P$, we also managed to represent the game as a Markov process with the finite number of states.

In case where the Markov representation is known, two methodologies were proposed to explain our observations. First, in the simplified case, the quenched disorder was neglected, i.e. fractions were assumed to be equal size. In this case, however, not all observations are properly explained. Behavior of predictability required extended methodology where the quenched disorder was used. Two payoffs, the steplike and linear, were separately analyzed. We showed that in case of the steplike payoff, the stochastic and deterministic transitions were possible, whereas for the linear payoff, all transitions were deterministic.

We argued that the Markov process representation of the game completely defines and explains the dynamics of the game in the stationary regime, and allows for the calculation of state occupancies. If the transition probabilities in the Markov chain are known, the phenomenology also becomes understandable. For example, the Markov representation provides an explanation of the
periodicity and preferred levels of the aggregate demand $A(t)$. In practical terms, this approach is tough for $m>1$ due to the large number of states but the whole reasoning remains valid in general. We failed to find any relation between the memory length $m$ and the total number of states.

Neither the simplified concept of state nor the Markov process description seem to be correct if the initial preference was given to any strategy. The definition of the stability was introduced in Sec.~\ref{sec:stability}, in order to better understand asymmetries observed for aggregated variables. The stability mechanism appeared to be sensitive to the payoff function. In case the steplike payoff was considered, then the frequency of opposite signs of $A$ after any $\mu$ had to be preserved. In case of the linear payoff, the negative and positive values of A had to compensate mutually. As a result, depending on the payoff, both the $H_a$ and $H_A$ were equal to zero in the herd regime.

Differences between systems with equal-size fractions and those with unequal-size fractions were more likely for larger numbers of agents. This was particularly reflected in distortions of: (i) the transition probabilities, in case of the steplike payoff, and (ii) attractors, in case of the linear payoff. In order to quantify this distortion, the measure of distance between two Markov processes was introduced.

We studied games with the full, maximal strategy space. Some authors, e.g. in Refs.~\cite{challet97PhysicaA246,li00PhysicaA284}, reduced the strategy space and reproduced many features of the full MG, e.g. behavior of variance per capita. The drawback of their method is that it reduces the number of states in the Markov-chain description of the game and significantly affects its time evolution. The Markov representation is oversimplified by such reduction.

Some observables for the proportional payoff were explained without using the Markov process. For example, there was an observation of distinct peaks of the aggregated demand, exhibiting height equal to a half of the population, assuming $S = 2$. We showed that in the herd regime, there
always exists a history $\mu_C$ for which the fraction $1-\frac{1}{2^{S}}$ of agents reacts identically and this is seen in the peak $A(t)=N(1-\frac{1}{2^{S-1}})$. Apart from using the Markov chain technique, we found another, simpler one, where only two classes of strategies were used instead of all $2^P$ classes. This technique is not limited to the case $N S \gg 2^P$ but works in the whole herd regime. This approach was also successfully exploited in our analysis of the multi-market minority game~\cite{wawrzyniak09ACS12No4and5}.

Considering further research, it could be a significant achievement if a single, closed-form equation were found for the entire parameter region of the MG. So far, in literature, such equations have indeed been found in the crowded and random regions separately. But a unified description of the entire range of parameters is still lacking.

%% References
%%
%% Following citation commands can be used in the body text:
%% Usage of \cite is as follows:
%%   \cite{key}          ==>>  [#]
%%   \cite[chap. 2]{key} ==>>  [#, chap. 2]
%%   \citet{key}         ==>>  Author [#]

%% References with bibTeX database:

\bibliographystyle{model1-num-names}
\bibliography{biblio}

%% Authors are advised to submit their bibtex database files. They are
%% requested to list a bibtex style file in the manuscript if they do
%% not want to use model1-num-names.bst.

%% References without bibTeX database:

% \begin{thebibliography}{00}

%% \bibitem must have the following form:
%%   \bibitem{key}...
%%

% \bibitem{}

% \end{thebibliography}

\end{document}